\documentclass[structabstract]{aa}   % used in 4 May 2022 
\usepackage{graphicx}
\usepackage{url}
\usepackage{txfonts}
 \usepackage{amssymb}
 \usepackage{natbib}
  \usepackage{lscape}
  \bibpunct{(}{)}{;}{a}{,}{,}
  \usepackage{times}
  \usepackage{float}
\usepackage[caption = false]{subfig}
 \def\allgaiaentries{18\,057\,300}
\def\totyso{449\,849}
\def\totcl{14\,178}
\def\totclmerge{7\,323}
\def\protoclusters{52}
\def\toplevelyso{302\,730}
\def\toplevelcl{1\,450}
\def\unclassyso{147\,119}
\def\unclasscl{5\,887}
\def\youngstars{124\,440}
\def\totsfr{354}
\def\midagestars{65\,863}
\def\totmidagecl{322}
\def\oldstars{43\,936}
\def\totoldcl{524}
\def\flatcamd{68\,491}
\def\totflatcamdcl{250}
\def\orionab{14\,832}
\def\orionlambda{1\,576}
\def\kerrmch{9\,351}
\def\kerrmchsfr{4\,676}
\def\kerrmchmidy{3\,914}
\def\kounkelamch{38\,567}
\def\kounkelamchsfr{23\,071}
\def\kounkelamchmid{9\,494}
\def\kounkelbmch{42\,350}
\def\kounkelbmchsfr{25\,511}
\def\kounkelbmchmid{9\,559}

  \def\deg{${}^\circ$}
  
  \def\sec{${}^{\prime\prime}$}
  \usepackage{longtable}
\usepackage{longtable,lscape}
  \usepackage{lscape}
  
\begin{document}
\def\hii{\ion{H}{ii} }
   \title{Low mass young stars in the Milky Way unveiled by DBSCAN 
     and {\it Gaia} EDR3. Mapping the star forming regions within 1.5\,Kpc}
 %  \subtitle{Mapping low mass young stars in the Milky Way unveiled by DBSCAN 
 %    and {\it Gaia} EDR3}

   \author{L. Prisinzano\inst{1}
   		\and
        F. Damiani\inst{1}
    		\and
          S. Sciortino\inst{1}
          \and
    		E. Flaccomio\inst{1}
          \and
          M. G. Guarcello\inst{1}
           \and
          G. Micela\inst{1}
           \and
          E. Tognelli\inst{2}
                    \and
          R. D. Jeffries\inst{3}
          \and 
          J. M. Alcal\'a\inst{4}
}
   \institute{INAF - Osservatorio Astronomico di Palermo, Piazza del Parlamento 1, 90134, Palermo, Italy \\
              \email{loredana.prisinzano@inaf.it}
 	\and
% 	Department of Physics `E. Fermi', University of Pisa, Largo Bruno Pontecorvo 3, I-56127 Pisa, Italy
CEICO, Institute of Physics of the Czech Academy of Sciences, Na Slovance 2, 182 21 Praha 8, Czechia
	\and
         Astrophysics Group, Keele University, Keele, Staffordshire ST5 5BG, United Kingdom
                 \and
         INAF - Osservatorio Astronomico di Capodimonte, via Moiariello 16, 80131 Napoli,Italy 
%                 \and
%         INAF - Osservatorio Astrofisico di Arcetri, Largo E. Fermi 5, 50125, Florence, Italy 
% 	    Florence, Italy
%          \and
%         INAF - Osservatorio Astrofisico di Catania, via S. Sofia 78, 95123, Catania, Italy 
%         \and
%         Dipartimento di Fisica e Astronomia, Universit\`a di Catania, via S. Sofia 78, 95123, Catania, Italy 
%         \and
%         Instituto de Astrof\'{i}sica de Andaluc\'{i}a-CSIC, Apdo. 3004, 18080 Granada, Spain
%          \and
%          % Instituto de F\'isica y Astronomi\'ia, Universidad de Valpara\'iso, Chile
% 	  %  \and
%          Dip. di Fisica e Chimica, Universit\`a di Palermo, P.zza del Parlamento 1, I-90134 Palermo, Italy
%          \and
%                 INAF - Osservatorio Astronomico di Bologna, via Ranzani 1, 40127, Bologna, Italy
%          \and
%          European Southern Observatory, Alonso de Cordova 3107 Vitacura, Santiago de Chile, Chile
%  	    \and 
%          Astrophysics Research Institute, Liverpool John Moores University, 146 Brownlow Hill, Liverpool L3 5RF, United Kingdom
% 	\and
% 	Departamento de Ciencias Fisicas, Universidad Andres Bello, Republica 220, Santiago, Chile
% 	    \and 
%  	INAF - Padova Observatory, Vicolo dell'Osservatorio 5, 35122 Padova, Italy    
         }
   \date{ }
% \abstract{}{}{}{}{} 
% 5 {} token are mandatory
 
  \abstract
   {With  an unprecedented astrometric and photometric data precision, 
   {\it Gaia} EDR3 gives us, for the first time, the opportunity to systematically
   detect and map in the optical bands, the  low mass populations of the star forming regions 
    (SFRs) in the Milky Way.}   
   {We provide a catalogue of the {\it Gaia} EDR3 data (photometry, proper motions and parallaxes) 
   of the young stellar objects (YSOs) identified 
   in the Galactic Plane ($|b|<30$\deg) within about 1.5\,kpc. 
   The catalogue of the SFRs  to which they belong
   is also provided  
    to study the properties of the very young clusters 
    and  put them  in the context of the  Galaxy structure.
   % We exploit the potentiality of the {\it Gaia} EDR3 photometry 
   % to consistently  select these objects. 
    }
  {We applied the machine learning unsupervised clustering algorithm DBSCAN on a sample of {\it Gaia} EDR3 data
  photometrically selected  on the region where very young stars
   ($t\lesssim$10\,Myr) are expected to be found,  with the aim to identify co-moving and spatially consistent stellar
   clusters. A subsample of  \protoclusters\ clusters, selected among the \totclmerge\
   found with DBSCAN, has been used as template data set, to identify  very young clusters from the pattern of the observed 
   color-absolute magnitude diagrams through a pattern match process.
   }
   {We find \youngstars\, candidate YSOs clustered in \totsfr\, SFRs and stellar
    clusters younger than 10\,Myr and within  $\lesssim1.5$\,Kpc.  
   In addition, \midagestars\, low mass members of \totmidagecl\, stellar clusters located within $\sim$500\,pc
    and with ages 10\,Myr$\lesssim t\lesssim$100\,Myr were  also found. }
  %In addition, we find \ysomg\, YSOs  associated to \totmg\, moving groups. }
   {The selected  YSOs  are spatially correlated with
   the well known SFRs.  Most of them are associated with well
   concentrated regions or complex structures of the Galaxy and
    a substantial number of them have been recognized for the first time. The massive 
   SFRs, such as, for example, Orion, Sco-Cen and Vela, 
   located within 600-700\,pc  trace a very complex three-dimensional pattern, while
   the farthest ones seem to follow a more regular  pattern  along the  Galactic
   Plane.} 
   %YSOs associated to moving groups are  distributed along the Galactic
   %Plane and those within $\sim300$\,pc are mainly concentrated toward the Galactic Center. }

   \keywords{methods: data analysis -- stars: formation, pre-main sequence
   			 -- Galaxy: open clusters and associations: general -- catalogs -- surveys}
\titlerunning{Mapping low mass young stars in the Milky Way with DBSCAN 
     and {\it Gaia} EDR3}
   \maketitle
%
%________________________________________________________________

\section{Introduction}
 It is by now well known that stars originate from collapse of cold molecular clouds, 
and mainly form in over-dense structures and clusters usually designated as star forming regions
 (SFRs).
During the very early phases, young stellar objects (YSOs) can be 
 identified in the near, mid far infrared (IR) and radio  wavelengths
because of the presence of the optically thick infalling  envelope or circumstellar disk 
around the central star. 
In the subsequent pre-main sequence stage phase, 
they become visible also in the optical bands. % if the circumstellar disk is still optically thick. However,
But, when the final dispersal of the disk material occurs and 
non-accreting transition disks form, YSOs can no longer be
identified in IR or radio surveys \citep{erco21} 
and a complete census is possible only in the optical bands.

While a clean identification of YSOs is very hard  using only optical photometry,
an efficient way to systematically  single out  SFRs is by the identification of  
 kinematical stellar groups having a common space motion.
With an unprecedented astrometric precision and sky coverage, {\it Gaia} data offer the
possibility to recognise the SFRs as common proper motion groups, 
at least within the {\it Gaia} observational limits.

 Data from the {\it Gaia} mission are revolutionising our capability to map
the youngest stellar populations of the Milky Way 
in the optical bands, which  is one of the main core science goals for 
 an overall understanding of the Galactic components.
The youngest stellar component is crucial to 
%trace the still unknown ,
better characterize the Galactic thin disk,
and its spiral arms and to understand its origin.

The characterisation of individual SFRs and their dynamics 
are also fundamental to understand the local formation,  evolution and dispersion
of star clusters, as well as the star formation history and the Initial Mass Function.
 Finally, statistical studies of YSOs during the early years of their 
formation, when the proto-planetary discs are evolving and planets form, are crucial
to shed light on  planet formation theory.

With more than 1.3 billion stars with precise proper motions and
astrometric (positions and parallaxes) and  photometric measurements,
{\it Gaia} DR2 data allowed several studies aimed to identify clustered populations
of the Milky Way. Some of these studies have been dedicated to 
SFRs, associations and moving groups.
\citet{zari18} presented an analysis of the clustered and diffuse young populations
within 500\,pc, using a combination of photometric and astrometric criteria.
Analogously, \citet{kerr21} studied the solar neighbourhood
by applying the HDBSCAN clustering algorithm \citep{mcin17}.
 They found 27 young groups,
associations and significant substructures, associated to known clusters and SFRs,
and released a  catalogue 
including $\sim3\times10^4$ {\it Gaia} DR2 YSOs within 333\,pc.

\citet{cant18} started from a list of known clusters to assign them unsupervised 
membership and  parameters.
Other studies have been dedicated to systematically find  open clusters 
in the Galaxy. \citet{cast18} used the DBSCAN algorithm \citep{este96} 
 to select a list of candidate open clusters (OC)
 which they then refined to identify real OCs with a well defined main sequence (MS).
Other papers have been recently published to both discover new open clusters
and derive their parameters \citep[e.g.][]{cant20,cant20b,cast20,liu19}.

 A recent attempt to find Galactic Plane (GP) clustered populations, including SFRs,  
has been made by  \citet{koun19} and \citet{koun20}, again using {\it Gaia} DR2 data
and the HDBSCAN unsupervised algorithm in 5D space 
($l$, $b$, $\pi$, $\mu_\alpha*$ , $\mu_\delta$). In these works, the first limited to 
1\,Kpc and the second to 3\,Kpc, they found clustered populations but also associations,
moving groups and string-like structures, parallel to the GP,
spanning hundreds of parsec in length. Clusters aged between  10\,Myr and 1\,Gyr 
have been found, with an onion-like approach, i.e. using the entire catalogue
with different cutoffs in parallax and  progressively merging the different catalogues. 

A different approach has been adopted by \citet{bica19} who used infrared (IR) data from 2MASS, {\it WISE},
VVV, {\it Spitzer}, and {\it Herschel} surveys to compile a catalog of  10\,978 Galactic star
clusters, and associations, including 4\,234 embedded clusters.

With the advent of Gaia Early Data Release 3 (EDR3), based on 34 months of
observations\footnote{{\it Gaia} DR2 data were based on 22 months of observations},
available  photometric and astrometric measurements improved significantly. 
In particular, photometric improvements have been made in the calibration models,
in the different photometric systems and in  the treatment of the BP and RP local
background flux \citep{riel20}.

In this paper, we use {\it Gaia} EDR3 data to systematically identify the low mass component of SFRs
in the Galaxy, with ages approximately  $<10$\,Myr and
 within a distance limit of $\sim$1.5\,Kpc, imposed by our data
selection.  We focus our analysis on very young clusters, by exploiting 
the  significant progress achieved with {\it Gaia} EDR3 data. 
A full exploitation of the Gaia data and  the results presented here would
require further data, such as spectroscopic determination of individual stellar parameters,
such as effective temperatures, gravities, and stellar luminosities as well as rotational and
radial velocities, crucial to derive
masses, ages and 3D space velocities. Even though the results presented here
 cannot be used at this stage
to determine the IMF, Star Formation history and 3D kinematics
of the SFRs, they can serve to trace the very young Galactic stellar component 
within 1.5-2\,Kpc through a systematic method that homogeneously identifies the
bulk population of the SFRs. Such results can be used both for statistical as well as 
for individual detailed analyses.
The paper is organised as follows: we describe in Sect.\,2  the requirements adopted 
to select the {\it Gaia} EDR3 data and in Sect.\,3 the photometric selection applied to 
obtain the starting sample of the YSO candidates.
In Sect.\, 4 we describe the method adopted to identify 
SFRs and stellar clusters, the criteria adopted to validate them and
the age classification. 
Our results and the discussion
 are presented in Sect.\,5 and \,6, respectively; finally,
our summary and conclusions are presented in Sect\,7.  In the 
Appendix\,\ref{reddeningeffect} we show the effects of the reddening
in the {\it Gaia} color-absolute magnitude diagrams,
 in the Appendix\,\ref{binapp} we estimate the effect of multiplicity in the selection of the
  YSOs while in the  
Appendix\,\ref{literaturecompapp}
we describe the comparison of specific regions with the literature.
%__________________________________________________________________

\section{{\it Gaia} data  \label{data}}
In this analysis, we use the {\it Gaia} EDR3 data  \citep{prus16,brow21}
which provide precise astrometry and kinematics 
($l$, $b$, $\pi$, $\mu_\alpha*$ , $\mu_\delta$ ) as well as  
 excellent photometry in three broad bands ($G$, $G_{\rm BP}$, $G_{\rm RP}$).
Since our analysis is focused on the Galactic midplane, where most of the YSOs
are expected to be found, we selected sources within
$|b|<$30\deg. We limit our selection to 7.5$<G\leq 20.5$. 
  The limit $G=$7.5 has been chosen to discard objects
with magnitudes derived from saturated CCD images, while $G=$20.5 is the limit to include
most of the objects with magnitude $G$ uncertainties lower than 0.2\,mag.
This range  includes the young low mass populations ($0.1\lesssim M/M_\odot\lesssim 1.5$) of the known SFRs
within the distance set by the limiting magnitude.
In addition, we considered only positive parallax values. This choice does not introduce any bias
since we do not expect to investigate stars with very small parallaxes that could have negative values
 \citep{luri18}. Finally,
we imposed a relative parallax error lower than 20\%,
 to discard stars with a poorly
constrained distance, and,
to take into account  the {\it Gaia} EDR3 systematics, we also considered the
renormalized unit weight error (RUWE), \citep{lind21b},
expected to be $<1.4$ for sources where the single-star model provides a good fit
 to the astrometric observations.

To summarise, data of our interest
were  selected from the Astronomical Data Query Language (ADQL) interface
of  the ESA Gaia Archive\footnote{\url{https://gea.esac.esa.int/archive/}} 
using the following restrictions:
\begin{equation}
\label{reqdata}
    			\begin{cases}
    						|b|<30{}^\circ \\
    						7.5< G \leq 20.5 \\
   							 \pi > 0\, \rm{mas} \\
   							 \sigma(\pi)/\pi<0.2 \\
   							 RUWE<1.4
  					  \end{cases}
\end{equation}  

%To speed up the data download process and limit the contamination,
We included in the query also a  photometric condition
aimed to include the Pre-Main Sequence (PMS) region of the $M_G$ vs. $G-G_{\rm RP}$ 
color-absolute magnitude diagram  (CAMD) where
all very young stars (t$\lesssim$10\,Myr) are expected to be found.
We split our selection in two samples, namely bright and faint, 
according to the following criteria:
% it has been approximated by two  consecutive line segments, 
%that we used to split our selection in two samples, retrieved 
%using  the additional following requirements:
%/Users/prisinzano/GAIAEDR3/MACRO/retta_isocrona.pro
%/Volumes/EXTERN_ARCH/GAIAEDR3/ADQL/query_dr3.adql e 
%/Volumes/EXTERN_ARCH/GAIAEDR3/ADQL/query_dr3_faint.adql
\begin{equation}
    \textrm{Bright sample}=\begin{cases}
       						M_{G} < 7.64\,(G-G_{\rm RP})+0.22 \\
       						 5<M_{G} \leq 9 \\
   							(G-G_{\rm RP})>0.58 \\
  					  \end{cases}
\end{equation}  
 
\begin{equation}
    \textrm{Faint sample}=\begin{cases}
   							 M_{G} < 15.00\,(G-G_{\rm RP})-8.25 \\
   							 M_{G} > 9 \\
   							(G-G_{\rm RP})>0.58. \\
  					  \end{cases}
\end{equation}
These limits are drawn as solid blue and green lines in Fig.\,\ref{gesmggrp}.
We note  that in this work, for the reddening uncorrected
 absolute magnitudes, we adopted the definition $M_G=G+5\,\rm {Log}(\pi) -10$,
based on  the inverted {\it Gaia} EDR3 parallaxes, since,
 as shown in  \citet{piec21}, within $<$2\,kpc, the inverse-parallax method gives 
 results comparable to  distances derived by the Bayesian approach \citep{bail21}.

The minimum value $M_{G}=5$ was set to avoid the upper region of the color-absolute magnitude diagram,
where the overlap of the Upper MS or PMS stars of the SFRs with giants, MS or Turn-off stars,
is expected to be very high, especially if the reddening is not corrected. This implies a cut 
of the massive population of the SFRs but it does not represent an issue for our investigation 
since we are mainly interested in the  rich low mass component of these populations.

In order to further reduce the fraction of contaminants we used also the condition
 $G-G_{\rm RP}>0.58$, that is the minimum  expected unreddened color for  low mass
 (M$\lesssim$1.2M$_\odot$) PMS (age $\leq$ 10\,Myr) stars.

Our photometric selection as well as the subsequent analysis are
 based on the $G-G_{\rm RP}$ colors. This choice
allows us to avoid the use of the $G_{\rm BP}$ magnitudes that for $G\gtrapprox$20
are strongly affected by
the application of the minimum flux threshold,  
which overestimates the mean BP flux. 
This issue also affects the RP flux, but with a considerably lower effect 
 in $G_{\rm RP}$ than in $G_{\rm BP}$
 \citep{riel20}. 
 
Once the data have been retrieved by the ESA Gaia Archive, 
parallax values were corrected by the zero point bias
reported in \citet{lind21}, using the Python code available to the 
community\footnote{\url{https://gitlab.com/icc-ub/public/gaiadr3\_zeropoint}},
that is a function of source magnitude, colour, and 
 celestial position.
 
In addition, we performed a further data filtering by considering only objects with 
error in the  $G-G_{\rm RP}$ smaller than 0.14\,mag. % See pre_classification_gaiadr3.pro
% corresponding to a mean error in the $G$ and $G_{\rm RP}$
%magnitudes smaller than 0.1\,mag. 
Standard errors in the magnitudes were computed by using the
propagations of the flux errors with the formulas:
\begin{equation}
\sigma(G) = \sqrt{(-2.5/ln(10)\sigma(FG)/FG)^2 + \sigma(G_0)^2}
\end{equation}  
 \begin{equation}
%\sigma(GBP) = \sqrt{(-2.5/ln(10)\sigma(FG_{BP})/FG_{BP})^2 + \sigma(G_{\rm BP_0})^2}
\sigma(G_{\rm BP}) = \sqrt{(-2.5/ln(10)\sigma(FG_{\rm BP})/FG_{\rm BP})^2 + \sigma(G_{\rm BP_0})^2}
\end{equation}  
\begin{equation}
%\sigma(GRP) = \sqrt{(-2.5/ln(10)\sigma(FG_{RP})/FG_{RP})^2 + \sigma(G_{RP_0})^2}
\sigma(G_{\rm RP}) = \sqrt{(-2.5/ln(10)\sigma(FG_{\rm RP})/FG_{\rm RP})^2 + \sigma(G_{\rm RP_0})^2}  
\end{equation}  
where $FG$, $FG_{\rm BP}$ and $FG_{\rm RP}$ are the mean fluxes in the $G$, $BP$ and $RP$ bands,
respectively, and $\sigma(G_0)=0.0027553202$, $\sigma(G_{BP_0})= 0.0027901700 $ and
$\sigma(G_{RP_0})= 0.0037793818$, are the  {\it Gaia} EDR3 zero point
 uncertainties\footnote{See \url{https://www.cosmos.esa.int/web/gaia/edr3-passbands for further details}}.

\section{Photometric selection of the input sample\label{photselsect}}

%RESTORE,'../../MODELS/PISA_GAIAEDR3/SAVEFILES/gaia_all.save',/v
%pisa_gaiadr3=pisa_gaia
%age_yng2=10e+6
%isodr3_yng2=where(pisa_gaiadr3.logt EQ age_yng2 and $
%               pisa_gaiadr3.phs GE 1 and pisa_gaiadr3.phs LE 5)
% print,interpol(pisa_gaiadr3[isodr3_yng2].mmsun,$
% pisa_gaiadr3[isodr3_yng2].G-pisa_gaiadr3[isodr3_yng2].RP,0.58)
%      1.22535

In this section, we describe and discuss how we performed the final photometric selection of the 
sample used as input for the subsequent clustering analysis, that is  based on 
the astrometric and kinematic {\it Gaia} EDR3 parameters,
as will be described in Sec.\,\ref{method}.

By considering the typical complexity of the environment of young stars and
the dependence of the reddening law from the stellar effective temperature due to
the large spectral range covered by the Gaia bands \citep{ande19},
we do not attempt to correct colors and magnitudes  for reddening and absorption
but we use their observed values. This is certainly one of the main source of contamination
by older stars that will be overcome as  will be discussed later.

\begin{figure}
   \centering
\includegraphics[width=9cm]{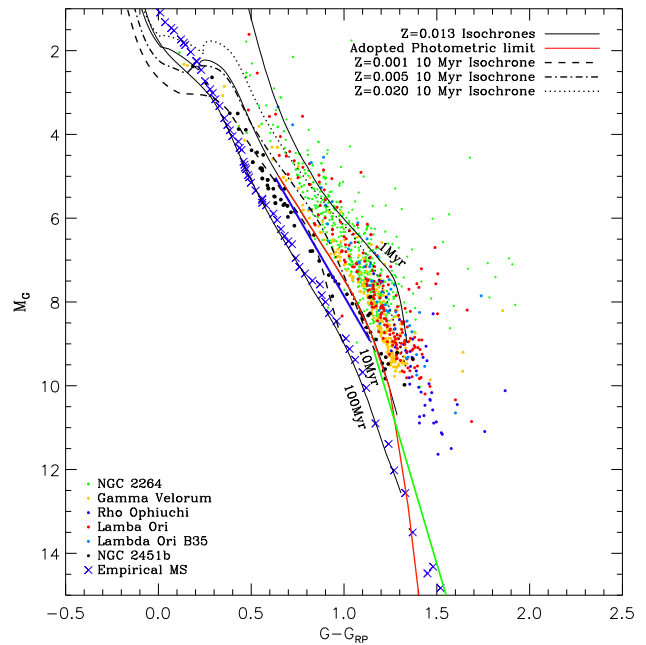}
      \caption{CAMD of YSOs of some representative young clusters with membership probabilities $>$0.90 assigned
      by combining spectroscopic and {\it Gaia} EDR3 criteria \citep{jack22}. 
      Blue x symbols  trace the empirical sequence by \citet{peca13}.
       Members of the  clusters Gamma Velorum (18\,Myr old) and 
      NGC2451b (50\,Myr old)  are also shown. Black solid   lines are the theoretical solar metallicity
        Pisa isochrones while the red solid
      line is the complete photometric limit adopted in this work including the low mass extrapolation. 
      Dashed, dashed-dotted and dotted lines are the 10\,Myr isochrones at different metallicities.
      Blue and green solid lines represent the limits described by the equations\,2 and 3.  }
         \label{gesmggrp}
   \end{figure}

 Our goal is to start from a complete sample, including all potential YSOs
with ages $<10$\,Myr,
at least in the photometric range set as described in Sect.\,\ref{data}.
In particular, we selected the objects with $M_G$ falling on the red side of the solar metallicity 
10\,Myr isochrone computed using the PISA models \citep{dell12,rand18,togn18,togn20},
in the $M_G$ vs. $G-G_{\rm RP}$ diagram shown in Fig.\,\ref{gesmggrp}.
To check if the selected photometric limit is compliant with our requirements, we 
compared it with the reddening uncorrected CAMD of some SFRs and young clusters for which membership has been 
recently derived by \citet{jack22}, based on the 3D kinematics of the spectroscopic targets.
%The members of the two clusters  Gamma Velorum and NGC2451b, with ages
%18 and 50\,Myr, respectively, are also shown for comparison.
We find that the adopted 10\,Myr isocrone  delimits  
the PMS region of clusters, such as
NGC\,2264, Lambda Ori, Lambda Ori B35 and Rho Ophiuchi, 
that are in the age range  ($t<10$\,Myr)  of our main interest.   
However, also members of $\sim$20\,Myr old clusters,  such as Gamma Velorum,
  fall completely in the selected
photometric region, while members of $\sim$50\,Myr old clusters,  such as NGC\,2451b,
 fall partially in the selected photometric region at $M_G\gtrsim 9$. Going to clusters 
 with ages $ t>50$\,Myr the overlapping region occurs at fainter magnitudes.

 Since the adopted isochrone is limited to 0.1\,M$_\odot$,
corresponding to $M_G$=10.7, the photometric limit at fainter magnitudes has been extrapolated
using a linear extrapolation. 
To check the position of such extrapolation, we compared it with the empirical sequence
by \citet{peca13} for which mean stellar colors and effective temperatures are given 
 down to  M and L  spectral types and that can be used as upper limit
 to the region we are interested on. Our photometric limit  approaches such sequence and 
crosses it at $M_G\sim13$.
This ensures us to set an inclusive  photometric selection  close to the MS at the lowest mass
tail. In fact, even though this implies the inclusion of stars older than 10\,Myr, on the other hand, it
avoides a bias against the selection of very young stars. 
 
We note that for the photometric selection, the minimum and maximum
$M_G$ associated to each observed star have been computed by considering the 1\,$\sigma$ parallax 
uncertainties, that are dominating  with respect to the magnitude
uncertainties.
The photometric selection with respect to the reference isochrone has been performed
by considering the compatibility of $M_G$ magnitudes, with respect to their minimum and maximum 
values, i.e. they were selected if either their minimum or maximum value lie inside the selection region.
At the end of this selection we remain with a catalogue of \allgaiaentries\, {\it Gaia} EDR3 entries.

Performing a photometric selection as inclusive as possible, as we have done,
implies the introduction of a significant contamination by old field or 
open cluster stars,
mainly due to the uncorrected reddening , 
binarity or to the overlapping photometric region,
at the low mass range, where the sensitivity of the $G-G_{\rm RP}$ colors in distinguishing 
PMS or MS stars becomes very low.
  
  However,
the contamination by field stars  does not represent a strong issue for our clustering analysis, 
 since  they are not expected to share similar  astrometric
 and kinematic properties. In addition,
since we aim at investigating the  low mass component of the SFRs, that is also the most dominant
\citep[$\gtrsim80$\%][]{lada06}, the statistical contrast with respect to field
contaminants is expected to be favorable to detect them.

A more tricky effect of our inclusive photometric selection
 is that also clusters older than 10\,Myr  
can   partially fall  in the selected region and be recognized as candidate 
clusters in the subsequent analysis. 
As shown in Fig.\,\ref{gesmggrp}, 
at faint absolute magnitudes ($M_G>9$),
 the model-computed isochrones
  are not very sensitive to stellar ages, and tend to overlap, especially in the 
$M_G$ vs. $G-G_{\rm RP}$ diagram.  In addition,
spectral synthesis of M dwarf stars suffers from the accuracy of the adopted  atmosphere models
and/or from incomplete molecular data.
The model-predicted colors of very low mass stars are therefore uncertain.
% and this can  
%hamper the prediction of reliable intrinsic colors of 
%very low mass stars or can
%imply a systematic bias in the  predicted
%intrinsic colors from stellar effective temperatures.
A further complication is the observed discrepancy between 
radii and colors of low mass stars, likely due to  the distorting effects of magnetic activity
and starspots on the structure of active stars \citep{some20,fran21}.
All these effects cause a spread of the low mass MS and can bring
magnitudes and colors of $\sim100$\,Myr old stars 
to fall in the region selected by us as compatible to stars with $t<10$\,Myr.
For all these reasons, as discussed for example in \citet{jeff17},
the ages judged from "standard" isochrones are almost certainly underestimated due
to a systematic bias.

 At faint magnitudes, the fraction of old cluster members falling in the adopted photometric
 region decreases with cluster ages.
 Hence  clusters 
of about 20-30 Myrs will be almost completely included in our
 selected sample, while at the age of 100-500 Myr only 
 the low mass tail will be included. However, because of
  the adopted photometric limit, the low mass tails will
   be included only for relatively close clusters ($d<$500\,pc).

As already mentioned before, a partial contamination by old cluster members in our photometric sample 
can occur also for bright stars ($M_G<9-10$) 
if their reddening or a binary status gives them  observed magnitudes and colors 
compatible with the selected photometric region. 
 As shown in the Appendix\,\ref{reddeningeffect},
the effects of using colors and magnitudes uncorrected for reddening are expected
to be more severe for reddened stars with spectral types earlier than G,
 in comparison with later spectral  types,
 in the sense that
the selected sample is expected to be contaminated 
mainly by these objects, falling in the
brightest part of the photometric region adopted in this work. 
The implications of this contingency will be discussed 
in the following sections.

Finally, we also considered the possible  effects due to the metallicity on the selection  by considering 10\,Myr
 isochrones for metallicity lower or higher than solar. The comparison shows that while YSOs 
 with over solar (Z=0.020, [Fe/H]=0.2) or sub solar 
(Z=0.005, [Fe/H]=-0.45) metallicities would fall in the selected photometric region, very metal poor
YSOs (Z=0.001, [Fe/H]=-1.10) would remain outside. However,  as recently found by \citet{spin17}
at Galactocentric radii from $\sim$6.5\,kpc to 8.70\,kpc, young open clusters and SFRs have
 close-to-solar or  slightly sub-solar metallicities and therefore we conclude that no SFRs 
  are expected to be missed for 
 metallicity effects with our photometric assumptions. 
 
 Based on the adopted photometric selection, our data set
encompasses substantially all YSOs with age $t\lesssim 10$\,Myr and observed  $M_G>5$, 
including the most reddened ($A_V<3-4$) that can be detected with Gaia. 
Even YSOs with accretion \citep[e.g.][]{gull98}  or
 seen in scattered light \citep{boni13} or 
 flares in M-type stars
\citep[e.g.][]{mitr05} are expected to be included in our sample. In fact, these phenomena
 affect the $G_{\rm BP}-G$ or the $G_{\rm BP}-G_{\rm RP}$ colors, causing the stellar colors becoming bluer than
their photospheric colors,  while, on the contrary,
their effect on the  $G-G_{\rm RP}$ colors goes in the same direction of the reddening,
 causing these latter colors to become redder. 
 
 We stress, however, that the constraint $M_G<5$, adopted to strongly reduce the contamination
 due to reddened Turn-off or  MS stars, makes 
 the selected photometric sample incomplete for the  massive 
 stellar component of the SFRs. A further expected missing stellar component is
that of  binary systems of the clusters, due to 
 the restriction of the {\it Gaia} data to RUWE$<$1.4 (see Appendix\,\ref{binapp}). 
 In addition, since available data do not allow us obtaining reliable corrections
 for the reddening affecting  colors and magnitudes of the selected YSOs, accurate stellar parameters
 such as individual stellar ages and masses will not be derived in the subsequent analysis. 
 However, even though the results we aim to achieve are not suitable for investigations based on
 complete young populations or accurate stellar parameters, they are expected to trace the dominant
 component of the SFRs, 
 i.e. their low mass population, and will be crucial to have an overall systematic view   
of the Galactic SFRs located within 1-2\,Kpc from the Sun, 
as well as for detailed individual or statistical investigations of these YSOs.
%%%%%%%%%%%%%%%%%%%%%%%%%%%%%%%%%%%%%%%%%%%%%%%%%%%%
\section{Method\label{method}}
%\subsection{Starting sample caveats}
\subsection{Clustering with DBSCAN\label{dbscansect}}
This section describes the methodology used to search for candidate clusters 
with an unsupervised algorithm 
as overdensities in the  five-dimensional (5D) {\it Gaia} EDR3 astrometric and
kinematics parameters ($l$, $b$, $\pi$, $\mu_\alpha*$ , $\mu_\delta$).

Starting from the data set selected as described in the  Sect.\,\ref{photselsect},
we performed a clustering analysis using the  
DBSCAN  code \citep{este96}, within the scikit-learn machine learning package in Python.
First of all, we preparared a grid of 5\deg$\times$5\deg\, boxes, covering the entire 
range of the Galactic
longitudes $l$ and for $|b|<30$\deg. In this step, we took into account 
 the discontinuity  at $l=0$\deg. To homogenise the variables having different dimensions  
  to comparable values,
   the five parameters ($l$, $b$, $\pi$, $\mu_\alpha*$ , $\mu_\delta$) within each box 
were first re-scaled using the {\tt RobustScaler} Python code based on
a statistics that is robust to outliers, according to the interquartile  range.
  
 The DBSCAN algorithm requires only two input parameters ($\epsilon$, $minPts$). 
  It identifies candidate clusters as overdensities in a multi-dimensional space (5D in our case)
  in which the number 
 of sources exceeds the required minimum number of points $minPts$,
  within a neighborhood of a particular linking length $\epsilon$, for all the five parameters,
  using a statistical distance, assumed to be  Euclidean. 
%The limiting radius   $\epsilon$ is computed as the Euclidean distance. 
DBSCAN does not require to know an {\it a priori} number of clusters and it is able to 
detect arbitrarily  shaped clusters. This is crucial for our analysis  aimed to 
find  SFRs that can  be characterised by circular or elongated or asymmetric shapes,
reminiscent of the native molecular clouds.
In order to determine the best input parameters 
($\epsilon$, $minPts$) to give as input to DBSCAN,
we  experimented several values in the direction of well known SFRs
and we noted that in the same direction more than a combination of the two parameters
is needed to reveal different real clusters located at different distances. 
This is due to the fact that close candidate clusters, such as associations and co-moving groups,
 can appear spatially (in $l$ and $b$)
 sparse while they are definitively clustered in distance and proper motions, while
 in the same direction it is possible to
 identify distant, but spatially concentrated candidate clusters.
 In the two cases the choice of two different  $\epsilon$ values 
 rather than a single $\epsilon$ 
 is required  to detect these kinds of clusters. 
  
Based on this  preliminar empirical analysis, we decided to run the DBSCAN codes 
in the entire Galactic
Plane, by adopting a total of 900 combinations of ($\epsilon$, $minPts$) values 
with $\epsilon$ ranging from 0.1 to 9 in steps of 0.1 and $minPts$ ranging from 5 to 50
in steps of 5. In addition, to account for candidate clusters falling in the borders of the
defined boxes, we defined  another 4 sets of grids, by shifting the original boxes
by $\delta l$=$\delta b$=[0.1\deg,\,0.2\deg,\,0.3\deg,\,0.4\deg] with respect to the original boxes.
In the following we will refer to  the 5 sets of grids as spatial configurations.
At the end, we run DBSCAN within a total of 360/5$\times$ 60/5$\times$ 5=4320 different boxes
with 900 combinations of parameter sets ($\epsilon$, $minPts$).

\subsection{Candidate cluster validation\label{validationsect}}
One of the most challenging phases of this analysis has been the validation 
of the  recognized candidate clusters.
In fact,  DBSCAN  is an unsupervised density-based algorithm
%https://towardsdatascience.com/dbscan-clustering-algorithm-how-to-build-powerful-density-based-models-21d9961c4cec
and, as a consequence, it   picks up not only overdensities which correspond to real OCs,
but  also overdensities only in statistical terms. For this reason, our a posteriori validation approach
has been based on the exploitation of two astrophysical constraints,
based on  the typical properties of the SFRs, by avoiding  the introduction of strong biases.

SFRs are not characterised by well defined age sequences  and they are typically observed
in the Hertzsprung-Russell (HR) diagrams as ensemble showing an apparent luminosity spread, often associated
to an age spread \citep[e.g.][]{pall99,pall05}. On the other hand, such spreads have also
been  ascribed to  complex phenomena affecting their photometry, such as variability, 
accretion and outflows, extinction, binarity 
and to our inability to quantify their contribution \citep{sode14}.
%used to infer the duration of the star formation process,
%and, as consequence, the physical mechanisms involved \citep[e.g][]{tan06,tass04}.

Neverthless, SFRs are usually observed with a typical mass distribution,
that can be shaped by a standard (or closely resembling) Initial Mass Function (IMF), 
characterised by an increasing fraction of members going towards  decreasing masses,
at least until masses of $\sim 0.3$M$_\odot$ \citep[e.g.][]{salp55,scal98,chab03}.

Since we are exploiting the excellent {\it Gaia} EDR3 results down to $G=20.5$,  within  
reasonable reddening values (A$_V\lesssim1$), with our data set, we expect to detect  YSOs 
with spectral type down to
 M-type, at distance  $\lesssim 1.5$\,kpc. This is the case, for example, of the cluster 
 NGC\,6530, located at around 1.3\,kpc, for which the low mass population down to 0.4\,M$_\odot$
 has been detected at V$\sim 20$ \citep{pris05}, roughly corresponding to our  $G$ magnitude
 limit.

% the following selection has been performed with dbscan_select_mg_g_mag.pro  
Based on these considerations a physical recognisable candidate cluster should 
incude its tail of low mass members. Hence, we imposed a minimum threshold of 
  10 objects with $M_G>7.7$, that means to require candidate clusters having at least 10 stars 
  with M$\lesssim$0.5\,M$_\odot$,
  assuming the isochrone of 10\,Myr from the Pisa models.
 %  We note that this does not 
 % correspond to an additional
 %  cut in   $M_G$ that remains $M_G>5$, according to our initial requirements 
 %  (see Sect.\,\ref{data}),
 %  but simply that each cluster must include a minimum subsample of 10 members 
 %   with $M_G>7.7$.
   
A further parameter that  we considered as an indicator of
reliability for the candidate cluster validation is the dispersion of the distances of each cluster.
The observed total distance dispersion is a combination of the intrinsic dispersion plus the 
contribution due to the measurement errors. While the intrinsic dispersion does not depend
on the distance, the contribution  due to the measurement errors
becomes dominant at large distances,
since {\it Gaia} EDR3 parallaxes become much more uncertain. 
Thus, among the  parameters used to find over-densities  by DBSCAN,
 the observed standard deviation of the distances is the most critical parameter to be constrained
 for the identification of real clusters.
 To this aim, for the cluster validation, we constrained the maximum allowed observed dispersion.
 For distances $<$1\,kpc, the constrain is set on the ratio  between the standard 
 deviation of the distances of the putative members and the derived mean distance
  for the given candidate cluster. For a valid candidate cluster  the above ratio
   has to be $<0.2$. For more distant candidate clusters, we adopted the more 
   stringent constrain that the standard deviation should be smaller than 200 pc.
This limit has been chosen  by considering that
for NGC\,2244 located at $\sim1.6$\,Kpc, one of the most distant  clusters that we  detect,
the distance dispersion is  about 175\,pc and therefore we do not expect to find physical real clusters
with a distance dispersion  larger than this threshold.
These choices may limit our ability to detect clusters 
at distance $\gtrsim1.5$\,kpc for which we could, in principle, detect, at the magnitude limit of
our data set, the massive component of the clusters 
down to $\sim1$\,M$_\odot$ regime. However, since 
the accuracy of {\it Gaia} EDR3 parallaxes and kinematic data beyond this limit becomes very low,
we prefer to maintain our constraints at the price of limiting our analysis to smaller  distances.  

The adopted constrains on the distance dispersion of cluster members have shown 
to be very effective in rejecting a large number of (unexpected) candidate massive
 clusters recognized by DBSCAN,  typically with more than
 $\sim$1000 members, located
  at distances $\gtrsim1$\,Kpc, that 
   do not include M-type stars, but only earlier stars and that are characterised by
   very large dispersions in distance. These structures are likely 
   those   identified as strings  in \citet{koun19,koun20}.
   However, since we do not recognise these structures as standard clusters, any further
investigation of them is beyond the scope of this work. 

The final cluster  member selection has been performed only for candidate clusters that satisfy
the  previous constraints.  As a result of our choice of the 
DBSCAN input parameters (see Sect.\,\ref{dbscansect}) and of the adopted spatial configurations, 
 a given candidate cluster can be identified by 
adopting similar input parameters, with possible small differences in the cluster membership.
In addition, a given candidate cluster can be identified in more than one box, with the same
membership result, if the candidate cluster is spatially small enough to be completely identified, for a given
couple of input parameters, in two or more overlapping boxes. Alternatively, it can be completely detected
within one box and only partially detected in a box where the candidate cluster falls
at the borders. In order to assign the most likely membership for a given cluster, we proceeded
adopting the following strategy. 

% see dbscan_compare_par_stat.pro NB. il tag si chiama mean e rms ma ho calcolato mediana e mezza ampiezza
We first considered the candidate clusters detected within the same spatial configuration but with different
set of parameters ($\epsilon$, $minPts$). 
For each of the selected candidate clusters, we computed the median values of the 5 parameters
($l$, $b$, $\pi$, $\mu_\alpha*$ , $\mu_\delta$) 
and then we selected all the candidate clusters that were simultaneously  compatible in these 5 parameters
 i.e. if the two compared distributions of each parameter overlap around the median, 
  within half  of the total width.
% see dbscan_compare_par_abc.pro dbscan_compare_par_d.pro dbscan_compare_par_e.pro dbscan_compare_par_fgk.pro 
Among the compatible candidate clusters, we selected
%as {\it best} input parameters, those for which a given member belongs to 
the most populated and discarded the others. This strategy 
allowed us to identify the most persistent candidate clusters at different scales.

%see dbscan_compare_master.pro and dbscan_def_selection.pro
In the subsequent step, we compared  the candidate clusters identified 
in each of the five spatial configurations  to select the {\it best} configuration,
or likewise, the {\it best} box in which the spatial coverage of the candidate cluster is maximised.
Since we can have more than one detection of the same cluster,
for each member, 
% presumably found in more than one candidate cluster, 
 we selected only the configuration
in which it is associated to the most populated candidate cluster and that member was removed from the 
less populated clusters in which it was identified by DBSCAN.
 % see dbscan_assign_cl_id.pro
The peripheral members of candidate clusters covering a spatial region larger than the 
area of the box (5\deg$\times$5\deg),
 left out from the richest centered candidate cluster, 
 were considered as additional candidate clusters only if they include at least
10 elements,\footnote{For this reason, our catalogue  includes cases in which a single 
physical cluster is identified  by more than one DBSCAN cluster.}
 as also assumed in other similar works, \citep[e.g.][]{cast18,kerr21}. 
 This selection strategy allowed us to include also likely members  at the candidate cluster's periphery,
 providing data for further investigations on the dynamics  of these stellar clusters.
 At the end of this process, we are left with a total of \totyso\,  
 detected stars within \totcl\, single candidate clusters.

Many SFRs are associated to giant molecular clouds and thus they can have a
spatial extension larger than the box of 5\deg$\times$5\deg\, used for our analysis.
In order to merge candidate clusters belonging to the same  complex, we proceeded
as follows: 
 we  computed the median and the 16\% and 84\% percentiles  
% that roughly set the width at 1\,$\sigma$,  
of the distance 
 and  proper motion distributions. 
 Then, we merged all neighboring clusters 
  for which distances
 and proper motions were compatible within 1\,$\sigma$.
The total number of  merged clusters
 is \totclmerge.
  %%###########################################%%%%%%%%%%%%%%%%%
\subsection{Cluster age  classification }
\begin{table*}[!ht]
\caption{Clusters used as template data set to select SFRs and other stellar clusters.
Flag is the value assigned to each cluster to characterise a given observed CAMD shape.
r$_{50}$ is the radius in which half of the identified members are concentrated,
$d$ is the distance obtained by inverting the median value of the member parallaxes  and
N is the number of members.
 \label{printtexprototipi}}
\centering
\begin{tabular} {c c c c c c c c c c}
\hline\hline
Literature Name & Flag & Reference &  $l$     & $b$     & r$_{50}$ & $d$  & logt & N  \\
                &      &           &  [deg]   & [deg]   &  [deg]   & [pc] & [yr] &    \\
\hline
$[\rm LK2002]$\,Cl10 &    1 &\citet{ledu02} &     79.867 &     -0.908 &      0.886&  1557&     &   167\\
65.78-2.61 &    2 &\citet{aved02} &     66.153 &     -3.123 &      1.194&  1324&     &   134\\
Rosette &    3 &\citet{zuck20} &    206.438 &     -1.903 &      2.025&  1571&   7.1&   810\\
NGC\,6530 &    4 &\citet{dias02} &      6.060 &     -1.287 &      1.020&  1364&     &   635\\
NGC\,6531 &    5 &\citet{dias02} &      7.585 &     -0.338 &      1.634&  1350&   8.6&   804\\
UBC\,386 &    6 &\citet{cant20} &    100.562 &      8.694 &      1.147&  1280&   6.8&   193\\
Ass\,Cyg\,OB\,9 &    7 &\citet{sitn03} &     78.753 &      1.778 &      2.293&  1339&   8.1&   616\\
Serpens\,South\,molecular\,cloud &    8 &\citet{fern14} &     29.364 &      2.870 &      0.976&   920&     &   123\\
CygOB7\,CO\,Complex\, &    9 &\citet{dutr02} &     92.653 &      2.529 &      0.950&  1123&     &    46\\
BRC\,27 &   10 &\citet{rebu13} &    224.621 &     -2.244 &      3.027&  1233&   6.9&  1709\\
$[\rm DB2002b]$\,G352.16+3.07 &   11 &\citet{otru00} &     -7.866 &      3.002 &      4.764&  1169&   7.0&  2357\\
IC\,1396 &   12 &\citet{zuck20} &     99.236 &      4.733 &      7.407&   945&   7.4&  3140\\
$[\rm MML2017]$\,2399 &   13 &\citet{mivi17} &     33.890 &      0.643 &      2.543&   609&     &   130\\
Chamaeleon\,II &   14 &\citet{zuck20} &    -56.363 &    -14.720 &      2.452&   200&     &    41\\
Cepheus &   15 &\citet{zuck20} &    108.911 &      4.359 &      9.748&   923&   8.2& 11445\\
NGC\,7039 &   16 &\citet{cant20} &     88.350 &     -1.717 &      5.322&   767&   7.3&  1048\\
$[\rm YDM97]$\,CO\,14 &   17 &\citet{yone97} &    104.508 &     13.950 &      3.039&   350&     &   124\\
Serpens &   18 &\citet{zuck20} &     28.783 &      3.082 &     10.166&   455&   7.2&  2388\\
IC\,348 &   19 &\citet{cant20} &    160.790 &    -15.812 &     11.430&   334&   7.4&  2661\\
Chamaeleon\,I &   20 &\citet{zuck20} &    -62.781 &    -15.444 &      3.099&   192&     &   156\\
Taurus &   21 &\citet{zuck20} &    172.114 &    -15.302 &      4.551&   131&     &   112\\
Ophiuchus &   22 &\citet{zuck20} &     -8.024 &     18.781 &     12.655&   144&     &  2398\\
Corona\,Australis &   23 &\citet{zuck20} &     -0.132 &    -17.592 &      3.291&   155&     &   107\\
$[\rm DB2002b]$\,G302.72+4.67 &   24 &\citet{dutr02} &    -57.143 &      4.739 &      5.854&   112&     &   235\\
Pozzo\,1 &   25 &\citet{cant20} &    261.858 &     -8.321 &     13.343&   398&   8.3&  6001\\
ASCC\,32 &   26 &\citet{cant20} &    237.327 &     -9.186 &      9.878&   818&   8.4&  4416\\
Lac\,OB1 &   27 &\citet{chen08} &     96.762 &    -15.032 &     11.268&   548&   7.4&  2367\\
RSG\,8 &   28 &\citet{cant20} &    109.331 &     -1.212 &     12.055&   468&   7.4&  2900\\
NGC\,2451B &   29 &\citet{cant20} &    253.198 &     -7.499 &      9.513&   401&   7.6&  2826\\
NGC\,2232 &   30 &\citet{cant20} &    215.533 &     -7.983 &     13.427&   372&   7.2&  1703\\
Sco\,OB2\,UCL &   31 &\citet{deze99} &    -29.000 &     16.813 &     15.052&   145&     &  1189\\
IC\,2602 &   32 &\citet{cant20} &    -70.259 &     -5.011 &      6.825&   151&   7.6&   315\\
NGC\,2516 &   33 &\citet{cant20} &    -86.236 &    -15.931 &      6.881&   427&   7.6&  1156\\
Melotte\,20 &   34 &\citet{cant20} &    147.504 &     -6.461 &      8.867&   174&   7.7&   414\\
Melotte\,22 &   35 &\citet{cant20} &    166.573 &    -23.406 &      5.882&   137&   7.9&   296\\
NGC\,2422 &   36 &\citet{cant20} &    230.995 &      3.061 &      6.238&   500&   8.0&   347\\
Alessi\,12 &   37 &\citet{cant20} &     67.678 &    -11.723 &      3.977&   546&   8.1&   127\\
NGC\,3532 &   38 &\citet{cant20} &    -72.815 &      2.279 &      4.851&   561&   8.6&    88\\
IC\,6451 &   39 &\citet{cant20} &    -19.939 &     -7.821 &      1.257&  1068&   9.2&    86\\
NGC\,6087 &   40 &\citet{cant20} &    -32.077 &     -5.426 &      2.532&  1007&   8.0&    77\\
Alessi\,62 &   41 &\citet{cant20} &     53.676 &      8.773 &      3.561&   622&   8.4&    87\\
UPK\,33 &   42 &\citet{cant20} &     27.965 &      0.108 &      3.931&   518&   8.4&   111\\
NGC\,1647 &   43 &\citet{cant20} &    180.355 &    -16.861 &      2.141&   606&   8.6&   272\\
NGC\,6124 &   44 &\citet{cant20} &    -19.205 &      6.078 &      5.404&   648&   8.3&  1102\\
NGC\,6494 &   45 &\citet{cant20} &      9.714 &      2.980 &      5.537&   755&   8.6&   680\\
IC\,4725 &   46 &\citet{cant20} &     14.022 &     -4.595 &      4.807&   669&   8.1&   788\\
Alessi\,44 &   47 &\citet{cant20} &     37.075 &    -11.510 &      7.285&   587&   8.2&   637\\
Stock\,2 &   48 &\citet{cant20} &    133.371 &     -1.160 &      8.292&   384&   8.6&   727\\
NGC\,2168 &   49 &\citet{cant20} &    186.647 &      2.327 &      2.616&   928&   8.2&   118\\
DSH\,J2320.1+5821A &   50 &\citet{kron06} &    111.248 &     -2.785 &      2.394&  1131&     &   243\\
UPK\,143 &   51 &\citet{cant20} &     91.810 &      0.514 &      1.752&   934&   8.4&   262\\
Collinder\,421 &   52 &\citet{cant20} &     79.429 &      2.527 &      1.061&  1265&   8.4&   154\\
\hline
\end{tabular}
\tablefoot{Flag=[1, 28] are assigned to clusters with ages $t\lesssim10$\,Myr,
Flag=[29, 36] are assigned to clusters with ages $10\lesssim t/$Myr$\lesssim 100$,
Flag=[37,52] are assigned to clusters with ages $t\gtrsim100$\,Myr.}
\end{table*}
 
    \begin{figure*}
   \centering
\includegraphics[width=\textwidth]{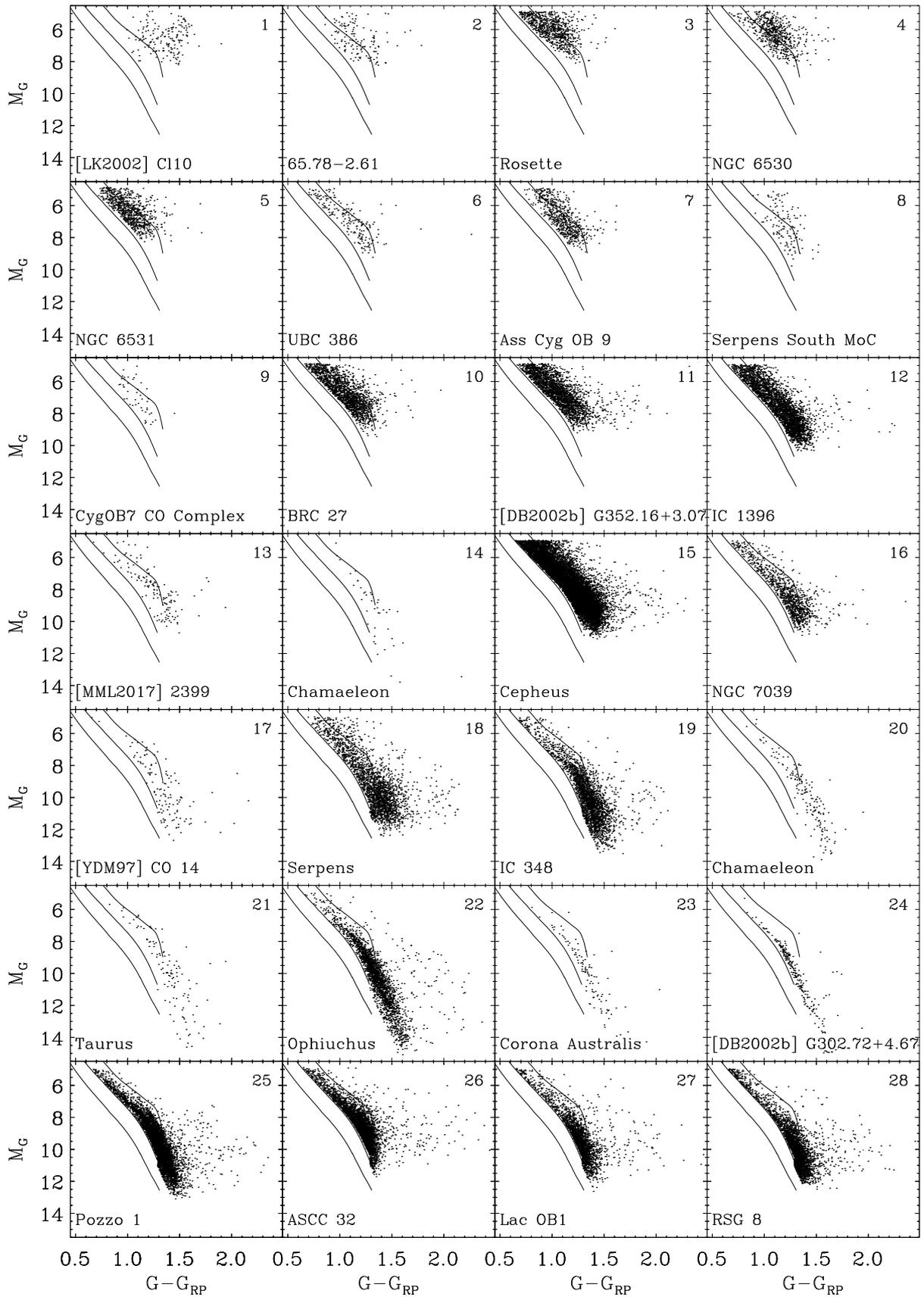}
      \caption{CAMD of YSOs identified in the clusters with ages $t\lesssim 10$\,Myr included in the
      template data set. 
      Black solid   lines are the theoretical solar metallicity
        Pisa isochrones of 1, 10 and 100\,Myr isochrones (from right to left). The number on the top right 
        edge of each panel is the flag assigned to each cluster.}
         \label{camdsfr}
   \end{figure*}

    \begin{figure*}
   \centering
\includegraphics[width=\textwidth]{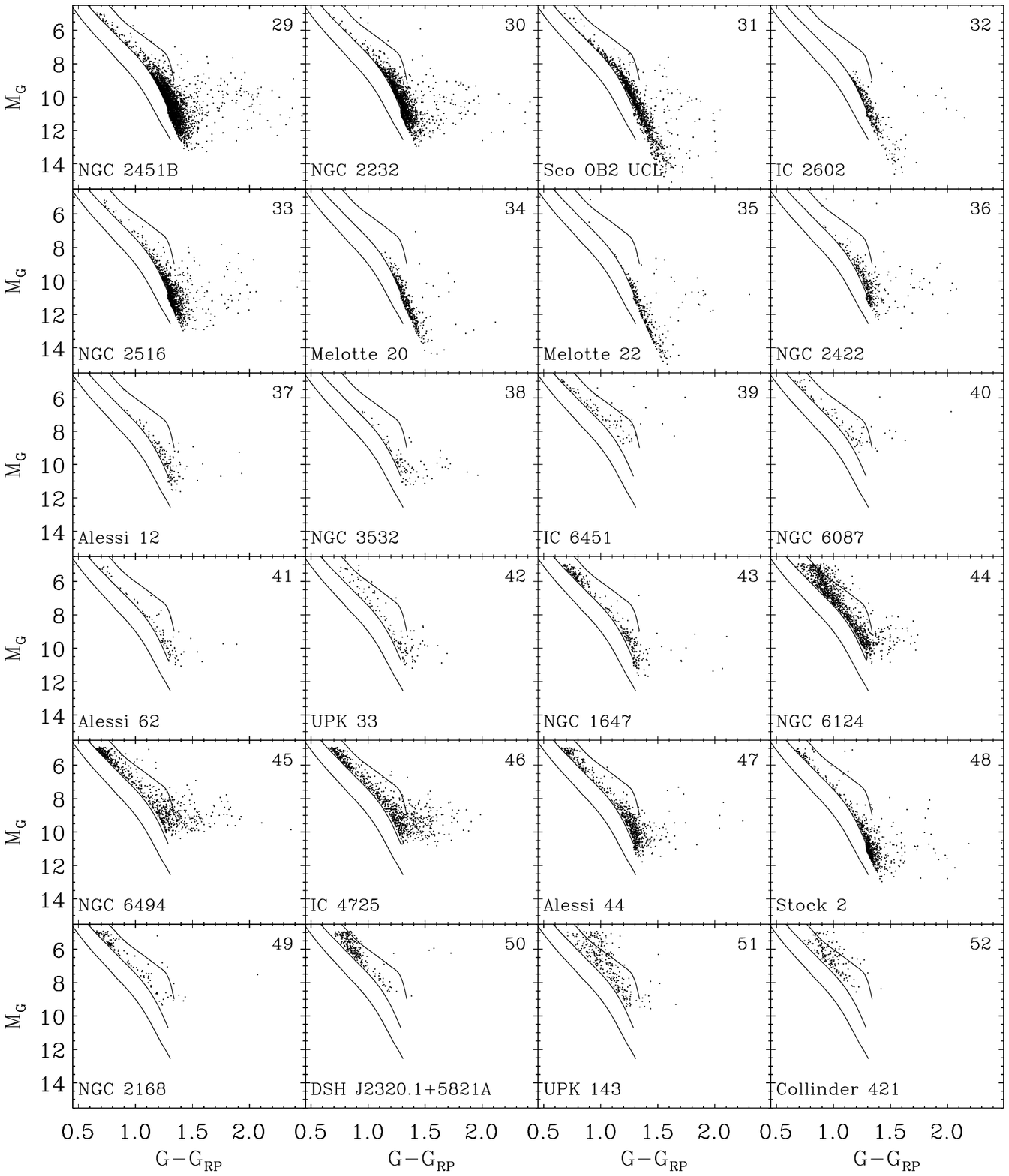}
      \caption{CAMD of clusters with ages 10\,Myr$\lesssim t\lesssim$100\,Myr,
      flagged from 29 to 36, and with ages $t\gtrsim 100$\,Myr, flagged from 37 to 52, included in the
      template data set. 
      %In the bottom right panel an example of peculiar clusters that will 
      %be discarded from our analysis.
      Black solid   lines are as in  Fig.\,\ref{camdsfr}.
      The number on the top right  edge of each panel is the flag assigned to the clusters.}
         \label{camdold}
   \end{figure*}
From a visual inspection of the photometric properties of the clusters found with this analysis, 
we noted that, while, as expected, for most of the recognized clusters their 
selected members of any mass  stay in the PMS region of the CAMD, there is a
 fraction of recognized clusters for which only the low mass members stay in that PMS region. 
 This is, for example, the case of clusters with low or moderate 
extinction ($A_V\lesssim1$) and  age 10\,Myr$\lesssim t\lesssim50$\,Myr
such as IC\,2602, Melotte\,20, NGC\,2451\,A, NGC\,2451\,B, where part of the MS or PMS 
 low mass tail  ($M_G\gtrsim9$) 
 overlaps the  photometric region considered here. 
 For clusters with ages $t\sim$100-200\,Myr, such as for example Melotte\,22 (Pleiades), NGC\,2422, NGC\,2516 
  a smaller fraction of  the MS low mass tail,
   likely composed by reddened members, cluster binaries
  or PMS members, is selected. 
  
  Further reddening effects or poorly constrained magnitudes or parallaxes, 
  can bring colors or   magnitudes of  members of even older clusters within 
   the PMS photometric region considered in this work.
For clusters with  extinction $A_V\gtrsim1$, the MS of  $t\gtrsim100$\,Myr old clusters in the range
 $5<M_G\lesssim8$ fall to the right of the unreddened 10\,Myr isochrone. Thus, depending on the cluster age,
  binaries or reddened members of clusters with ages $t>10$\,Myr
   can also fall in the selected photometric region.
   Since these objects share the same proper motions and are at the same distance, they are 
   recognized as belonging to a cluster  
   and are therefore included in our catalogue. 

To distinguish SFRs from old clusters, we adopted a pattern match procedure based on the extraction of the
different patterns that characterise the observed CAMD of clusters of different ages. 
Among the clusters identified as described in the previous sections,
we selected those listed in Table\,\ref{printtexprototipi} (\protoclusters\,  in total) 
and we used them as template data set.

In the template data set, we identified 28 clusters, shown in Fig.\,\ref{camdsfr},
 that have been used by us as proxy for clusters with ages
$t\lesssim 10$\,Myr. Such clusters
 were selected since most of them show a consistent
luminosity spread, typical of the SFRs, starting  from our brightest limit M$_{\rm G}$=5.
% with the exception of the close clusters 
% Corona Australis and [DB2002b]\,G302.72+4.67, which show a quite well
%defined sequence compatible with ages $t<5$\,Myr.
 However, their general shape is also set by the
reddening and the distance, with the observed M$_{\rm G}$ maximum limit that increases  as distance decreases.
All these cases have been included in the template data set 
to retrieve all the possible patterns observed in the CAMD 
due to different ages, distances, reddening and cluster richness.
For each of these clusters, we assigned an increasing flag from 1 to 28, aimed to represent
the different shapes of the observed CAMD shown in Fig.\,\ref{camdsfr}. 
 
We identified  also 8 clusters as representative for the ages $10\lesssim t/$Myr$\lesssim 100$,
flagged from 29 to 36, according to the ages given in  \citet{cant20}. The observed CAMD
of these clusters are shown in Fig.\,\ref{camdold}.
These clusters
show an evident PMS region that is mainly populated in the range M$_{\rm G}\gtrsim 8$ (e.g. NGC\,2451B, NGC\,2232),
due to our photometric selection.
Such region becomes thinner and thinner for older clusters such as Melotte\,20 and Melotte\,22.   
Finally, we selected 16 clusters, flagged from 37 to 52, as proxy for clusters with ages $t\gtrsim 100$\,Myr,
in agreement with \citet{cant20}.  
 Most of these clusters have been included in the template sample 
to take into account the non-uniform distribution
of the absolute  magnitudes  of their members in the observed CAMD. In fact, while for very young clusters
it is uniformly populated, accordingly to their age and the IMF, for these reddened and old clusters,
the  population is not entirely identified.
For example, the clusters with flag from 43 to 52 are characterised in the CAMD by an overdensity of members 
with M$_{\rm G}\lesssim 9$. Most of them are quite distant clusters (d$\gtrsim$500\,pc) and thus very
 likely  affected by reddening. As shown in Appendix\,\ref{reddeningeffect}, 
the effect of the reddening for the Gaia bands, depends on
the stellar effective temperature \citep{ande19}, and for high mass stars such effect is larger than for low mass stars.
This would explain  the presence of the peak at higher masses in the observed magnitudes of the CAMD
for most of these clusters.
 Depending on the cluster distance, also part of the low mass tail is detected, 
 but the overall non-uniform pattern
 of their CAMD is different from that expected for young clusters.

 Since most of the clusters show asymmetric structures,
to evaluate their extension we estimated the radius in which half of the identified members are concentrated,
as r$_{50}=0.5\times \sqrt{(width^2+height^2)}$, as done in \citet{cant20}.

In our final catalogue,  we also  noted   
the presence of  other photometrically unphysical aggregates 
 including mostly only faint stars (with $G>18.5$), 
 %likely low or very low mass stars,
  with  very red 
$G-G_{\rm RP}$ colors,  with an horizontal distribution in the CAMD, 
likely compatible with those of giant stars and with
M$_{\rm G}$ nearly constant. 
 Since most of these peculiar clusters
are in the direction of the Galactic Centre, we infer that they correspond to
very distant giants for which {\it Gaia} EDR3 parallaxes are systematically wrong due to
the strong effects of crowding and high extinction in the direction of the Galactic Center.

To separate
these aggregates from SFRs or stellar clusters, we included in our template data set further 
27 cases of these peculiar aggregates, flagged from -27 to -1, with median M$_{G}$ from 7.6 to 15.8,
covering their observed magnitude values. 

According to the known ages of the clusters of the template data set, we defined the  
three age bins, $t\lesssim 10$\,Myr, $10\lesssim t/$Myr$\lesssim 100$ and $t\gtrsim 100$\,Myr, 
including the clusters with flags in the ranges [1, 28], [29, 36] and [37, 52], respectively.
Then we used a python implementation of the  
two-dimensional version of the Kolmogorov-Smirnov (KS)
test\footnote{available at \url{https://github.com/syrte/ndtest}},
developed by \citet{peac83} and generalised by \citet{fasa87}, to identify
for  each of the \totclmerge\, clusters, 
the most similar amongst the chosen template clusters in the CAMD,
 i.e. the one for which the KS statistic is minimum.

The  procedure does not intend to derive any best fitting parameter but it is aiming only 
 to assign a flag to each cluster and then  a "coarse" age range to which it belongs to.
 At the end, we selected only
the  \toplevelcl\,clusters with more than 20 members 
(corresponding to  \toplevelyso\, objects),
for which the KS test statistic is $<0.2$. % see definition toplevelyso
 
In conclusion, we classified \youngstars\, candidate YSOs that belong to \totsfr\, 
structures with  $t\lesssim$10\,Myr, distributed  within  $\lesssim1.5$\,Kpc.
 From now on we will indicate these structures
as SFRs, meaning  regions that can include at least one
very young cluster and mostly consistent of YSOs with $t\lesssim$10\,Myr.
In addition, we classified
\midagestars\, low mass members of \totmidagecl\, stellar clusters, 
mainly located within $\sim$500\,pc
    and ages  10\,Myr$\lesssim t\lesssim$100\,Myr,  
and, finally, \oldstars\, members of \totoldcl\, clusters
    with  $t\gtrsim$100\,Myr. 
The objects that belong  to photometrically unphysical aggregates 
are  \flatcamd. The results are summarised in Tab.\,\ref{classtab}.
 From our catalogue we reject all clusters with ages 
$t\gtrsim$100\,Myr, the photometrically unphysical aggregates  and those 
that remain unclassified since are mainly poorly populated with
a CAMD that does not allow to properly classify them.

\begin{table}[!ht]
\caption{Results of the cluster age classification}             % title of Table
\label{classtab}      % is used to refer this table in the text
\centering                          % used for centering table
\begin{tabular}{c c c c }        % centered columns (4 columns)
\hline\hline                 % inserts double horizontal lines
Classification 					 & \# Stars & \# clusters 			& Flag \\    % table heading
\hline                         
$t\lesssim$10\,Myr     					 & \youngstars  & \totsfr 	& [1, 28]   \\      % inserting body of the table
$10\lesssim t/$Myr$\lesssim$100   & \midagestars & \totmidagecl    	& [29, 36]  \\
$t\gtrsim$100\,Myr   			 & \oldstars    & \totoldcl			& [37, 52] \\
Phot. unphysical aggregates 		 & \flatcamd    & \totflatcamdcl    & [-27, -1]  \\
Unclassified   					 & \unclassyso  &  \unclasscl  	 	&  \\
\hline                                   %inserts single line
\end{tabular}
\end{table}
SFRs and stellar clusters with ages t$\lesssim$100\,Myr are listed in 
Tab.\,\ref{printtexclusters},
while cluster members are given in Tab.\,\ref{printtexmembers}.
Most of the clusters listed in the table are
very extended complex regions including several subclusters 
known in the literature, merged here within single structures. 
Since the  aim here is to detect these Galactic young structures,
 the literature cluster names
given  in Tab.\,\ref{printtexclusters},  mainly taken 
 from \citet{cant20} or \citet{zuck20} or from Simbad, are only indicative of the region. 
%The authors recommend to identify any cluster of interest  by using coordinates and parallaxes.

\section{Results}
\subsection{Photometric completeness}
Within the magnitude range explored in this work and assuming the restrictions on {\it Gaia} data
defined in Sect.\,\ref{data},
the photometric cluster completeness,  for clusters with $t\lesssim10$\,Myr, 
is expected to be near 100\% for not embedded YSOs, since, as shown in Fig.\,\ref{gesmggrp},
all members detectable in this age range and in the optical bands,
 are expected to lie in the selected photometric region.
 
 Nevertheless, the adopted restriction $RUWE<1.4$ introduced a bias in the selection of multiple members
in the SFRs. To estimate the fraction of missed binary members with 
the Gaia-based selection used in this paper, we used as reference the Tau-Aur binary-star list 
by \citet{krau12}. Details about the comparison of this list with our catalogue and {\it Gaia} EDR3 data
are given in Appendix\,\ref{binapp}. This comparison shows that, due to the $RUWE$ restriction,
in SFRs at distances similar to Tau-Aur, we have lost
about 72\% of  their binary populations. Assuming a binary frequency of $\sim$50\% 
\citep{math94}, a loss of $\sim$35\% of PMS members can be expected. 
However, at large distances, the projected binary motions become smaller,
and  therefore we expect a less significant binary member loss for the farther-out SFRs.
 
For clusters with ages $t\gtrsim10$\,Myr, the cluster completeness decreases with ages and strongly depends
on the cluster distance. In fact,  clusters with 10\,Myr$\lesssim t\lesssim$100\,Myr (indexed from 29 to 36), 
are mainly in the solar neighborhood 
($d<500$\,pc). For these clusters, 
even though we are not
able to detect the entire cluster population, we are, however, able to detect part of the very low mass tail component. 
The fraction of the detected very low mass tail component decreases with age  and, in fact, for clusters with $t\gtrsim 100$\,Myr
(indexed from  37 to 52), mainly concentrated at $d\gtrsim 500$\,pc, the completeness is very low.
These latter have  been discarded from our final catalogue since  they include only a small 
fraction of the cluster members and are not in the age range of interest for this work. 
Clusters with ages 10\,Myr$\lesssim t\lesssim$100\,Myr have been included in our catalogue, since the age transition
to  the clusters with t$\lesssim$10\,Myr is not sharply defined and, in addition, there are structures such as Sco OB2
that include clusters in both age ranges, that very likely belong to correlated star forming processes.
%Since the low mass tail of these clusters include the highest fraction of the entire population, likely around
%80\% \citep{lada06}, they can be used for several analysis with the caveat 
%%%%%%%%%%%%%%%%%%%%%%%%%%%%%% 
\subsection{Spatial distribution}
\begin{figure*}[!h]
\centering
\begin{minipage}{0.63\linewidth}
\includegraphics[width=13cm,angle=0]{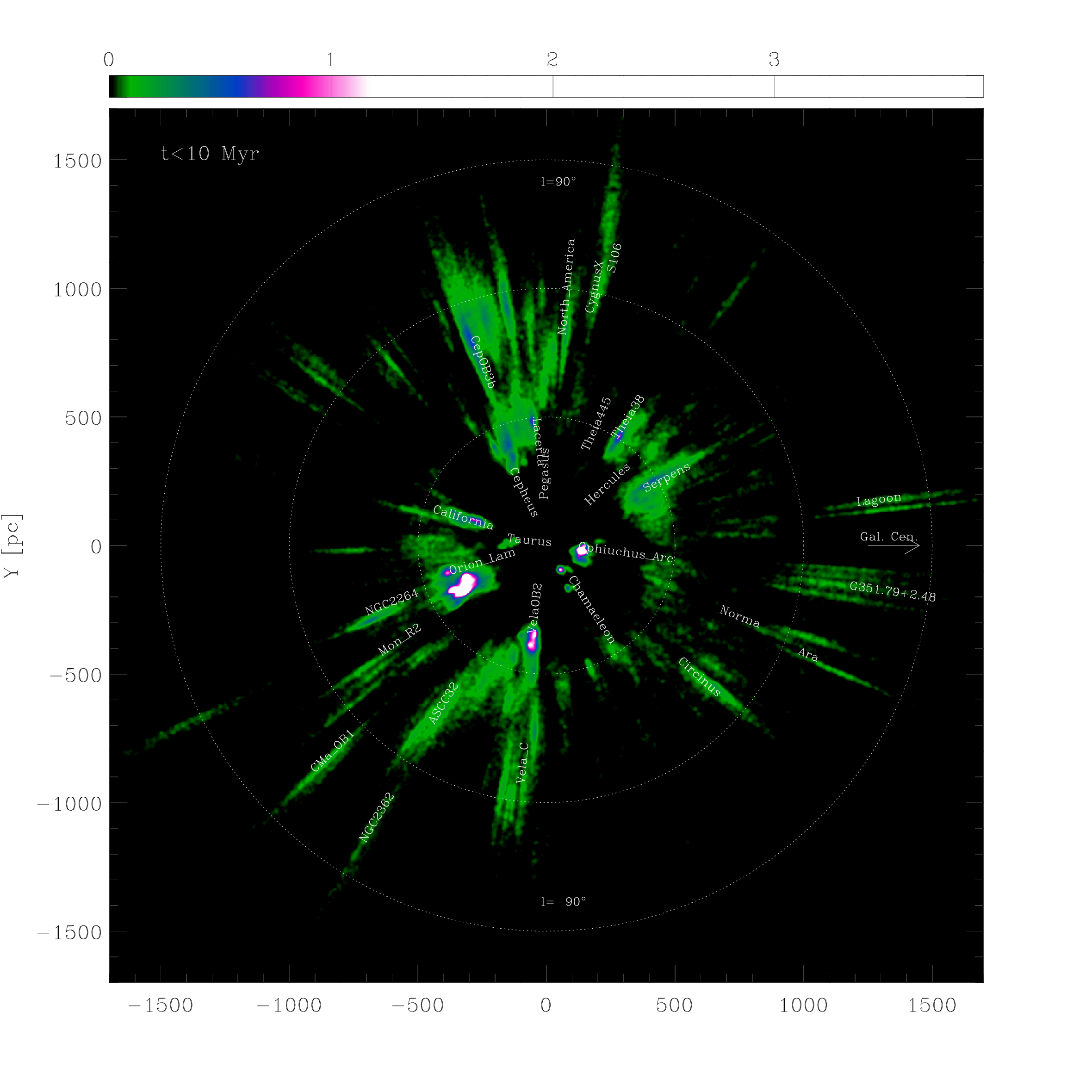} 
\end{minipage}
\begin{minipage}{0.33\linewidth}
\includegraphics[width=13cm,angle=90]{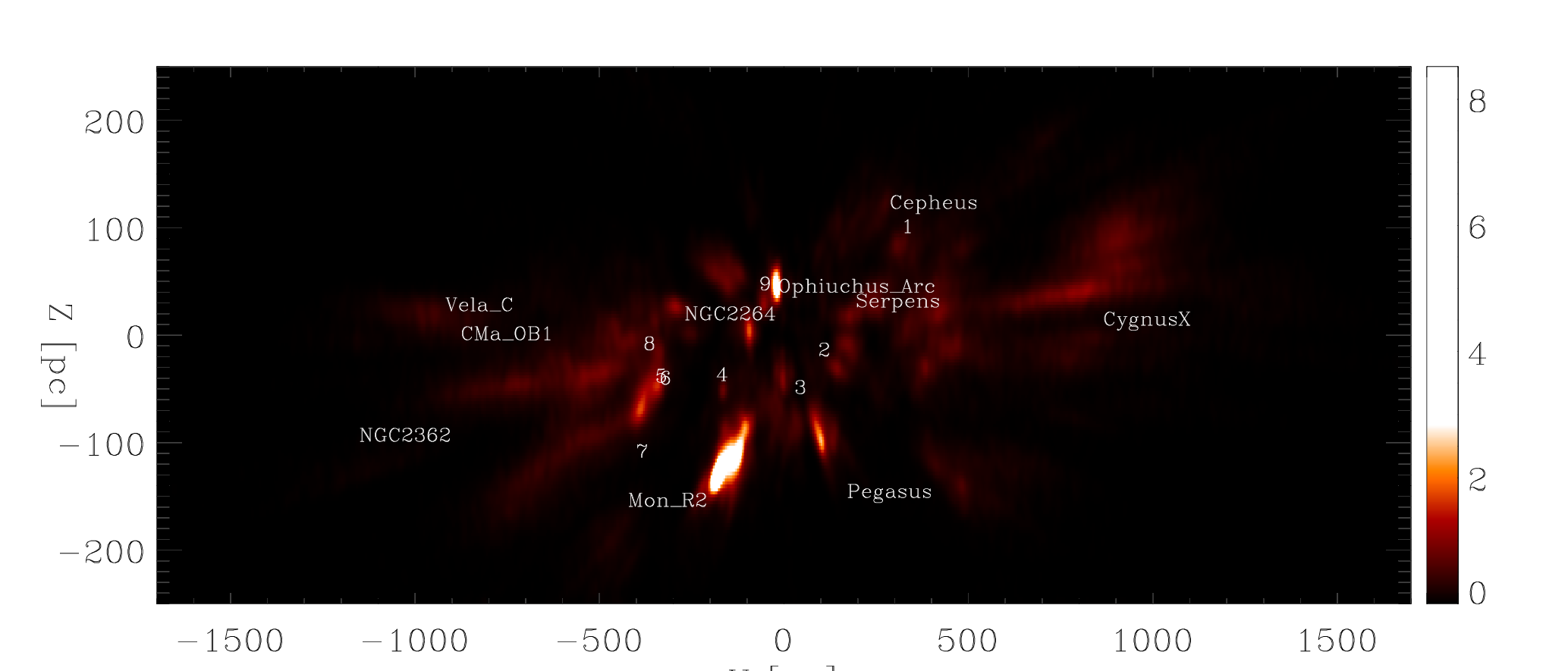} 
\end{minipage}
\begin{minipage}{0.97\linewidth}
\includegraphics[width=13cm,angle=0]{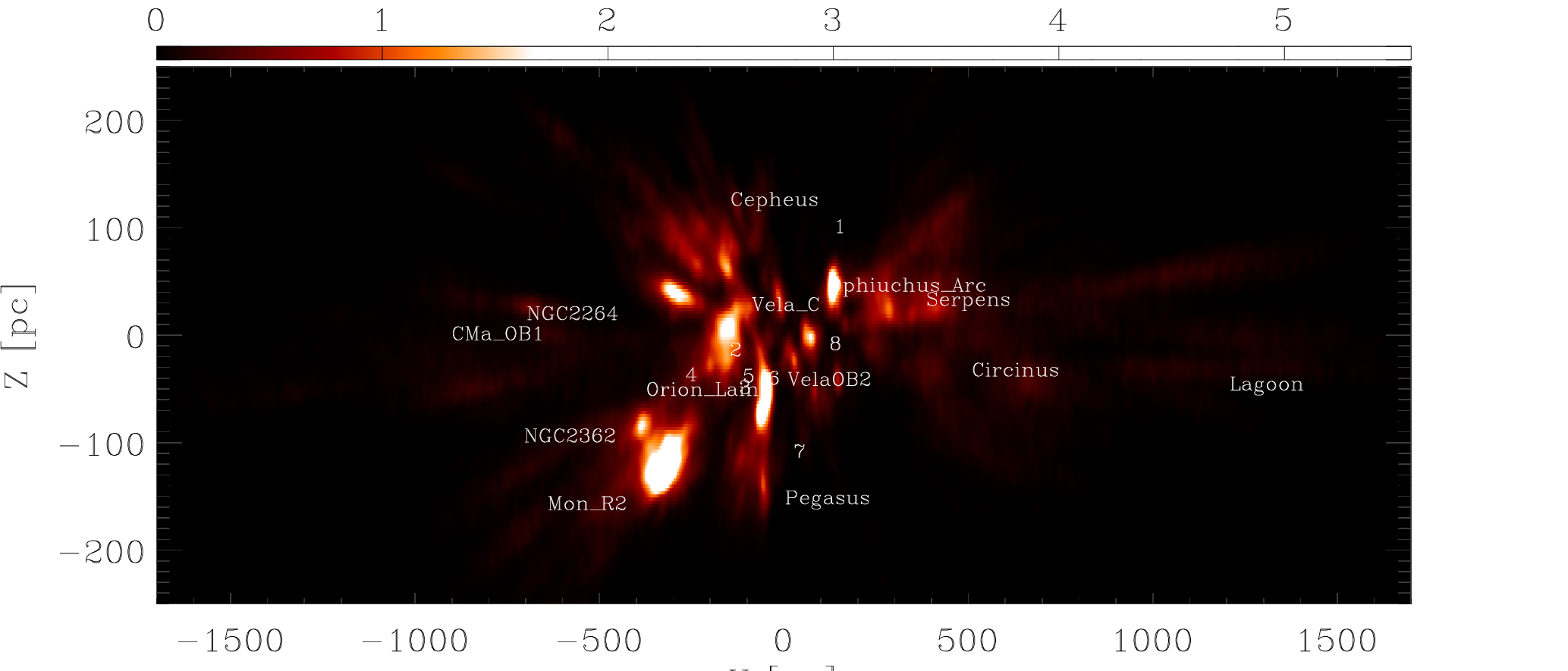} 
\end{minipage}
\caption{Density map on the three orientations of the GP of the YSOs associated to SFRs with
ages $t\lesssim 10$\,Myr. 
      In the upper left panel, the Sun is at (0, 0), the x-axis is directed towards the Galactic Centre, 
      and the y-axis towards
      the direction of the galactic rotation. Dashed white circles are drawn at distance steps of 500\,pc.
      In the upper right and lower left panels, the z-axis is perpendicular to the GP.
      Color bars indicate the surface densities, i.e. the number of stars per bin and per pc$^2$.
      Some known SFRs are indicated. The numbers from 1 to 9 in the upper right and bottom panels,
      indicate the position of the clusters indicated in Fig.\,\ref{xyzmapidage} (upper left panel).}
\label{xyzmapcl}
\end{figure*}
%%%%%%%%%%%%%%%%%%%%%%%%%%%%%% 
 %%%%%%%%%%%%%%%%%%%%%%%%%%%%%%   
\begin{figure*}[!h]
\centering
\begin{minipage}{0.63\linewidth}
\includegraphics[width=13cm,angle=0]{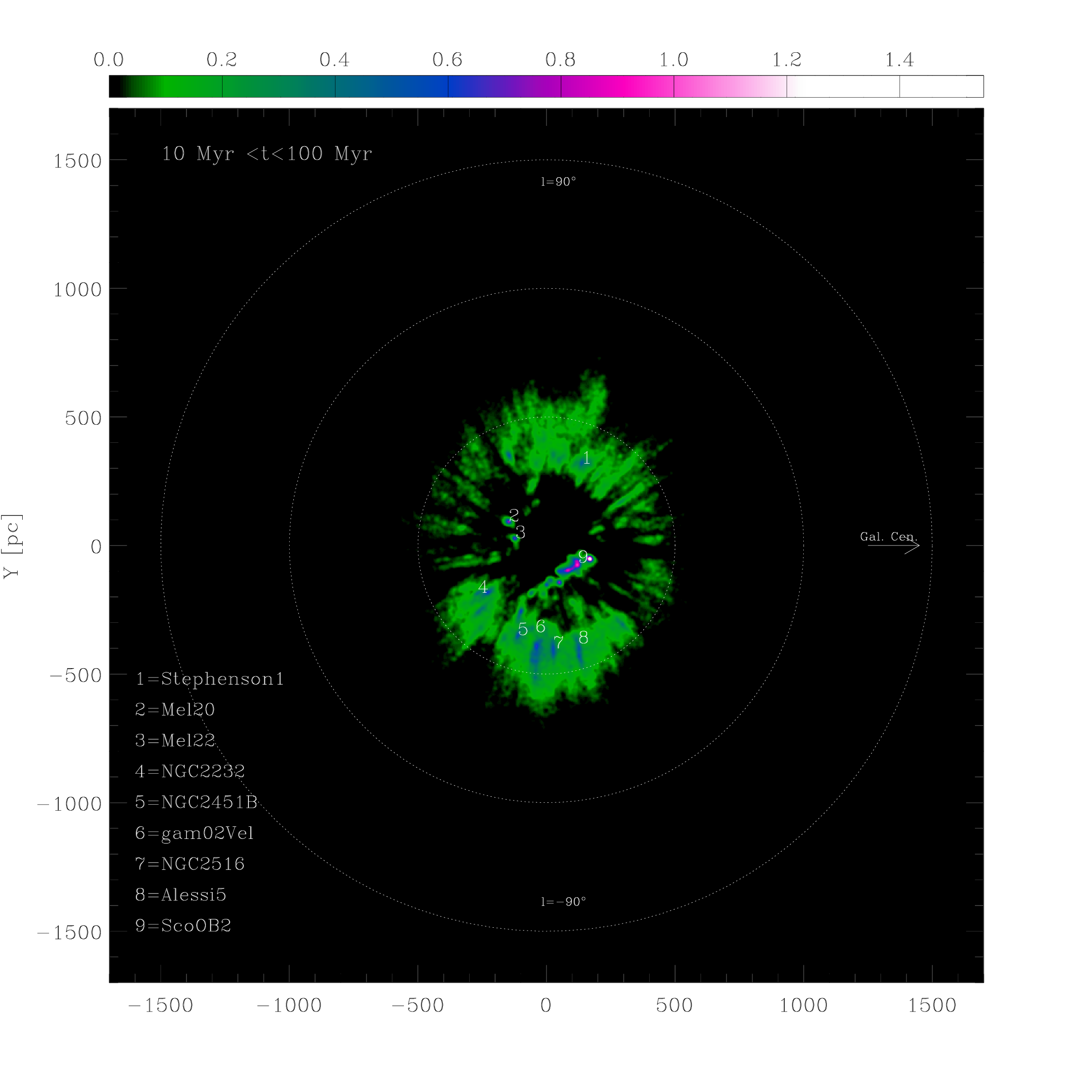} 
\end{minipage}
\begin{minipage}{0.33\linewidth}
\includegraphics[width=13cm,angle=90]{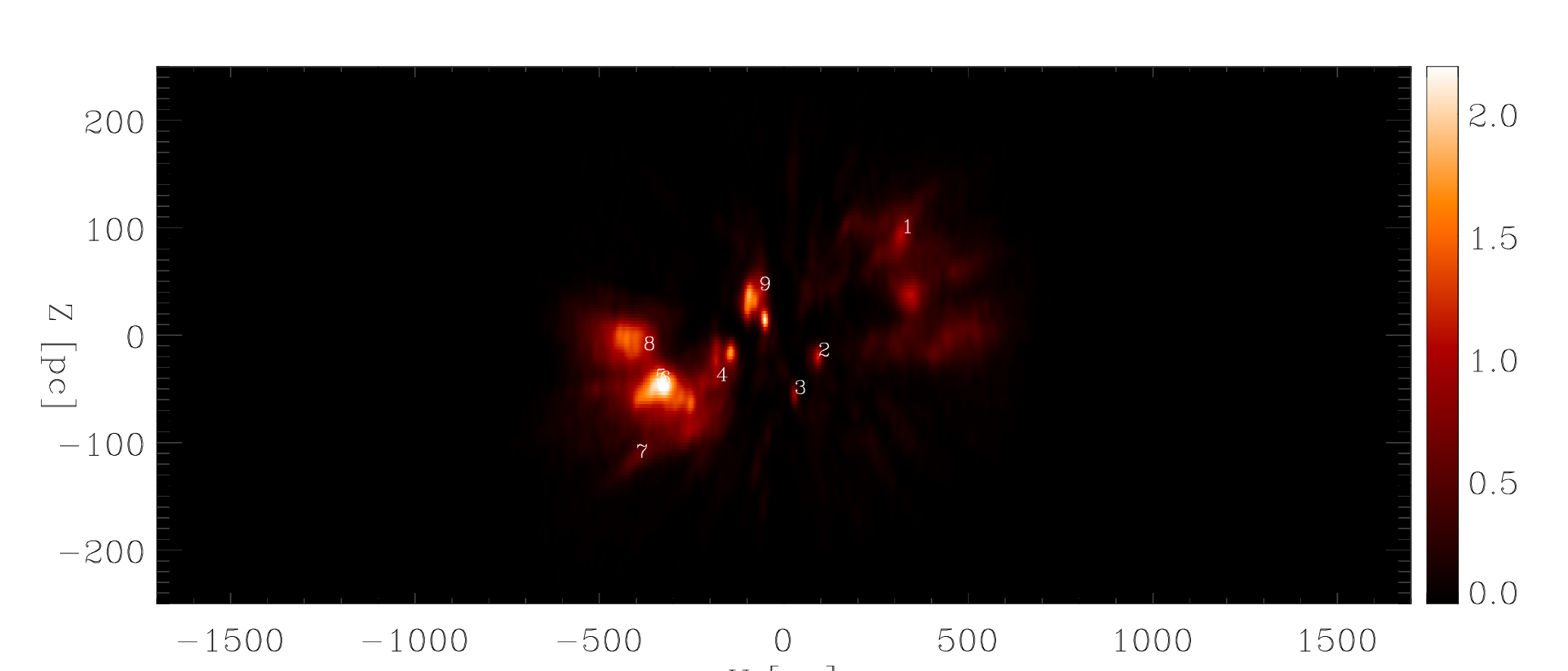} 
\end{minipage}
\begin{minipage}{0.97\linewidth}
\includegraphics[width=13cm,angle=0]{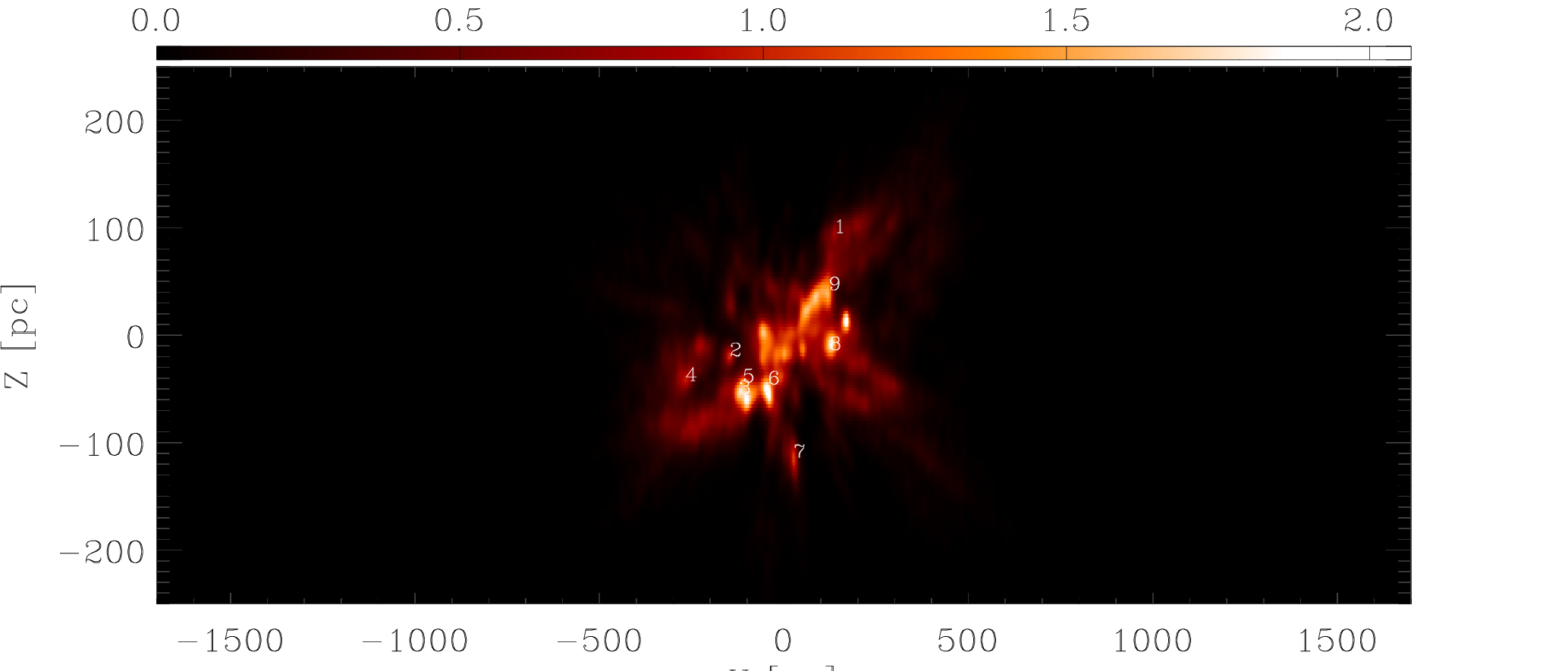} 
\end{minipage}
\caption{Same as Fig.\,\ref{xyzmapcl} but for stars  associated to clusters with ages 10\,Myr$\lesssim t\lesssim100$\,Myr.}
\label{xyzmapidage}
\end{figure*}
%%%%%%%%%%%%%%%%%%%%%%%%%%%%%%
\longtab{
\begin{longtable}{c c c c c c c c}
\caption{SFRs with $t\lesssim 10$\,Myr (flag from 1 to 28) and  young  stellar clusters
with 10\,Myr$\lesssim t \lesssim 100$\,Myr (flag from 29 to 36)  found in this work.
r$_{50}$ is the radius in which half of the identified members are concentrated,
$d$ is the distance obtained by inverting the median value of the member parallaxes  and
N is the number of members.
 \label{printtexclusters}}\\
\hline\hline
ID &  $l$     & $b$     & r$_{50}$ & $d$  & N  & Flag & Name \\
   & [deg]    & [deg]   &  [deg]   & [pc] &    &      &      \\
\hline
\endfirsthead
\caption{continued.}\\
\hline\hline
ID &  $l$     & $b$     & r$_{50}$ & $d$  & N  & Flag & Name \\
   & [deg]    & [deg]   &  [deg]   & [pc] &    &      &       \\
\hline
\endhead
\hline
\endfoot
   1 &   -96.449 &      0.072 &     12.478&   472&  2275 &   36 & Trumpler\,10\\
   2 &   -94.538 &     -8.490 &     11.040&   465&  2290 &   30 & NGC\,2547\\
   3 &   -94.465 &     -7.524 &      6.055&   181&   119 &   32 &\\
   4 &   -93.624 &     -3.647 &      5.498&   727&  1041 &   16 & IC\,2395\\
   5 &   -92.666 &    -13.246 &      4.699&   260&    59 &   32 &\\
   6 &   -91.409 &     -6.554 &      7.230&   360&   534 &   32 & UPK\,535\\
   7 &   -90.205 &      3.832 &      6.919&   530&   265 &   36 &\\
   8 &   -89.938 &     -0.530 &      1.120&  1157&    60 &    7 &\\
\end{longtable}
}

\begin{table*}
\caption{Gaia EDR3 data of members of SFRs with $t\lesssim 10$\,Myr (flag from 1 to 28) and  young  stellar clusters
with 10\,Myr$\lesssim t \lesssim 100$\,Myr (flag from 29 to 36)  found in this work.
The full table is available as an electronic table via the CDS. 
 \label{printtexmembers}}
\centering
\begin{tabular} {c c c c c c c c c c c c c c }
\hline\hline
Gaia EDR3 ID &  $l$     & $b$  & $\pi$  & $\sigma_\pi$ & $\mu_\alpha*$ & $\sigma_{\mu_\alpha*}$ & 
$\mu_\delta$  & $\sigma_{\mu_\delta}$ & $G$    & $G_{BP}$ & $G_{RP}$ & Flag & Cl. ID \\
              & \scriptsize{[deg]}    & \scriptsize{[deg]}&  \scriptsize{[mas]} & \scriptsize{[mas]}&
 \scriptsize{[mas yr$^{-1}$]} & \scriptsize{[mas yr$^{-1}$]}          & 
\scriptsize{[mas yr$^{-1}$]} & \scriptsize{[mas yr$^{-1}$]}         & \scriptsize{[mag]} &
\scriptsize{[mag]}  & \scriptsize{[mag]}  &      &            \\
\hline
\scriptsize{      5307477283092480512} & \scriptsize{ 276.491} & \scriptsize{  -2.559} & \scriptsize{   1.925} & \scriptsize{   0.071} & \scriptsize{ -13.112} & \scriptsize{   0.093} & \scriptsize{   8.232} & \scriptsize{   0.079} & \scriptsize{  17.27} &\scriptsize{  18.56} &\scriptsize{  16.13} &\scriptsize{ 30} &\scriptsize{ 1} \\
\scriptsize{      5310444177789318784} & \scriptsize{ 275.554} & \scriptsize{  -3.171} & \scriptsize{   1.923} & \scriptsize{   0.265} & \scriptsize{ -12.888} & \scriptsize{   0.288} & \scriptsize{   8.518} & \scriptsize{   0.327} & \scriptsize{  19.40} &\scriptsize{  20.85} &\scriptsize{  18.02} &\scriptsize{ 30} &\scriptsize{ 1} \\
\scriptsize{      5310482871168676352} & \scriptsize{ 276.316} & \scriptsize{  -2.671} & \scriptsize{   1.846} & \scriptsize{   0.036} & \scriptsize{ -14.232} & \scriptsize{   0.044} & \scriptsize{   8.038} & \scriptsize{   0.039} & \scriptsize{  16.31} &\scriptsize{  17.24} &\scriptsize{  15.27} &\scriptsize{ 30} &\scriptsize{ 1} \\
\scriptsize{      5310525477245693696} & \scriptsize{ 275.730} & \scriptsize{  -2.877} & \scriptsize{   1.901} & \scriptsize{   0.030} & \scriptsize{ -12.997} & \scriptsize{   0.034} & \scriptsize{   8.875} & \scriptsize{   0.029} & \scriptsize{  15.63} &\scriptsize{  16.57} &\scriptsize{  14.66} &\scriptsize{ 30} &\scriptsize{ 1} \\
\hline
\end{tabular}
\end{table*}

Figure\,\ref{xyzmapcl} shows the maps of the \youngstars\, YSOs associated to the \totsfr\, SFRs with 
ages $t\lesssim10$\,Myr, 
while Fig.\,\ref{xyzmapidage} shows the maps of the \midagestars, stars  associated to the stellar clusters
with ages  10\,Myr$\lesssim t\lesssim100$\,Myr.
Each map has been obtained as two-dimensional histogram smoothed with a Gaussian kernel at 3$\sigma$,
adopting a pixel size of 3\,pc$\times$3\,pc.

Most of the overdensities in Fig.\,\ref{xyzmapcl}
are associated to known SFRs, some of which are labeled in the figure.
With the exception of those within 200-300\,pc, all clusters present a 
radial elongated shape, tracing the increasing uncertainties in the distances. 
  
The clusters with ages  10\,Myr$\lesssim t\lesssim100$\,Myr are mainly limited within $\sim600$\,pc
(see Fig.\,\ref{xyzmapidage})
and show a much more  diffuse spatial distribution. 
Very rich clusters such as, for example, NGC\,2232, NGC\,2451B, Gamma Velorum,  NGC\,2547,
NGC\,2516, Alessi\,5 at distance of $\sim 400$\,pc, seem to belong to a common
giant complex, mostly lying in the third Galactic quadrant.
\subsection{Literature comparison} 
In this section, we present the comparison of our results with those previously obtained in the literature 
for two particular regions, Sco OB2 and NGC\,2264.  
These comparisons will be used to estimate our completeness and the contamination level,
at least when the completeness of the comparison sample enable us to do it.
We note that we will consider here each of the merged clusters as a unique ensemble.
 A  detailed subclustering  analysis of  them, with the identification  of  possible
substructures with age gradient or kinematic
 subclusters will be deferred  to a future paper. A detailed comparison with the literature 
 for other SFRs is presented in the Appendix\,\ref{literaturecompapp}, where we 
 also compare the whole catalogue with other all-sky catalogues, mainly derived with {\it Gaia} DR2 data. 
\subsubsection{The Sco-OB2 association}
   \begin{figure}
   \centering
\includegraphics[width=9cm]{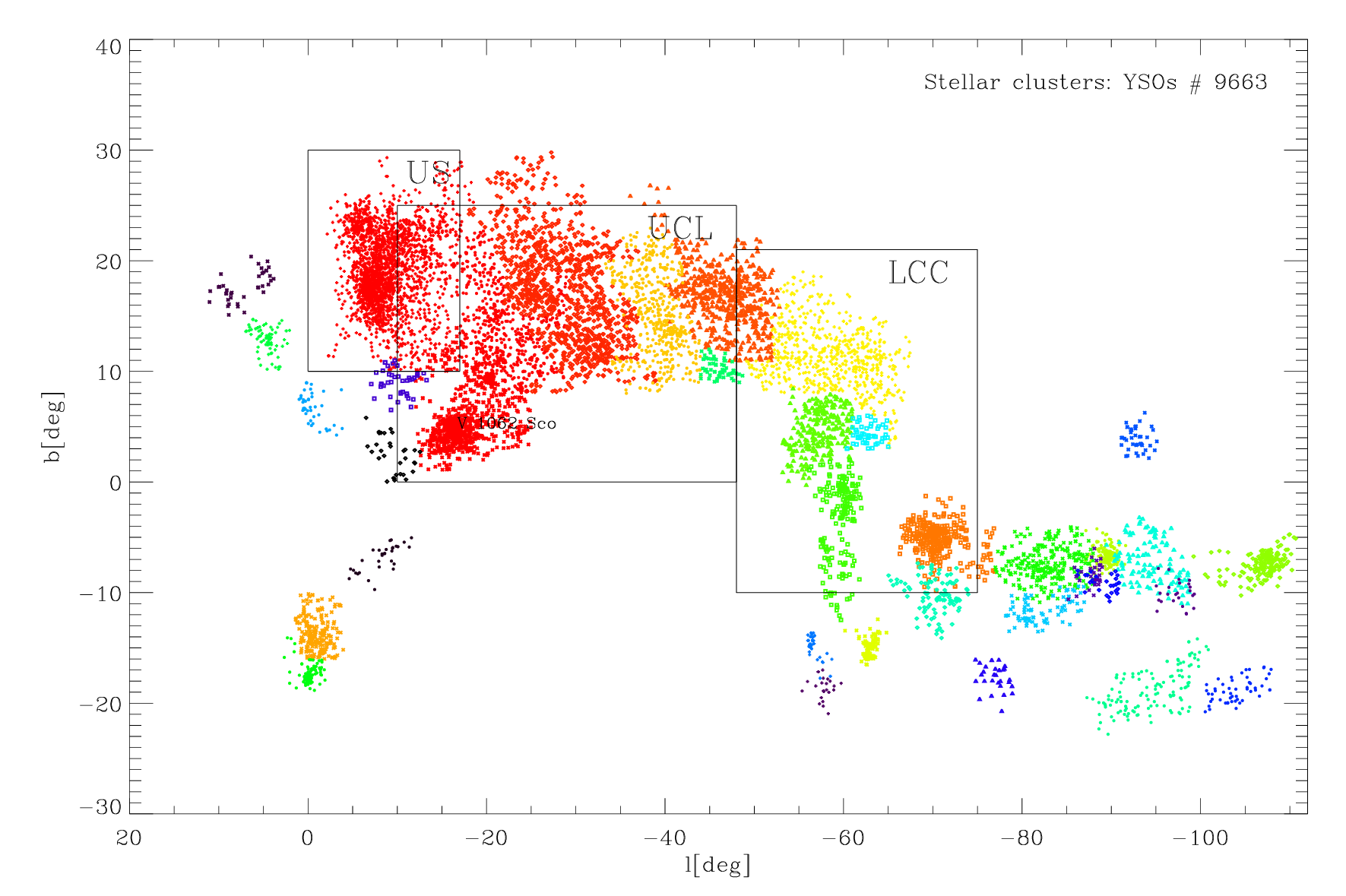}
      \caption{Spatial distributions in  Galactic coordinates of the 
       YSOs associated to the Sco-OB2 Association. The \citet{deze99} subregions of Upper Sco (US),
       Upper Centaurus-Lupus (UCL) and Lower Centaurus-Crux (LCC) are shown. The different colors
    indicate all the different substructures found in this region.}
         \label{lbscoob2}
   \end{figure}
%%%%%%%%%
\begin{figure*}[!h]
\centering
\begin{minipage}{0.32\linewidth}
\includegraphics[scale=0.3,angle=0]{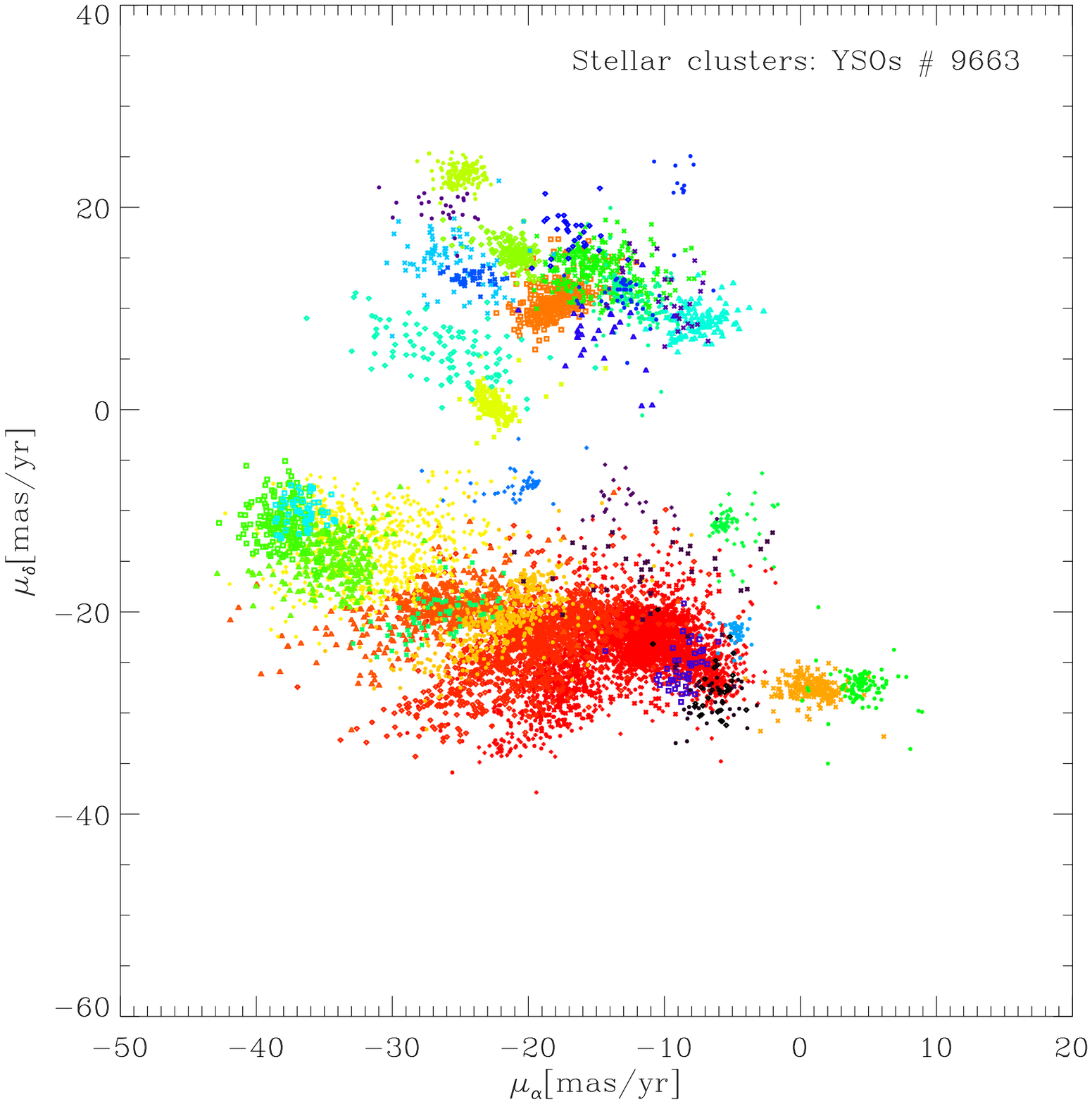} 
\end{minipage}
\begin{minipage}{0.32\linewidth}
\includegraphics[scale=0.3,angle=0]{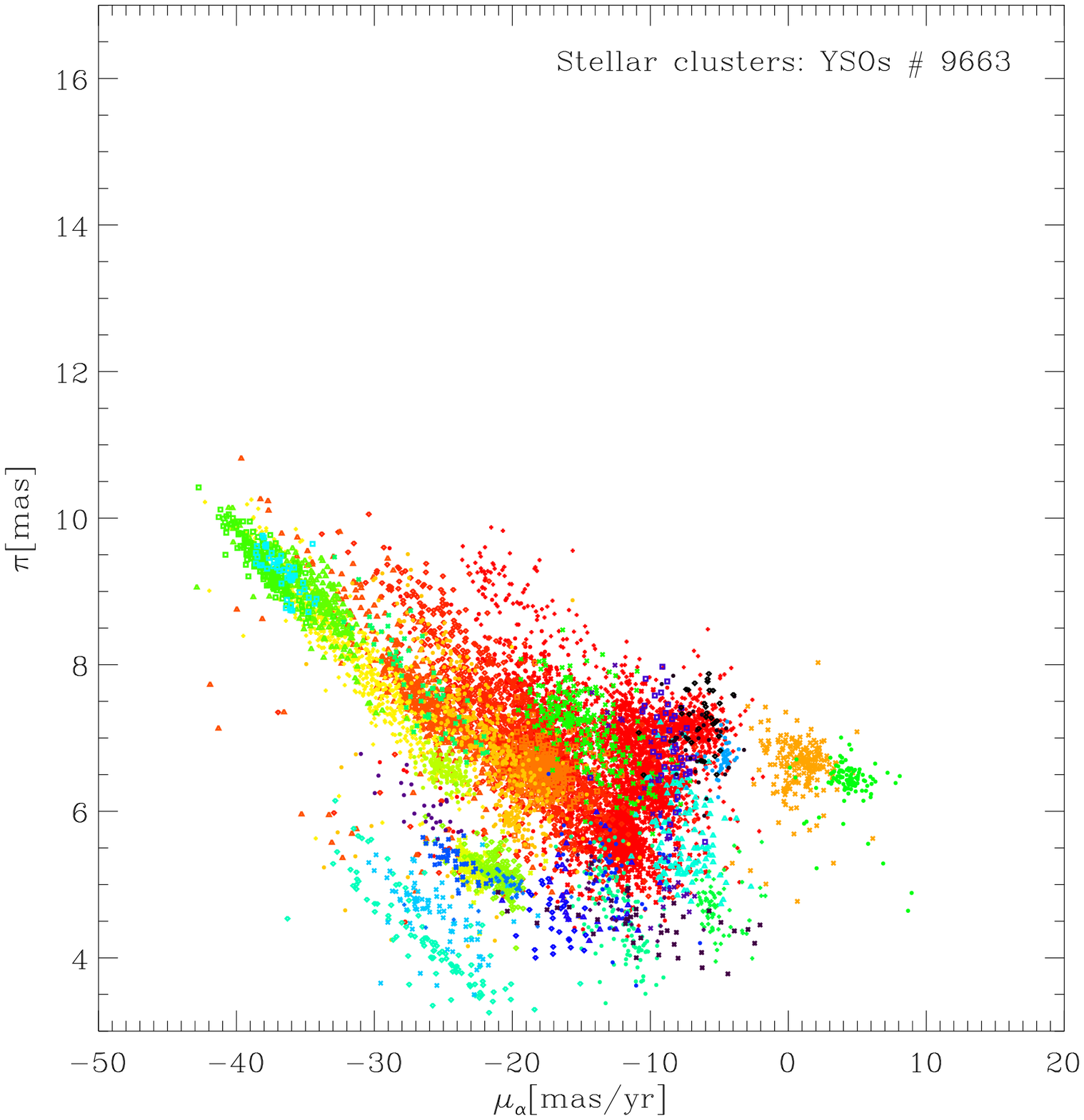} 
\end{minipage}
\begin{minipage}{0.32\linewidth}
\includegraphics[scale=0.3,angle=0]{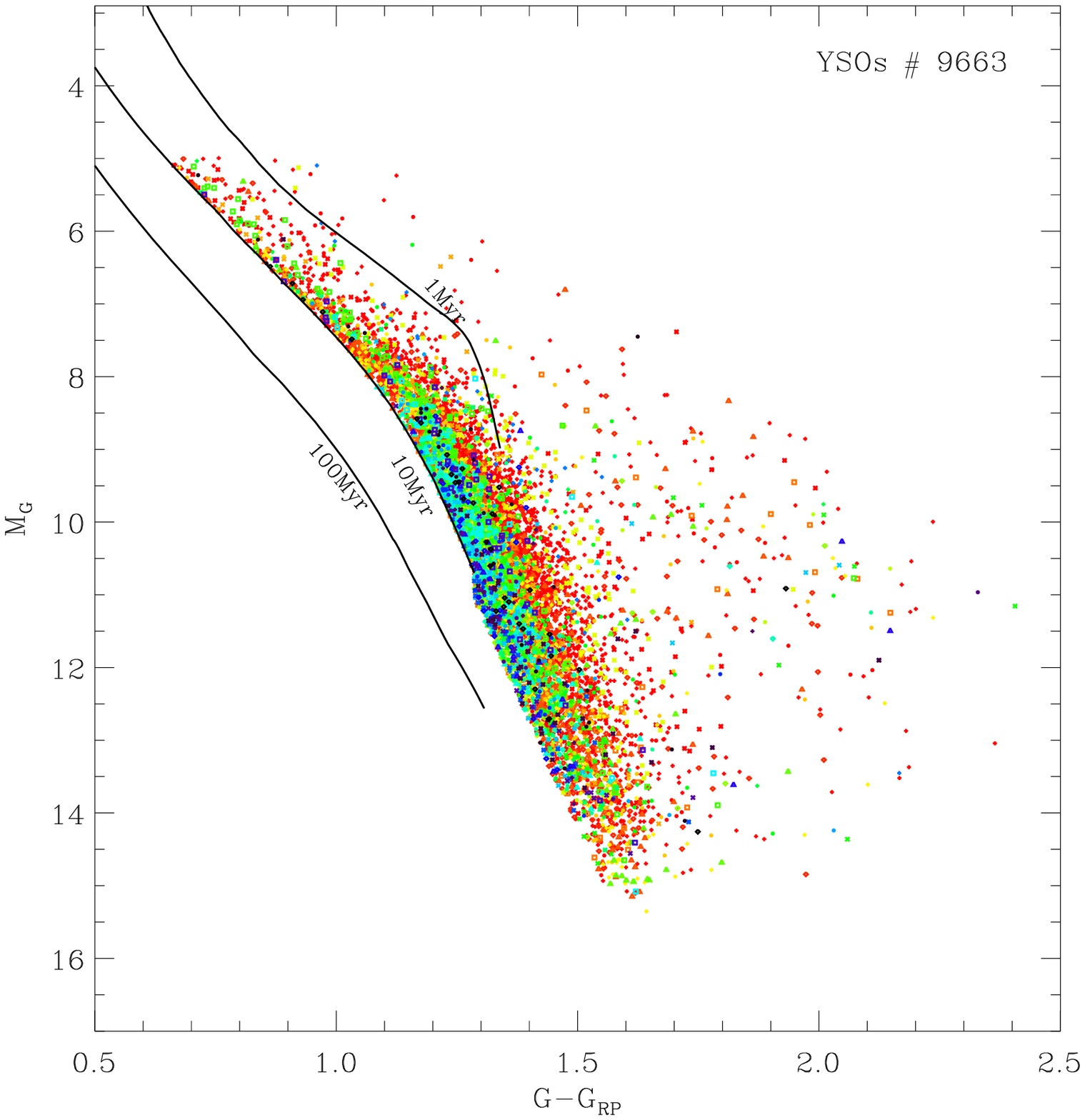} 
\end{minipage}
%\begin{minipage}{0.32\linewidth}
%\includegraphics[scale=0.3,angle=0]{scoob2_pmra_pmdec_mov_g.ps} 
%\end{minipage}
%\begin{minipage}{0.32\linewidth}
%\includegraphics[scale=0.3,angle=0]{scoob2_pmra_plx_mov_g.ps} 
%\end{minipage}
%\begin{minipage}{0.32\linewidth}
%\includegraphics[scale=0.3,angle=0]{scoob2_mg_grp_mov_g.ps} 
%\end{minipage}
\caption{Proper motions in RA and Dec, parallaxes and CAMDs of the groups of 
YSOs associated to the Sco OB2 association. The symbol colors are as in Fig.\,\ref{lbscoob2}.
 Three representative  solar metallicity isochrones from the Pisa models
 are also shown as solid lines in the right panels.}
\label{scoob2all}
\end{figure*}   
   
The Sco-Cen or Sco-OB2 association is a very  extended SFR ($\sim$120\deg$\times$60\deg)
quite close to the Sun ($d\sim150$\,pc) that, in the last years, has been the subject of 
several studies focused on the low mass population.
By exploiting available all-sky surveys, these studies finally allowed to study the entire region and its complexity
 \citep[e.g.][]{zari18,dami19,koun19,kerr21,luhm22}.

To select the members of this region, 
we performed a spatial selection by considering all  stars with 
$-102$\deg$<l<10$\deg,  $-30$\deg$<b<40$\deg\,
%scoob2_lb.pro
%lmin=-102
%lmax=10
%bmin=-30
%bmax=40 
and, as assumed in \citet{dami19}, distance $d<200$\,pc.
% scoob2_lb.pro 
% ind_sfr
We end up with a total of 9\,663 YSOs with ages  $t\lesssim$100\,Myr, 
distributed as in Fig.\,\ref{lbscoob2}.
In the ($l,b$) plane,  the pattern of the YSOs 
associated to Sco-OB2 is that already found in the literature \citep[e.g.][]{deze99,dami19,kerr21}.
%help,VERY_YOUNG      LONG      = Array[4232]
Among the selected objects, 4\,232 YSOs have been classified in the range $t<10$\,Myr.
Those concentrated in the Upper Sco (US) region are 2\,472. % see flag 22
They correspond to the youngest subpopulation of Rho Ophiuchi.
Another prominent subpopulation, classified  in the range 10\,Myr$\lesssim t \lesssim$100\,Myr (flag 31), 
includes  3\,741 YSOs, falling in the Upper Centaurus-Lupus (UCL) and Lower Centaurus-Crux (LCC)
regions. It then represents the first generation of stars of the Sco OB2 region, in agreement with  recent results
 \citep[e.g.][]{dami19,luhm22}.

Proper motions, 
parallaxes and the CAMDs of the different subpopulations detected in the Sco OB2 association are shown in   
Fig.\,\ref{scoob2all}.
 The proper motions of the YSOs
associated to Sco-OB2 show a quite complicated pattern, confirming the complex  kinematic structure 
 of this association.
 The values of parallaxes of YSOs in  Sco-OB2 are 
 mostly enclosed between $\sim5$\,mas and $\sim10$\,mas, corresponding to a mean distance of 152\,pc
 and standard deviation $\sigma=28$\,pc. %see output scoob2_lb.pro
  Finally,  
 in the CAMD,  we can recognise a usual distribution of YSOs in the PMS region. As already noted,
 the census of stars that belong to the first generation of stars in the Sco OB2 association
 is likely incomplete since it is expected to lie in the region of the CAMD that has not been considered
 in this work. 
 
 To estimate the completeness level of our census, we compared our list of Sco-OB2
  YSOs with the ones recently published by \citet{dami19} and
\citet{kerr21}, based on {\it Gaia} DR2 data and by \citet{luhm22}, based on {\it Gaia} EDR3 data. To perform  these comparisons 
we used the Gaia identification number of YSOs in our catalogue, retrieved as 
described in Sect.\,\ref{compareallsky}. We find that the YSOs in common with the \citet{dami19}
catalogue, that includes a total 10\,839 members, are 6\,492. % see scoob2_lb.pro aa_data 
 i.e. about 60\% % 6492./10839.
of the \citet{dami19} list.
%restore,'../SAVEFILES/scoob2_edr3_damiani.save'
%help,sub_sco ;  il match con tutta la mia lista da 6542
% see scoob2_lb.pro aa_data 6492  sono quelle in comune e classificate da noi con flag tra 1 e 36.
Among the 9\,663 YSOs selected by us in the  Sco-OB2 association, those falling in the spatial region
and magnitude range $G<19.5$ covered by  \citet{dami19} are 7\,553. % IND_SFR_SPA in scoob2_lb.pro.
Therefore, the objects in common are  % see aa_data
 86\% of our sample in the same field. % 6492./7553.
Many of the remaining  1\,061 YSOs (14\%) % 7553-6492
not selected by \citet{dami19} show a spatial distribution
consistent with that of the other members and thus we discard the hypothesis that they are contaminants,
and suggest that they are likely  YSOs missed by \citet{dami19}, that is based on the less complete 
{\it Gaia} DR2 catalogue. 

Adopting the same spatial constraints, we retrieved in the
Sco-OB2 region 9\,083 objects, % see sel scoob2_plot_kerr21.pro
 selected as candidate YSOs in the \citet{kerr21} catalogue,
independently from their clustering type of classification.
 Among these, 5\,203 % comm_all  in scoob2_plot_kerr21
are in common with our catalogue but those classified as YSOs are 
5\,109  % comm_yso in scoob2_plot_kerr21
i.e. $\sim56.2\%$ of the \citet{kerr21} sample\footnote{Using the {\it Gaia} DR2 number, we cross-matched 
the \citet{kerr21} and \citet{dami19} lists and found 6\,423  objects in common.}.

% see scoob2_plot_kerr21

The \citet{luhm22} catalogue includes a total of 10\,509 YSOs but to be consistent with 
our selection, we selected those with $M_G>5$, 7.5$<G<$20.5, $RUWE<1.4$ that are in total 7\,925.
%see scoob2_luhman.pro
%sel_l=where(luhman.gmag LT 20.5 and luhman.gmag GT 7.5 and mg_l GE 5 and luhman.ruwe LT 1.4 and $
%		    luhman.gmag-luhman.rpmag GT 0.58)
Using the {\it Gaia} EDR3 id, we found that the YSOs in common with our catalogue 
are 6\,341 % scoob2_luhman.pro sub_sco
representing 80\% of the \citet{luhm22} catalogue  % 6341./7925.   0.80
and 85.6\% of our list of YSOs in the Sco Cen, by considering  the 
7\,408 % see selspa  scoob2_luhman.pro sub_sco
counterparts falling in the region covered by \citet{luhm22}.

These percentages can not be strictly used to estimate our
level of completeness or contamination, since the catalogues have
been obtained starting from different initial constraints, both for the photometric 
 and the astrometric selection, that inevitably can introduce several biases. 
However, these comparisons are useful to confirm the membership for $\sim$85\%
 of the selected members. 
The remaining 1\,067 % 7408 -6341.
objects not retrieved by \citet{luhm22} but selected by us as YSOs 
show a spatial distribution consistent with that of the other members with two
 strong  concentrations of them in the US region and around V\,1062\,Sco. We therefore conclude
 that they are genuine members rather than contaminants, likely missed by \citet{luhm22} 
 in the   photometric selection based on the $G_{\rm BP}-G_{\rm RP}$ colors.

\subsubsection{The Monoceros OB1/NGC\,2264 complex and the Rosette Nebulas}
    \begin{figure}
   \centering
\includegraphics[width=9cm]{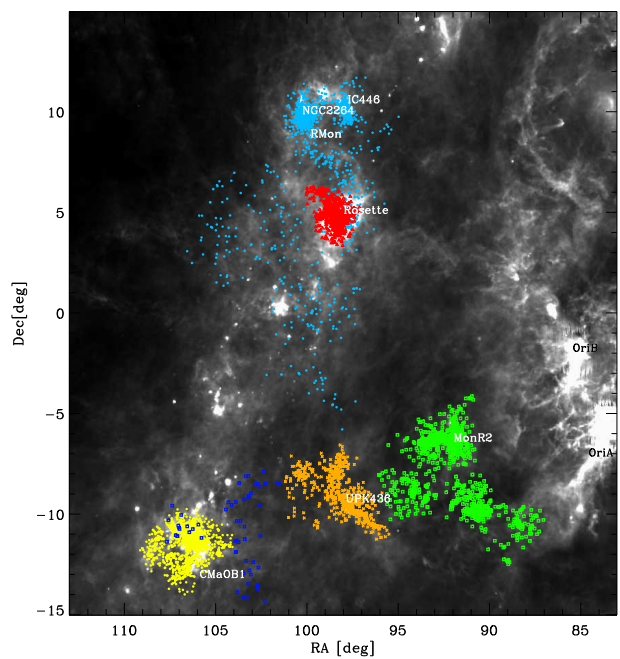}
      \caption{Spatial distribution in  Equatorial coordinates of the 
       YSOs associated to NGC\,2264, NGC\,2244, Mon R2, CMA OB1 and  UPK\,436.  YSOs are 
overplotted on a IRIS 100\,$\mu$m image \citep{mivi05}. For clarity, Orion members have not been
plotted.}
         \label{radecngc2264}
   \end{figure}

\begin{figure*}[!h]
\centering
\begin{minipage}{0.32\linewidth}
\includegraphics[scale=0.3,angle=0]{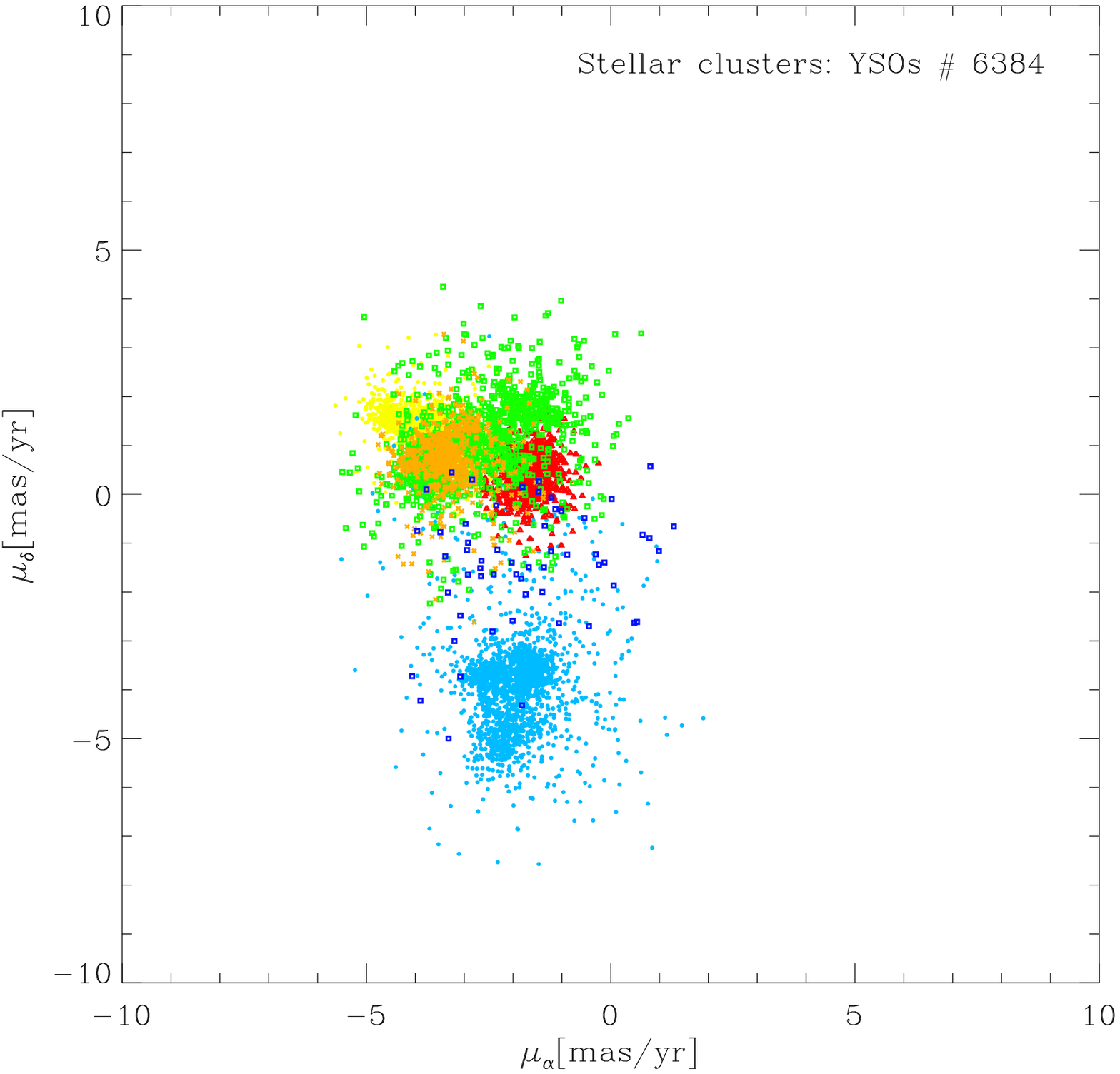} 
\end{minipage}
\begin{minipage}{0.32\linewidth}
\includegraphics[scale=0.3,angle=0]{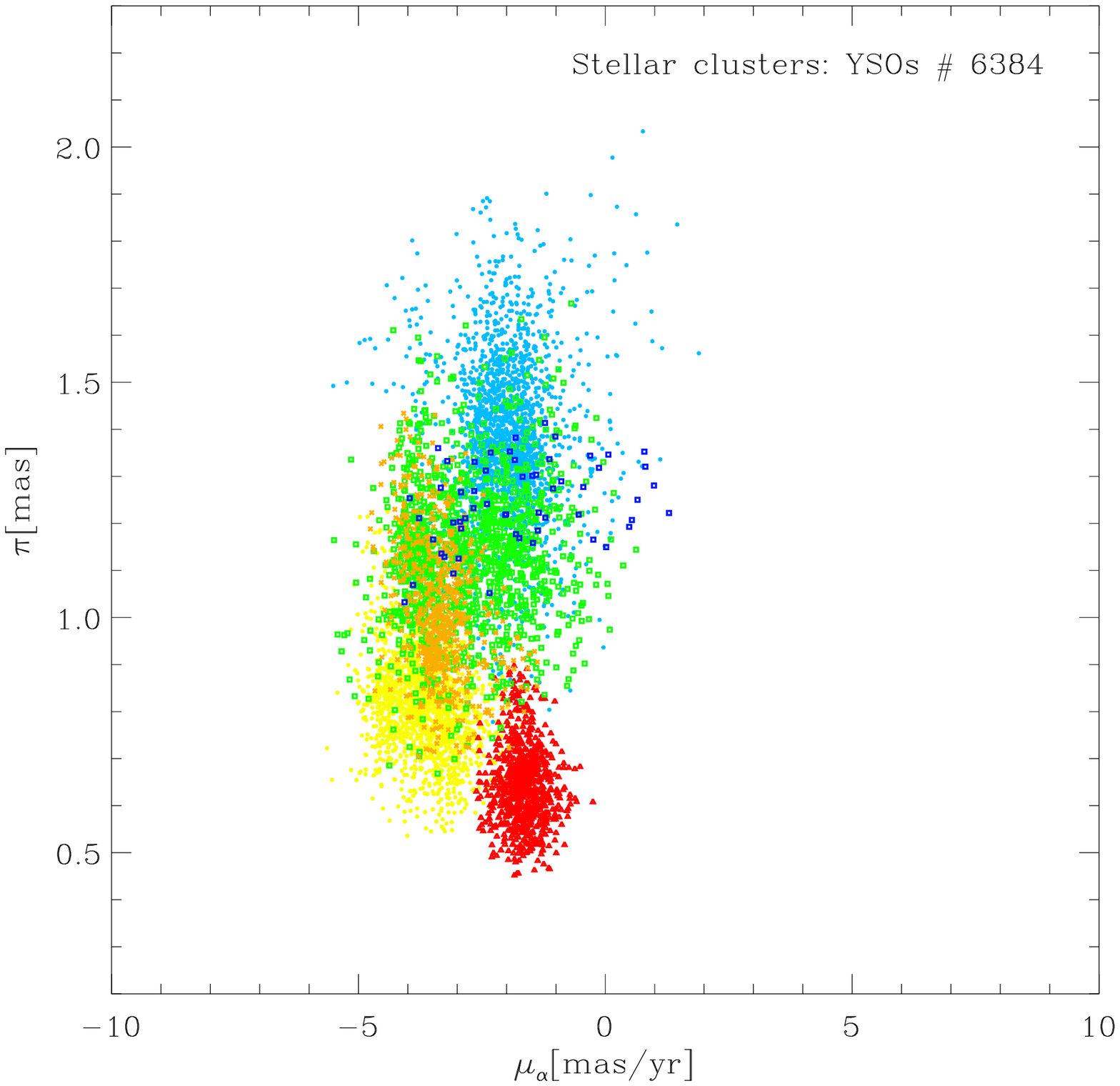} 
\end{minipage}
\begin{minipage}{0.32\linewidth}
\includegraphics[scale=0.3,angle=0]{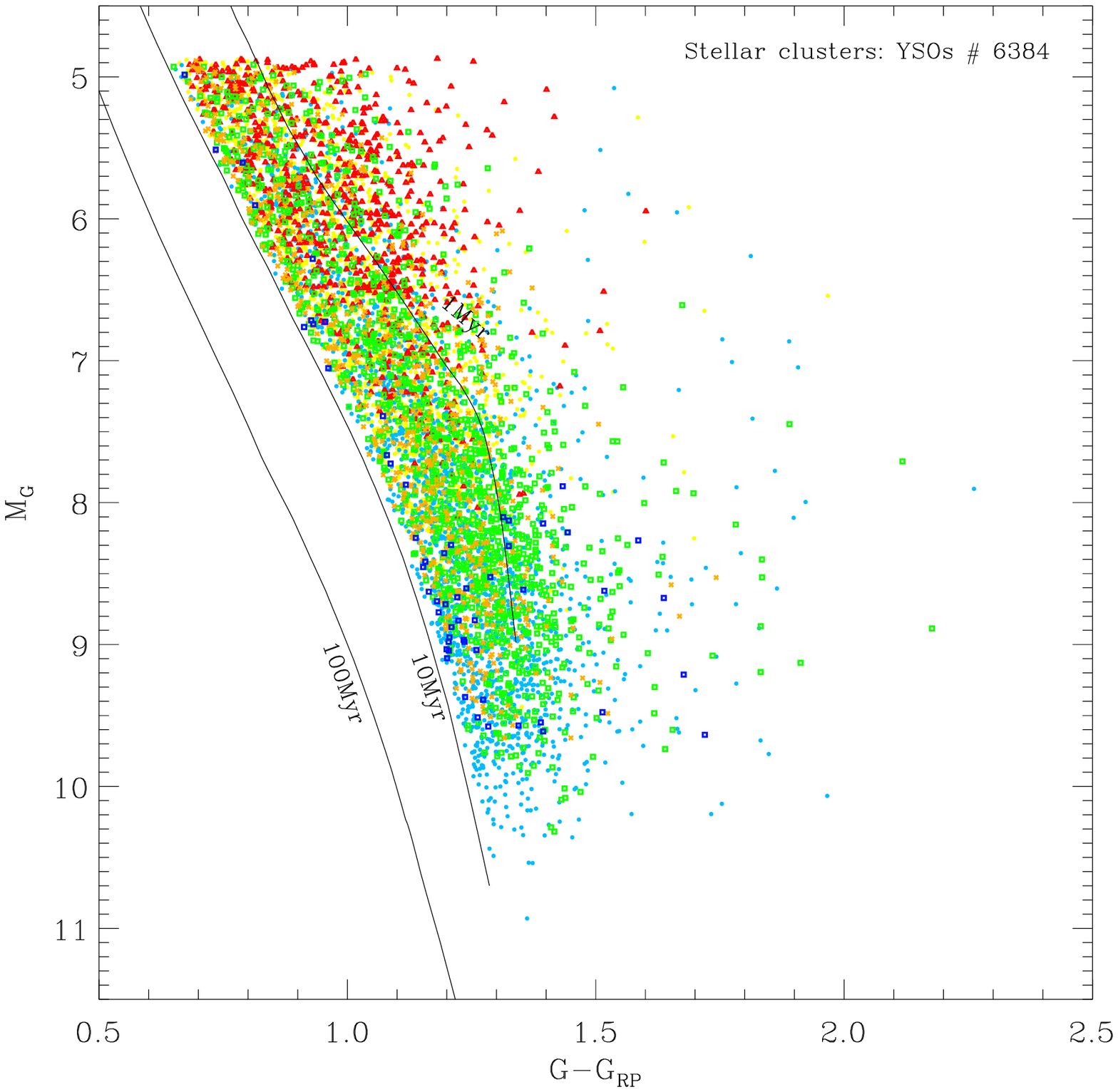} 
\end{minipage}
\caption{Proper motions in RA and Dec, parallaxes and CAMDs of the SFRs falling in the field 
of view of NGC\,2264. The symbol colors of the clusters are as in Fig.\,\ref{radecngc2264}.
 Three representative solar metallicity isochrones computed
 from the Pisa models are 
also shown.}
\label{ngc2264all}
\end{figure*}
Another well studied region that we used to test our results is the cluster NGC\,2264 in the
Monoceros OB1 complex. This relatively compact and close ($\sim720$\,pc from the Sun)
SFR, devoid of background and foreground emission,
has been the subject of many detailed studies, including, for example, X-ray observations \citep{flac06},
optical and near IR analysis of its low mass population \citep{venu19},
 coordinated synoptic investigations
  with Optical and Infrared light curves with CoRot and Spitzer \citep{cody14}.
Flaccomio et al. (2022, in preparation) compiled the most complete  data set
of NGC\,2264, based both on all-sky surveys ({\it Gaia} EDR3, 2MASS, VPHAS) and 
 dedicated observations falling in the region 
%ettore=mrdfits('../FLACCOMIO/211102_members_tab_2264_flaccomio.fits',1) 
%IDL> ra_lim=minmax(ettore.ra)
%IDL> de_lim=minmax(ettore.dec)
98.93\deg$<RA<101.47$\deg\,
and 8.45\deg$<Dec<10.95$\deg.  
The young structure identified by us in this field includes a total of 1\,916 YSOs, but
%IDL> restore,'../SAVEFILES/cluster_classification_prob.save'
%restore,'../SAVEFILES/ks2d_classification.save'
%IDL> sel_2264=where(data.cluster_id_merge EQ 12 )           
%IDL> help,sel_2264
%SEL_2264        LONG      = Array[1916]
% see radec_ngc2264
%help,gr1,gr2,gr3
%GR1             LONG      = Array[450] ; ROsette zone
%GR2             LONG      = Array[1062] ; 2264
%GR3             LONG      = Array[404] ;IC4462
only 1\,062 of them ($\sim55$\%) fall in the region investigated by Flaccomio et al. (2022, in preparation).
The remaining YSOs are in part (404 YSOs) concentrated 
in the region corresponding to the cluster IC\,446, while a further unknown 
group of 450 YSOs are sparsely distributed in the Southern region of NGC\,2264.
As shown in Fig.\,\ref{radecngc2264}, a sub-group of these latter form a visual bridge 
along a filamentary structure, clearly visible in the IR IRIS image, down  to the location 
corresponding to the more distant  Rosette Nebula,
 located at $\sim$1.5\,Kpc, hosting the SFR NGC\,2244. 
Thus, our finding is that the known cluster NGC\,2264 actually belongs to a 
structure larger than the $\sim$2\deg$\times\sim2$\deg region, typically considered in the literature for this SFR.
The mean distance of YSOs associated to the complex NGC2264-IC\,446
 is 731.86$\pm$95.5\,pc, % see radec_ngc2264.pro
 even though the proper motion distributions of the three subgroups suggest they share
 similar but not equal values.\footnote{A detailed kinematic analysis of these sub-regions
 is beyond the aims of this work.}
 
In the same region, we identified further 5 sub-structures with distance
 $>0.5$\,Kpc\footnote{This  limit has been adopted to avoid the Orion sub-structures}, 
the most populated being
the cluster in the CMa\,OB1 association, 
centered around RA=106.3\deg\, Dec=-11.47\deg, at a distance of  
1250$\pm$162\,pc,
associated to the Reflection Nebula IC\,2177, including 1709 YSOs. In addition,
we identified the cluster  NGC\,2244, including 810 YSOs, centered at RA=98.3\deg\, Dec=4.9\deg, at a distance of  
1580$\pm$199\,pc and the cluster associated to Mon R2, at a distance of 897$\pm$112\,pc,
including 1272 YSOs. In addition, we detected the cluster indicated in \citet{cant20} as UPK\,436
with 620 members and a minor sparse cluster in the region of CMa\,OB1 located at 807\,pc.

Figure\,\ref{ngc2264all} shows PM, parallaxes and CAMD of all these substructures,
where it is clearly visible that they are spatially and kinematically uncorrelated,
while in the PMS region of the CAMD they are indistinguishable since they consist of
similar age stars.

The membership defined in Flaccomio et al. (2022) includes two levels of confidences.
One based on the combination of several criteria derived by dedicated X-ray, spectroscopic and IR observations,
including % see ngc2264_ettore_mch.pro
%restore,'../SAVEFILES/gaia_ettore_nomch.save'
%help,nall_c,nall_cwide
2263  confirmed members (sample C) and where the fraction of false positives is negligible
and another list (sample C-Wide),  based exclusively on all-sky surveys, including 1542 YSOs,
where the membership has been deduced by a smaller number of criteria and thus 
the  number of false positives is expected to be higher.
We find that the YSOs of the sample C (sample C-Wide) in common with our list of YSOs in the NGC\,2264 region
%help,nmch_c,nmch_c,nmch_cwide
are 972 (960), corresponding to a fraction of 
% see ngc2264_mg_grp_nomch outpuy
43\% (62\%) with respecto to the Flaccomio sample. These fractions are considered
here as indicators of our level of completeness of the entire SFR population.
However, these results are strongly conditioned by the starting photometric selection ($M_G>5$) and 
the restrictions on the {\it Gaia} EDR3 data that we adopted in this work.
% 498 % 1958-n_elements(n_b) ; 1460=n_b quelli con dati Gaia
% of the 1958  % nall_b tutti i b
In addition, the Flaccomio et al. sample C  includes 497 % 2263-n_elements(n_c) ; 1766=n_c quelli con dati Gaia
 of the 2263  % nall_c tutti i c
 confirmed members that do not have a Gaia counterpart.

To estimate the efficiency of our method to recover  YSOs, we considered the members selected by
Flaccomio et al. with a Gaia counterpart, falling in the  photometric region considered in this work, and  compliant 
with our initial data restrictions (i.e. $RUWE<1.4$ and parallax relative error $<0.2$). 
Adopting this sample, the fraction of the YSOs selected by us in common with the Flaccomio et al. membership
% see output ngc2264_mg_grp_nomch.pro
is 95\%-96\%, considering both the samples C and C-Wide.
We note that this is the efficiency of our clustering method but is not 
the efficiency of the {\it Gaia} data. In fact, if for the two lists, we consider
 the members falling in the same photometric region but we do not consider
the restrictions in  $RUWE$ and in the parallax error, the  fraction of YSOs in common is 72\% 
for sample C and 77\% for sample C-wide. This suggests that
 23\%-28\% of genuine YSOs are missed by us due to remaining issues in the {\it Gaia} data, at least in the
current {\it Gaia} EDR3 release. 

Finally, we find that among the 1\,052 % spa_data of ngc2264_ettore_mch.pro
YSOs selected by us in the NGC\,2264 region, a total 
of 1\,034 % help,ind_mch_data
are included in the list of objects collected by Flaccomio et al. but 62 of them are
 not members in the more complete and less contaminated sample C.
 This means that about 92\% i.e. 
(1\,034-62)/1\,052 of the YSOs selected by us are confirmed members.
Hence we conclude that the contamination level of the sample that we have selected is 
 $\sim$8\%.

% see ngc2264_kounkel19.pro
% nella regione di 2264 lei trova 650 ma non da la parallasse dei membri, ma solo dei theia.
% i theia con distanza > 500 pc sono solo 3 e quindi ci sono due theia con
% 548  e 89 stelle stelle compatibili con 2264. Tra queste, 420 sono in comune con noi.
For comparison, in the same region, \citet{koun19} found 637 YSOs, belonging to the 
clusters named as Theia 41 and 189, with 548 and 89 objects, respectively.
Of them, 420, i.e. about 66\%, are in common with our list.

\section{Discussion}
  In the previous sections, 
 we described how overdensities in the 5D parameter space
 ($l$, $b$, $\pi$, $\mu_\alpha*$, $\mu_\delta$) have been identified, 
starting from a photometrically selected
sample, that covers the expected PMS region of YSOs with ages $t<10$\,Myr.

Since no attempt has been made to correct for interstellar reddening, the starting
sample was contaminated also from older reddened stars. Another possible
reason for the contamination   of older stars derives from the adopted strategy
to select the starting sample in the plane $M_G$ vs. $G-G_{\rm RP}$, where the  
sensitivity to stellar ages
of the available  isochrones is quite low for the low mass population.  
In fact, for faint and very low-mass stars, isochrones get closer and closer for ages 
larger than about 50-100\,Myr and, consequently, it is difficult to separate young
populations from the older ones.
%This means that 
%the MS of $\sim100$\,Myr old clusters crosses the 10\,Myr isocrone, adopted as photometric limit,
%and the low mass members of these clusters fall in our starting sample.   
As a result, the DBSCAN clustering algorithm, 
adopted to resolve spatially concentrated and/or co-moving stellar populations
located at the same distance, can select also clusters older than 10\,Myr.

A pattern match procedure has been adopted to disentangle SFRs and young clusters from older
and photometrically unphysical clusters. We found \totsfr\, SFRs with ages $t\lesssim 10$\,Myr
and \totmidagecl\, young clusters with ages approximatively in the range 10-100\,Myr.
We discuss here these validated findings in the context of the GP structure
within $\sim$1.5\,kpc from the Sun. 

      \begin{figure*}
   \centering
\includegraphics[width=9cm]{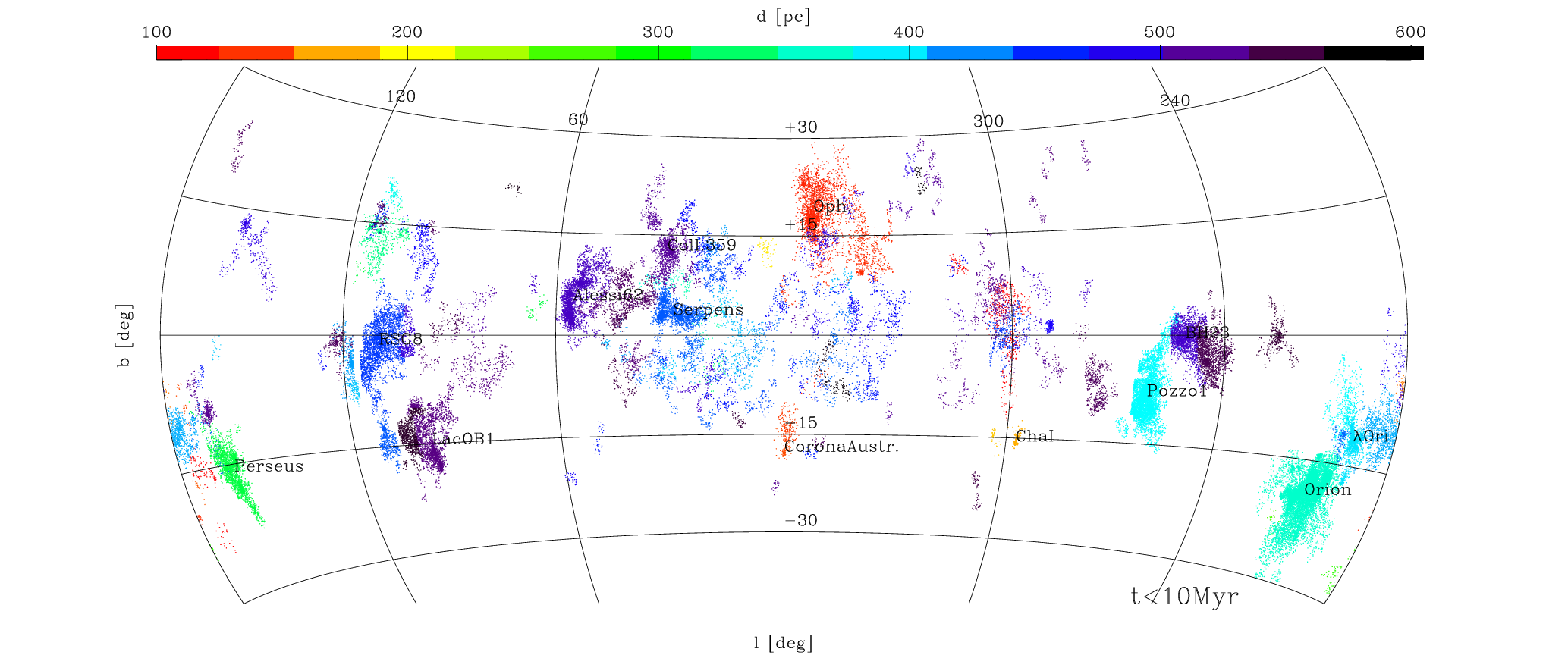}
\includegraphics[width=9cm]{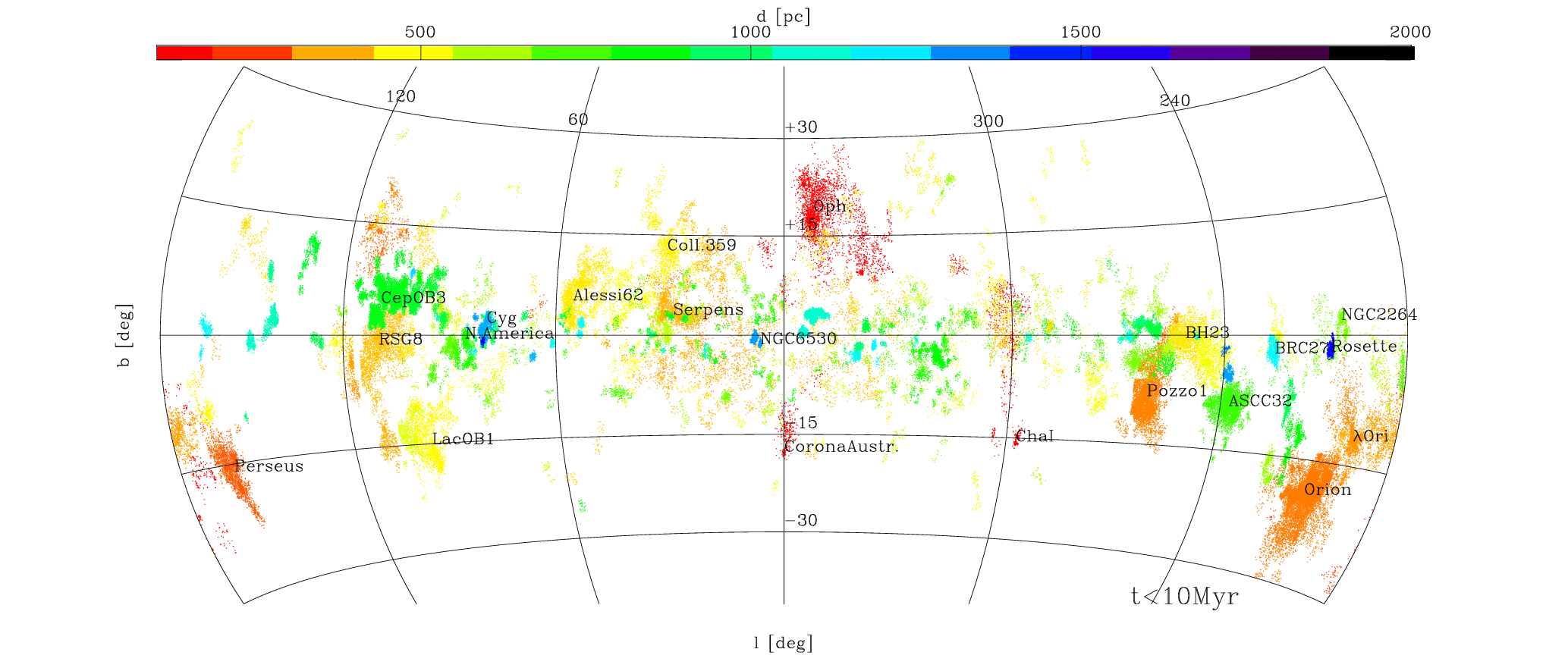}  
\includegraphics[width=9cm]{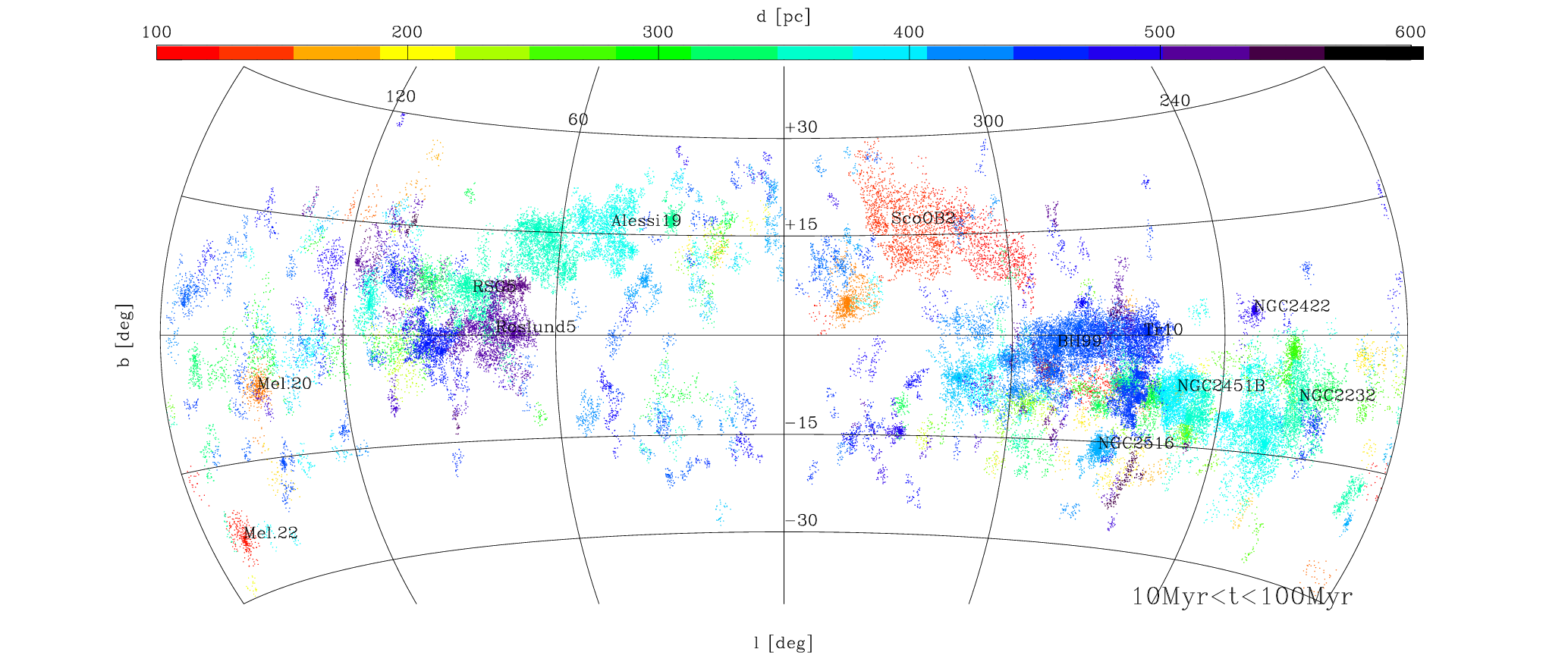}
\includegraphics[width=9cm]{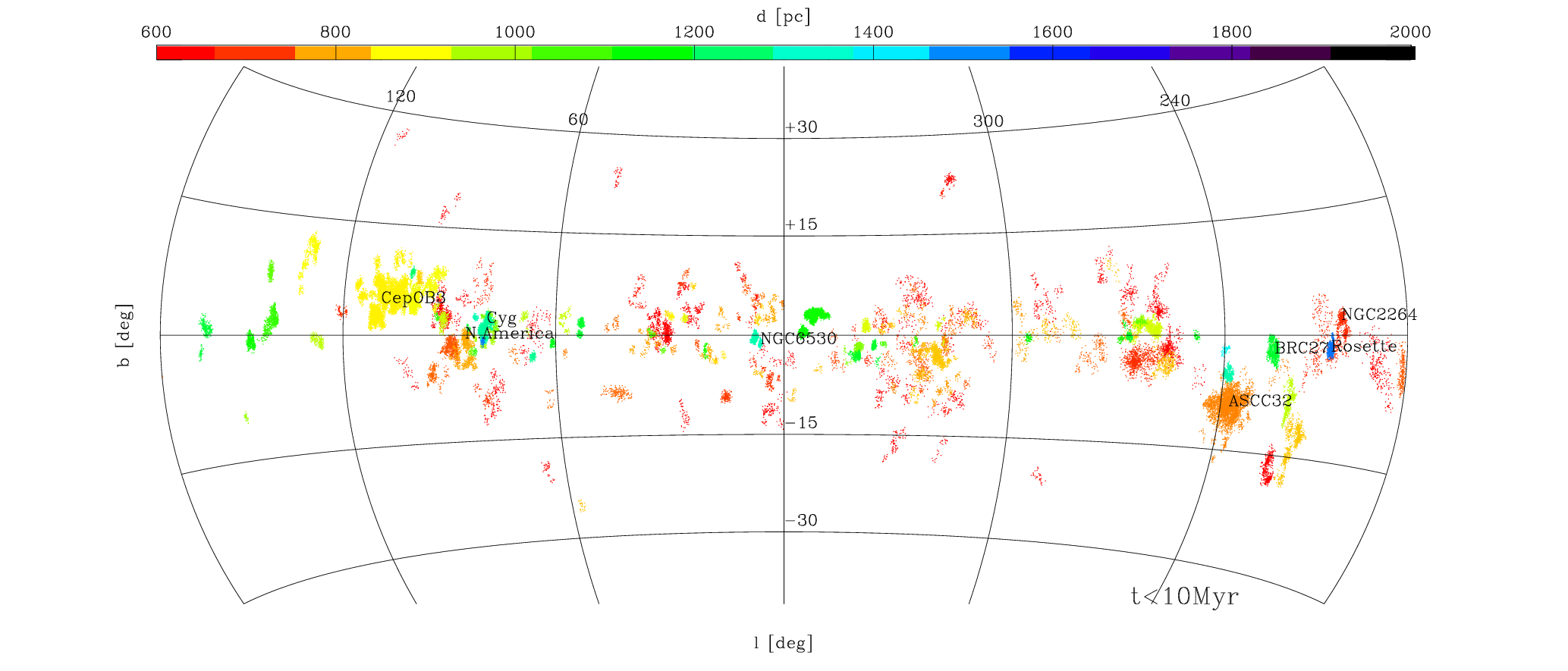}
\includegraphics[width=9cm]{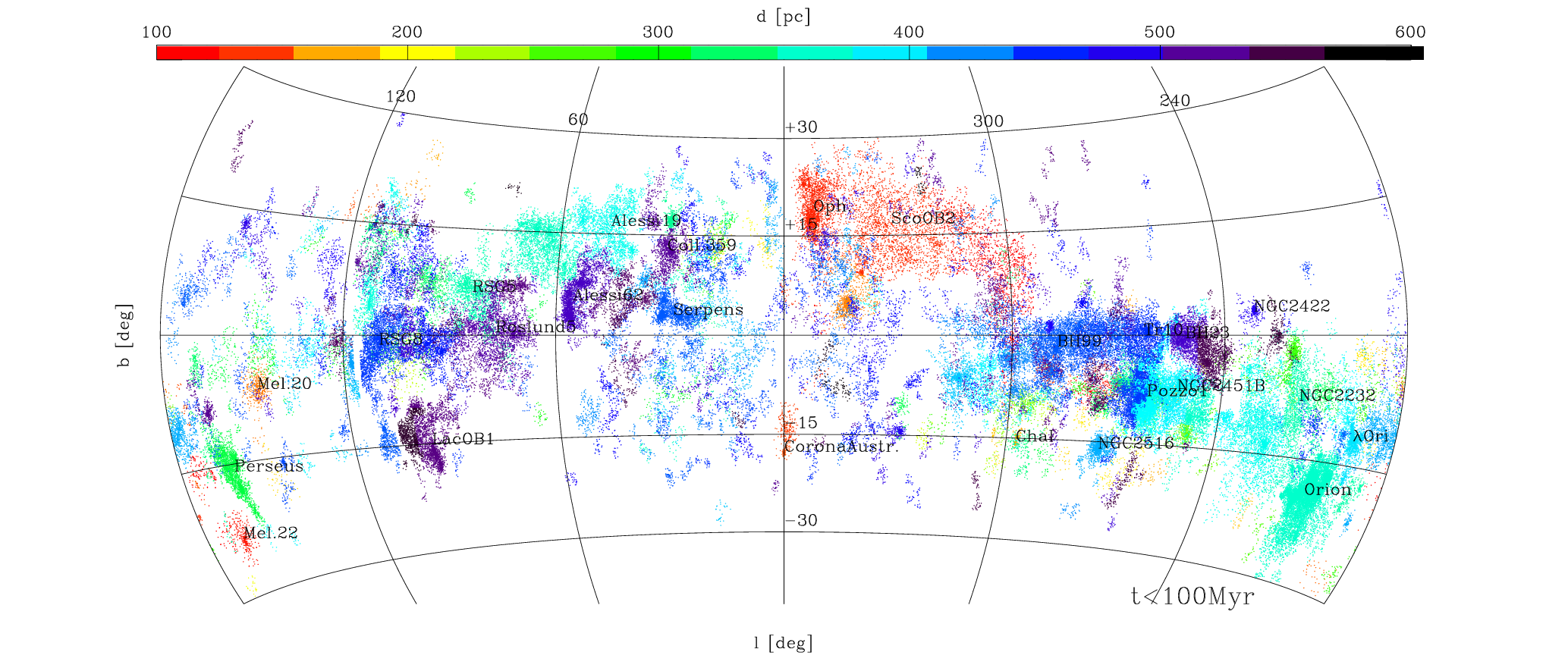}
\includegraphics[width=9cm]{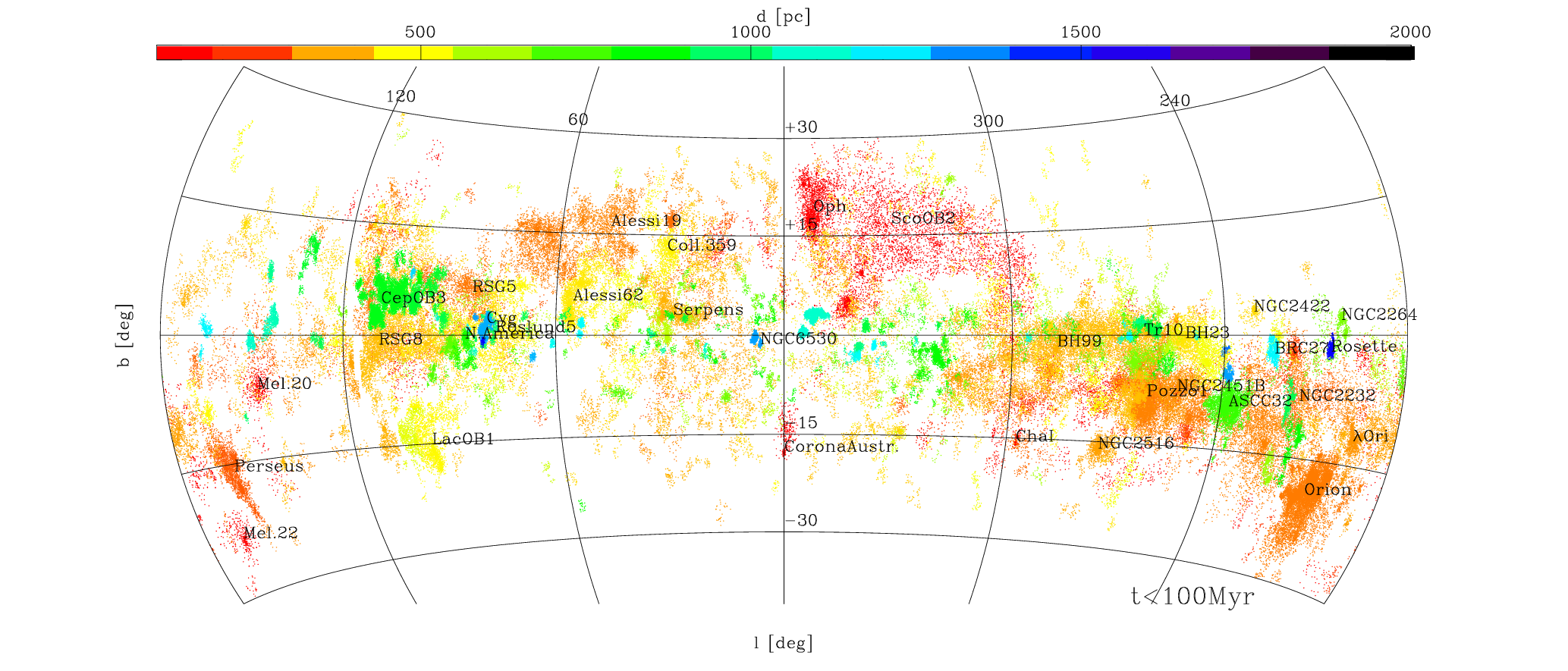}

      \caption{Aitoff projections in Galactic coordinates of the YSOs in the
      different  age bins ($t<10$\,Myr, 10\,Myr$<t<100$\,Myr and $t<100$\,Myr), with 
      distance in the range ([100, 600]\,pc (left panels),
       [600, 2000]\,pc (mid right panel) and
       [100, 2000]\,pc) (upper and bottom right panels). Color codes indicate stellar distances. }
         \label{galmap}
   \end{figure*}

The maps of the young stellar clusters recognised by the DBSCAN clustering algorithm, most of them already known in the literature,
 have been shown in the previous sections, and specific spatial and kinematic details
have been presented for some of them. 

To identify  clusters extended on scales larger than the 5\deg$\times$5\deg\, boxes
used in the analysis,
we merged adjacent clusters with consistent proper motions and distances. 
This procedure has been applied  to identify extended SFRs  as a whole,
as in the case of the Orion Complex or Sco OB2 UCL, with r$_{\rm 50}$ equal to $\simeq17$\deg and $\simeq15$\deg,
respectively, that are among the most extended structures resolved in this work.
% and therefore we provide two levels of clustering, a sub-level pinpointed by the  algorithm 
%and a second inclusive level of merged clusters. We note, however, that even the sub-level 
%clustering, derived with our strategy of  DBSCAN parameter selection, 
In several cases, it identifies
clusters that encompass  multiple populations, as in the case of NGC\,2264, that has been
identified as a unique structure including also the close cluster IC\,446 and other YSOs in the 
surrounding region. A more in depth analysis of  the two clusters shows that their proper motions 
can be distinguished into slightly different sub-populations and thus our overall procedure
to define clusters tends  to include multiple sub-populations sharing similar properties, likely associated to
the progenitor molecular cloud.
 
The question of the cluster and subcluster identification is a very complex issue that can be 
dealt at different spatial precision levels, required for a given analysis, 
as done, for example, for the MYStiX project, in \citet{feig18}, where a parametric
statistical regression approach, providing  hierarchical ellipsoid structures, has been adopted. 
The evidence of a wide range of central surface densities  found  in the MYStIX maps is 
in agreement with the different spatial morphology of the SFRs identified in this work.

Figure\,\ref{galmap}  shows the spatial distribution
of the young stellar clusters   found in this work, in three different bins of distance,
 [100, 600]\,pc, [600, 2000]\,pc and [100, 2000]\,pc. The young
  clusters are drawn by distinguishing  them
in the  age bins  $t<10$\,Myr,  10\,Myr$<t<100$\,Myr and $t<100$\,Myr. 
Note that clusters with 10\,Myr$<t<100$\,Myr 
have been found only in  the solar  neighbourhood ($<600$\,pc) and thus are shown 
only in the [100, 600]\,pc distance range.

The distribution of SFRs ($t<10$\,Myr)  within 600\,pc is dominated by the presence
of big young structures crossing the GP such as the Orion and Perseus Complexes,
Gamma Velorum (Pozzo\,1), Lac\,OB1, under the GP,  BH\,23
 \citep[corresponding to Theia\,80 in][]{koun19} 
and RSG\,8 close to the GP, Serpens, Alessi\,62, Collinder\,359 and  
Rho Ophiuchi, over the GP.
The clusters with ages 10\,Myr$<t<100$\,Myr in the same distance range
appear  definitely more diffuse. Apart from the well known Sco-Cen association covering 
$\sim$60\deg\, in longitudes, we detected as unique complex
the likewise huge association in the Vela-Puppis
region, including Trumpler\,10, $\gamma$  Velorum, NGC\,2457, NGC\,2451B,
as well as the associations around NGC\,2232,  Roslund\,5 and Alessi\,19.
 
Their positions appear to be connected to the clusters with  $t<10$\,Myr
since they follow a spatial pattern  crossing or very close to that of the  SFRs. 
This suggests that they likely belong to a common star formation process encompassing at least two
generations of YSOs, with the first generation including extended populations 
of dissolving young clusters and associations.

The large Sco-Cen association is connected to the Vela and Orion Complexes, confirming
 what already found 
by \citet{bouy15} with Hipparcos data. These three regions are described there 
as three large-scale stream-like structures. 
%As already noted by \citet{bouy15}, 
%the Orion stream includes YSOs  at distance $\sim200$\,pc
%that is a foreground population with respect to the known Orion Complex located at about 400\,pc.
%The closest or foreground populations of Cepheus and Cygnus are partially found 
%also in this distance bin.

%The distribution of YSOs with distances in the range [300-600]\,pc shows a more patchy structure.
%A very large area ($l$=[$\sim$0\deg,$\sim$60\deg]) falling in 
%the first Galactic quadrant  and in part of the fourth 
%($l$=[$\sim$330\deg,$\sim$360\deg]), is populated by YSOs
%of the Serpens and Aquila Rift. The other quadrants are dominated by the YSOs of the
% Cepheus, Cygnus, Lacerta, Vela and Orion SFRs.

Going towards
larger distances ($d\gtrsim600$\,pc), the SFRs  
show a more regular pattern, approximately parallel to the GP. 
The most prominent SFRs are ASCC\,32 and Cep\,OB3b in the Cepheus, respectively
under and over the GP at distance of $\sim$800-900\,pc.
Among the most distant
SFRs with more than 300 members and distance $\gtrsim1400$\,pc, 
we detected NGC\,2244, NGC\,6530, 
NGC\,6531 , NGC\,2362 and FSR\,0442.

The overall distribution of YSOs in SFRs with $d\lesssim600$\,pc traces 
a complex 3D pattern in the solar  neighbourhood.
In particular, in the Z vs. X edge-on Galactic  projection
(see bottom/left panel in Fig.\,\ref{xyzmapcl} and top/left panel in Fig.\,\ref{galmap})
we find evidence of a projected inclined structure, mainly traced by the Orion, Vela OB2 and 
Rho Ophiuchi star forming
 complexes  in the third and fourth Galactic quadrants
 and,  by the Serpens, Lacerta\,OB1  and
  Perseus in the first and second Galactic quadrants.
However, the SFRs falling in the Cepheus region do not follow this pattern.
 A global view of these structures and their spatial correlation with the surrounding 
nebular emission, suggests a pattern  consistent  with the results found in \citet{moli10},
where massive proto-clusters and entire clusters of YSOs in active SFRs are
associated to clouds that collapse into filaments.
  %can be associated to the {\it Gould Belt}, 
%The thickness of such structure increasing from the 
%narrow  going to the Orion complex, 
%that extends in distance from $\sim200$ to $\sim600$\,pc.

As already found in \citet{zari18}, 
current data reveal  a very complex 3D structure that 
cannot be simply described with  the  {\it Gould Belt},
 i.e. the giant  flat structure, inclined by $\sim20$\deg\, with respect to the Galactic
Plane,  first pointed out by \citet{goul79}.
This insight was already suggested by 
\citet{guil01}, who presented the first detection of the {\it Gould Belt} late type star population,
and who proposed the alternative scenario of a {\it Gould disk}. 
%While no evidence of the {\it Gould Belt} has been found by \citet{zari18}, 

A more detailed representation of the young Galactic component in the Solar Neighborhood
has been recently proposed by \citet{alve20},
who determined the 3D distribution of all local cloud complexes
by deriving accurate distances to about 380 lines of sight.
They suggested that such 3D distribution  could be described by a damped sinusoidal wave, that they call 
Radcliffe Wave, with an amplitude of $\sim160$\,pc and a period of $\sim2$\,Kpc. It crosses
Orion, around a minimum, Cepheus (crest), North America and Cygnus\,X. This structure
is separated and distinct from a second structure, indicated as "split", crossing Sco-Cen, Aquila and Serpens
clouds. They propose that the {\it Gould Belt} is a projection effect of two linear cloud complexes.
The spatial distribution of YSOs associated to SFRs that has been identified in our work  
shows much more complex and diffuse structures but, the 
 two elongated linear structures suggested by \citet{alve20} approximately cross the  borderline
of the two separated structures visible in the X, Y map of Fig.\,\ref{xyzmapcl},
delimited by the SFRs indicated by \citet{alve20}. This brings us to confirm that the local young
Galactic component is very complex. While our data are broadly consistent with the \citet{alve20} findings,
further investigations, including a more detailed analysis of the 
kinematics of the structures, based on the 3D  space coordinates ($X,Y,Z$) and velocities ($U,V,W$) 
\citep[e.g.][]{deze99}
are required to confirm the scenario and 
to find  additional insights about their origin.

To gain further insights on the  star formation history  of the SFRs, it is crucial to derive
more accurate stellar ages. However, 
we do not attempt to derive stellar ages of the selected YSOs, for several reasons:
 first of all,  the
lack of a suitable photometric system. In fact, the large {\it Gaia} EDR3  G and RP  photometric bands 
used for this work
are not sensitive to the fundamental stellar parameters (effective temperatures, stellar ages, 
etc...), especially for low mass stars. However, future Gaia releases, overcoming issues related to the BP 
bands at faint magnitudes, could be crucial to this aim.  Second, 
the lack of spectroscopic data  needed to derive  individual stellar reddening values,
to appropriately place these YSOs on the HR diagram. Alternative ways,
such as the use of 3D reddening  maps \citep[e.g.][]{bovy16,lall19}, 
require careful analysis, since the integrated
extinction tends to be underestimated in the molecular clouds, 
where SFRs are typically located. 
A detailed analysis is deferred to future works based on the combination of Gaia and 
spectroscopic data from available surveys, such as  Gaia-ESO \citep{gilm12,rand13},
 LAMOST \citep{zhao12} or APOGEE \citep{maje17},
or future surveys such as WEAVE \citep{dalt12} and 4MOST \citep{guig19}.

\section{Summary and conclusions}  
We used the machine learning unsupervised clustering algorithm DBSCAN to
systematically identify all SFRs with ages $t\lesssim10$\,Myr,
within $\sim$1.5\,Kpc from the Sun. The density-based  clustering algorithm
has been applied to the {\it Gaia} EDR3 positions, parallaxes  and proper motions of a
 photometrically-selected starting sample.  
 
 A pattern match procedure based on a template data set including typical 
 clusters detected within the photometric sample has been used to distinguish 
 very young clusters from the contaminant old clusters and from photometrically 
 unphysical clusters.
 We provide here a catalogue with the main parameters (positions, spatial extent, median distance and 
 number of members) of the \totsfr\, SFRs with ages  $t\lesssim10$\,Myr. 
 The parameters of the  \totmidagecl\, young clusters with ages  10\,Myr$\lesssim t\lesssim100$\,Myr are also
 given. 
We provide also   the  list of the \youngstars\, plus \midagestars\, YSOs 
found in the SFRs and the young clusters, respectively,
 including mainly late type K-M stars. A substantial number of YSOs
have been recognized for the first time.
Based on the comparison of our list of YSOs in the well known region
Sco-Cen and NGC2264, we roughly 
estimate that, within our observational limits, the completeness of the census  of cluster members obtained with our analysis 
is  $\gtrsim$85\%, at least in very rich and concentrated SFRs. 
For low density regions, such as the Taurus-Auriga association (see Appendix\,\ref{literaturecompapp}),
 this completeness figure is expected to be around 50\%. 
The mass function  coverage of each cluster, 
strongly depends on the cluster distance, and   is set by the observational limit.  

Compact regular clusters, as well as SFRs in large complexes such as, for example, 
Taurus, Orion, Sco-OB2, Perseus, and Cygnus, have been identified with high efficiency,
as estimated from the comparison with other available catalogues (see Appendix\,\ref{literaturecompapp} ).

The overall distribution of these clusters in the Galaxy context shows that they are 
distributed along a very complex 3D pattern that seems to connect them at least within
500-600\,pc. Outside of this distance, the clusters appear  to be more
regularly and  closely distributed along the GP. 

As far as we know, the catalogue of YSOs presented in this work is
the sole all-sky catalogue based on the most recent {\it Gaia} EDR3 data, which benefit of 
 major improvements with respect to {\it Gaia} DR2. This catalogue represents a step forward
 in the census of  SFRs and can be used, for example, for further detailed 
 interpretations of their spatial distribution
 in the context  of the spiral arm model \citep{reid19}, since it covers a substantial region
 crossed by the Local arm and, marginally, some regions of the Perseus and Sagittarius-Carina arms \citep{pogg21}.
 Future and photometric deep surveys, such as the Rubin Legacy Survey of Space and Time (LSST)
 will allow to extend these limits. 
 
We note that these results are not at this stage suitable for studies such as star formation history,
 cluster dynamics, based on the full space 3D velocity determination,
 or IMF, since the census of the SFRs is not complete and accurate masses and ages, as well as radial
 velocities can not
 be determined, until further data are available.
Nevertheless,  the dominant component of the SFRs has been detected and thus
these results   can be  used as
 driving samples for the extraction of complete populations from {\it Gaia} data, by relaxing 
 the stringent constraints adopted in this work. 
 
 Finally, the SFRs identified in this work are defined well enough to allow 
detailed studies of circumstellar disk evolution and direct imaging
of young giant planets, based on 
 multiband analisys of  available or future additional  observations (X-rays or IR or radio),
targeting some of the individual clusters.

\begin{acknowledgements}
This work has made use of data from the European Space Agency (ESA) mission
{\it Gaia} (\url{https://www.cosmos.esa.int/gaia}), processed by the {\it Gaia}
Data Processing and Analysis Consortium (DPAC,
\url{https://www.cosmos.esa.int/web/gaia/dpac/consortium}). Funding for the DPAC
has been provided by national institutions, in particular the institutions
participating in the {\it Gaia} Multilateral Agreement.
E.T. acknowledges Czech Science Foundation GA\u{C}R (Project: 21-16583M).
JMA acknowledges financial support from the project PRIN-INAF 2019
"Spectroscopically Tracing the Disk Dispersal Evolution.
The authors are very grateful to the anonymous referee, for 
providing constructive comments and suggestions 
which significantly contributed to improving this publication.
\end{acknowledgements}
\bibliographystyle{aa}
\bibliography{/Users/prisinzano/BIBLIOGRAPHY/bibdesk}
\begin{appendix} %First  appendix
\section{ Interstellar reddening  effects \label{reddeningeffect}}
 In this section we show the effects of the reddening on the
sample selected as described in Section\,\ref{photselsect}. 
As discussed in \citet{ande19}, for a generic passband $i$, extinction coefficients A$_i$/A$_{\rm V}$
depend on the stellar effective temperature. The subsequent dust
attenuated photometry of very broad photometric passbands, such as the {\it Gaia} EDR3 ones,
is not a simple linear function of A$_{\rm V}$, but it is also a function of
the source spectrum i.e. its effective temperature.

The  PARSEC  1.2S stellar models \citep{bres12,chen14,tang14} 
have been implemented to predict tracks and isochrones also at non-zero extinction.
As done in \citet{mont21}, in order to have an indication of the reddening which affects our data,
we used the CMD 3.3 input form web interface, and 
we constructed a grid of stellar models assuming  
 the 1\,Gyr isochrone, and  A$_{\rm V}$=[0.1, 0.5, 1.0, 1.5, 2.0, 2.5, 
 3.0, 3.5, 4.0, 4.5, 5.0, 6.0, 7.0, 8.0, 9.0, 10.0, 20.0, 30.0].
 
\begin{figure}
   \centering
\includegraphics[width=9cm]{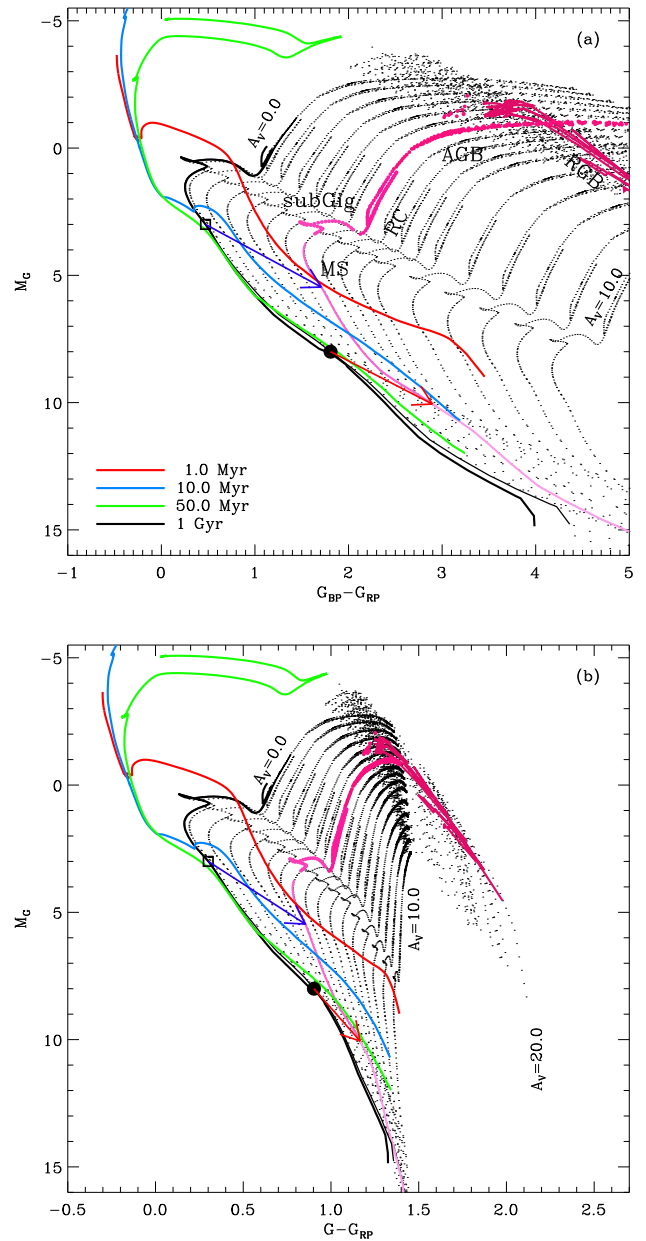}
      \caption{ PARSEC 1\,Gyr isochrone at solar metallicity 
      with extinction A$_{\rm V}$ ranging from 0 to 
      10.0 in the CAMD obtained by adopting the 
 different {\it Gaia} magnitudes (black dots).
 The 1\,Gyr isochrone at A$_{\rm V}=0$ is highlighted by a thick black solid line.
 The 1\,Gyr isochrone at A$_{\rm V}=3$ is highlighted by pink coloured lines of different shades 
 during the Red Giant Branch (RGB), Asymptotic Giant Branch (AGB),
  Red Clump (RC), sub Giants (subGig) and MS phases.
 The red, light-blue and green solid lines are the  1, 10 and 50 \,Myr 
 Pisa isochrones at solar metallicity. The empty square and the bullet in each panel represent a star of 
 6930\,K and 3945\,K, respectively, while the blue and red arrows are the reddening vectors
 corresponding to A$_{\rm V}=3$, for these two representative stars (see text).}
         \label{gaiacamdiso1e9av}
   \end{figure}

 Figure\,\ref{gaiacamdiso1e9av} shows how the 1\,Gyr isochrone changes 
 by increasing extinction A$_{\rm V}$ from 0 to 10, in the CAMD obtained by adopting the 
 different {\it Gaia} colors (panels a and b).
The 1\,Gyr isochrone at A$_{\rm V}=0$  is highlighted by a thick black line  while 
the 1\,Gyr isochrone at A$_{\rm V}=3$ is highlighted
by symbols with different shades of pink  in the different 
stellar evolution phases.

Note that the reddened isochrone is not linearly shifted  along a reddening direction,
as usually happens when adopting a
% i.e. the usually adopted 
 reddening vector.   For example,
% see output gaia_camd_iso1e9_av.pro
for an object at M$_{\rm G}=3$, corresponding to a star with 
(G$_{\rm BP}$-G$_{\rm RP})_0$=0.47, (G-G$_{\rm RP})_0$=0.30  and  effective temperature of 6930\,K at 1 Gyr
(black empty square in the Figure) with extinction A$_{\rm V}=3$, the reddening 
E(G$_{\rm BP}$-G$_{\rm RP}$) is equal to 1.24 and E(G-G$_{\rm RP}$) is equal to 0.55 (blue arrows in the Figure).
But, for an object at M$_{\rm G}=8$, corresponding to a star with 
(G$_{\rm BP}$-G$_{\rm RP})_0$=1.81, (G-G$_{\rm RP})_0$=0.90  and  effective temperature of 3945\,K at 1 Gyr
(black bullet in the Figure) with extinction A$_{\rm V}=3$, the reddening 
E(G$_{\rm BP}$-G$_{\rm RP}$) is equal to 1.09 and E(G-G$_{\rm RP}$) is equal to 0.26 (red arrows in the Figure).

Thus, at different temperatures, and for a fixed A$_{\rm V}$,  the shift in color
due to the reddening  is smaller  for the colder star. And this effect is higher in the G vs. G-G$_{\rm RP}$ diagram,
as can be deduced from the different slopes of the  blue and red arrows. In this case, 
for a $\sim$4000\,K star, the E(G-G$_{\rm RP}$)
value (equal to $\simeq$0.26)  is about half than the one ($\simeq$0.55) associated to a $\sim$7000\,K star.

This implies that while a reddened 1\,Gyr old star with effective temperature of $\sim 7000$\,K can 
be expected to be found  in the PMS region and mimic a star younger than 10\,Myr, a colder star of $\sim$4000\,K,
of the same age and affected by the same extinction, does not fall in the PMS region 
(see Fig.\,\ref{gaiacamdiso1e9av}, 
panel b).
In conclusion, the effect of uncorrected reddening in terms of   
contamination of our initial photometric sample by old stars  is larger for  stars with 
spectral type F and G,  than for K and M stars.

%M type stars are expected to fall in a strip, roughly on the faint side of the
%1\,Myr isochrone, that it has been interpreted as due to other
%effects, such as inflation, activity, age spread and etc... \citep[e.g.][]{morr19}.

%The grid shows also as reddened and very reddened stars in MS or at the Turn Off or subgiant
%phase are expected to be found in the  region of the CAMDs where PMS stars are 
%expected to be found in absence of extinction, i.e. in the brightest side of the 1\,Myr isochrone.

%As a result, if the reddening is not corrected, a mixture of
%overlapping regions are expected in the CAMD.

%----------------------------------------------------------- S_vib
%%%%%%%%%%%%%%%%%%%%%%%%%%%%%%%%%%%%%%%%%%%%%%%%%%%%%%%%%%%%%%%%%%%%%%%%%%%%%
\section{The effect of binarity or multiplicity on astrometric selections \label{binapp}} 
%We note that due the adopted restriction RUWE$<$1.4, our initial sample does not include 
%the majority of partially resolved astrometric binaries or tight astrometric binaries with a significant 
%orbit-induced displacement of the photocenter, 
%with low mass ratio between the components and orbital period of $\sim$1 year \citep{kerv22}.
%Using a sample of benchmark eclipsing binaries (EBs), \citet{stas21} confirmed that 
%the RUWE parameter tends to be sensitive to unresolved companions (i.e. tertiaries in the case of EBs)
%and that there is a smooth transition in the fraction of EBs with proper motion anomalies for RUWE values 
%in the range 1.0-1.4. 
At the level of astrometric sensitivity offered by Gaia, the orbital motions of binary 
(or multiple) stars become sometimes measurable, and also difficult to disentangle from 
proper motion. This holds both for resolved pairs, and for unresolved unequal-mass
pairs where the photocenter displays significant motion \citep[see][]{kerv22}.
If the binary period resonates with the Gaia sampling frequency,  parallax
measurements will be also affected. Perhaps the best-studied star-forming region in terms
of its binary-star population is Taurus-Auriga, and we may refer to the review by \citet{math94}
to get an idea of the expected range of system parameters. Taurus is one of the few SFRs where
lunar occultation techniques were feasible for detection of close pairs, down to separations
 of 0.009\sec \citep[][Table~A1 and references therein]{math94}.
 %(Mathieu 1994, Table~A1 and references cited). 
 Therefore, the projected binary separations range across a factor of $\sim 1000$, with no
``typical'' value. Correspondingly, their orbital periods span a range of a factor of $\sim 30\,000$.
% if a is the projected separation, the period is ~sqrt(a^3)

We have checked empirically if the Gaia-based selection used in this paper keeps 
the binary members of a SFR, by matching the Tau-Aur binary-star list in Table\,1 from
\citet{krau12} with the {\it Gaia} EDR3 catalogue, and with its subset selected in this paper. 
Out of 156 stars in Kraus et al., we found 142 Gaia-EDR3 counterparts within 0.5\sec, 
of which 40  selected  in this work using DBSCAN.

\begin{figure}
   \centering
\includegraphics[width=9cm]{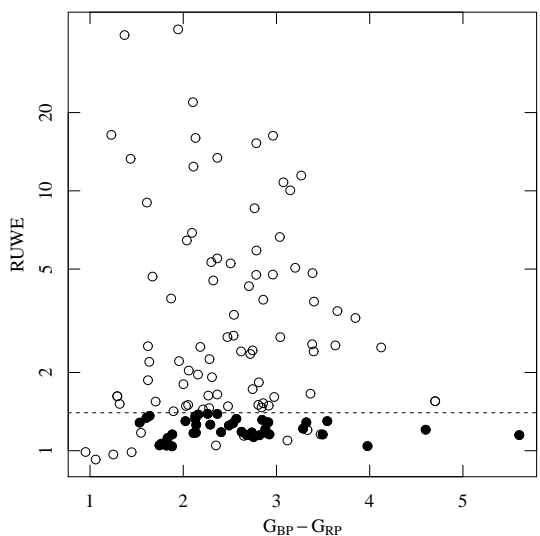}
      \caption{Diagram of the RUWE values vs. the Gaia  color G$_{\rm BP}-$G$_{\rm RP}$ 
      of the \citet{krau12} Tau-Aur binary-star list with {\it Gaia} EDR3 counterparts.
      Filled symbols are the binaries selected also in this work, while empty symbols
      are those rejected.}
         \label{bprpruwe}
   \end{figure}

\begin{figure}
   \centering
\includegraphics[width=8.5cm]{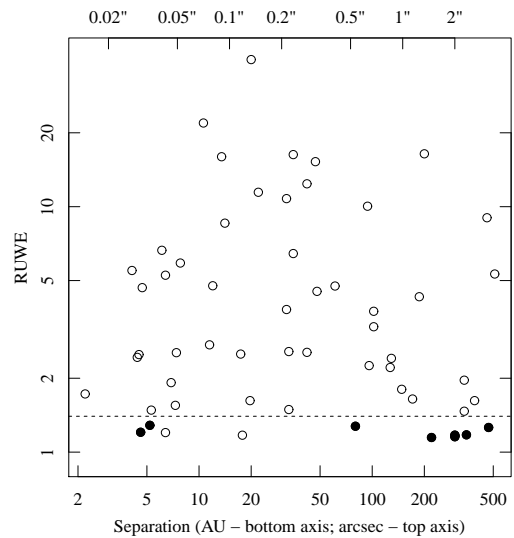}
      \caption{Diagram  of the RUWE values vs. the  binary separation of the \citet{krau12}
       Tau-Aur binary-star list with {\it Gaia} EDR3 counterparts. Symbols are as in Fig.\,\ref{bprpruwe}.}
         \label{sepruwe}
   \end{figure}

We have then compared the RUWE distributions of the selected vs. the unselected systems,
to gain insight on how binary motions impact RUWE and the subsequent selection. 
Figure\,\ref{bprpruwe} shows a diagram of RUWE vs.\, {\it Gaia} color G$_{\rm BP}-$G$_{\rm RP}$. 
The horizontal line indicates our maximum accepted RUWE value. 
Filled symbols are stars passing our selection, while empty symbols are the unmatched binaries,
i.e. those not retrieved in our catalogue. 
It should be remembered that the cut in absolute-$G$ magnitude rejects some stars
 which would have passed the RUWE constraint. Nevertheless, the vast majority of 
 unmatched stars have indeed RUWE values well above the chosen limiting value, 
 and were likely rejected for this reason. There is little or no dependence of
  RUWE on {\it Gaia} color, and therefore mass (although part of color spread is also
due to high extinction towards Tau-Aur). Also interesting is the diagram shown
in Fig.\,\ref{sepruwe}, showing RUWE vs. binary separation. Here,
the absence of any dependence of RUWE on projected separation (when measurable)
is very evident, including unresolved pairs, where only the photocenter is 
affected by orbital motion. This latter diagram contains only 6 out of 40 stars 
selected by us, since about one-half of the Kraus et al.\ pairs are spectroscopic
 binaries with no measured separation. We also point out  that out of the 76 binaries 
 with no measured separation, 32 pass our selection (42\%),
 while only 8 out of the 66 binaries with measured separation
 pass the selection (12\%), probably 
 because photocenter motion has a smaller effect on astrometry compared to the
  motion of resolved components.
Overall, extending this result from Tau-Aur to other SFRs at similar distances,
we would predict to have lost $\sim$~72\% of their binary populations 
due to our selection criteria.
Thus, if a binary frequency is as high as 
50\% \citep{math94}, a loss of $\sim$35\% of PMS members can be expected.
 However, since this work selects stars at distances up to $\sim$1500\,pc,
this estimate should not be naively extended to our whole sample: 
the larger the distance the smaller the projected binary motions 
hence, closer to our detection limit.
%at larger and larger distances, the projected binary motions become smaller, 
%and increasingly closer to the detection limit.
 We therefore expect a less
significant binary member loss for the farther-out SFRs. A more detailed 
study of these effects would however be far beyond the scope of this paper, 
also taking into account that the upcoming {\it Gaia} data release DR3 (foreseen June 2022)
 will contain orbital astrometric solutions for 135\,760 non-single 
 stars\footnote{See https://www.cosmos.esa.int/web/gaia/}.  
%%%%%%%%%%%%%%%%%%%%%%%%%%%%%%%%%%%%%%%%%%%%%%%%%%%%%%%%%%%%%%%%%%%%%%%%%%%%%%   
\section{Literature comparison  \label{literaturecompapp}}
\subsection{The Taurus-Auriga association}
\begin{figure}
   \centering
\includegraphics[width=9cm]{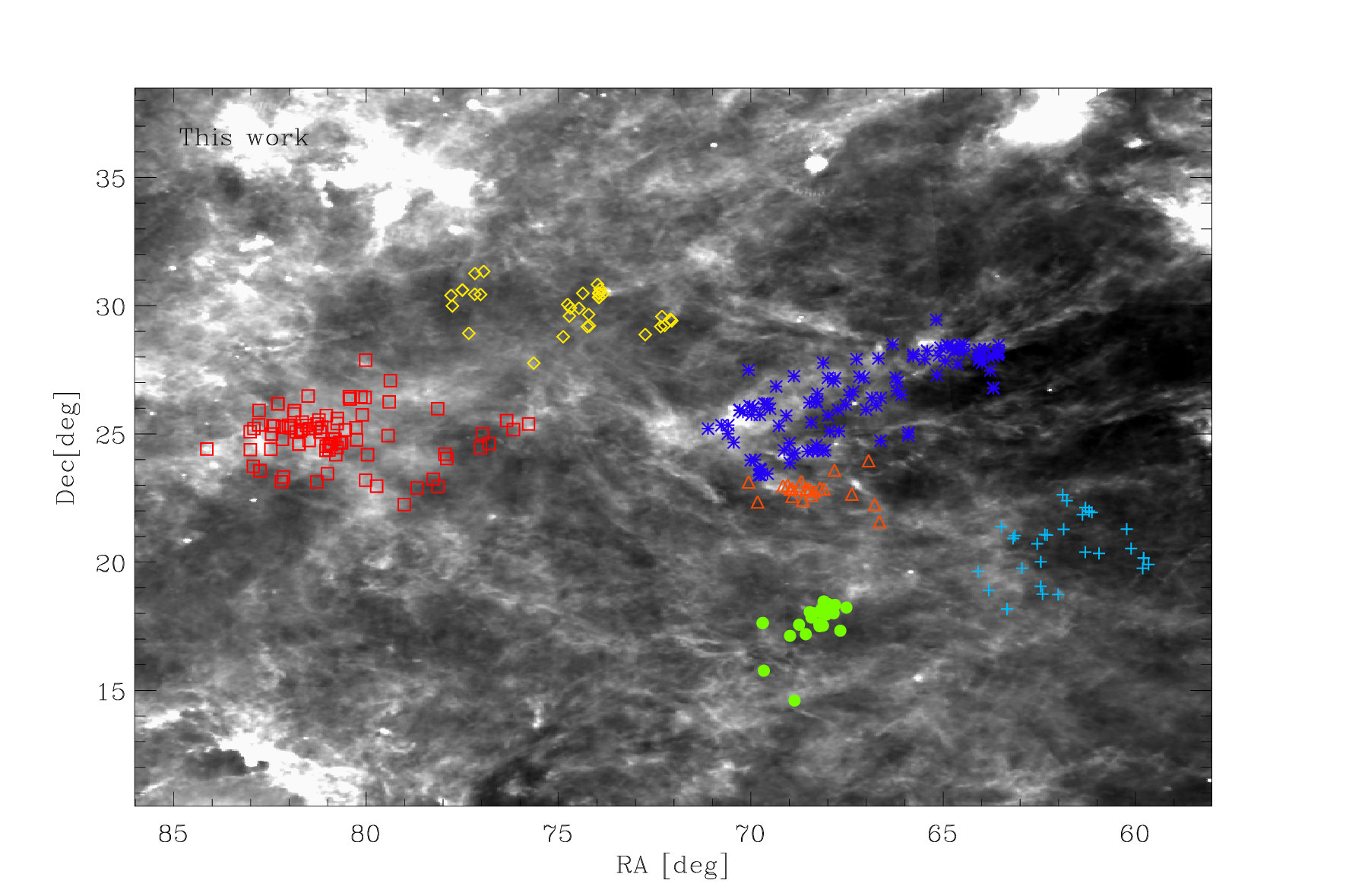}
\includegraphics[width=9cm]{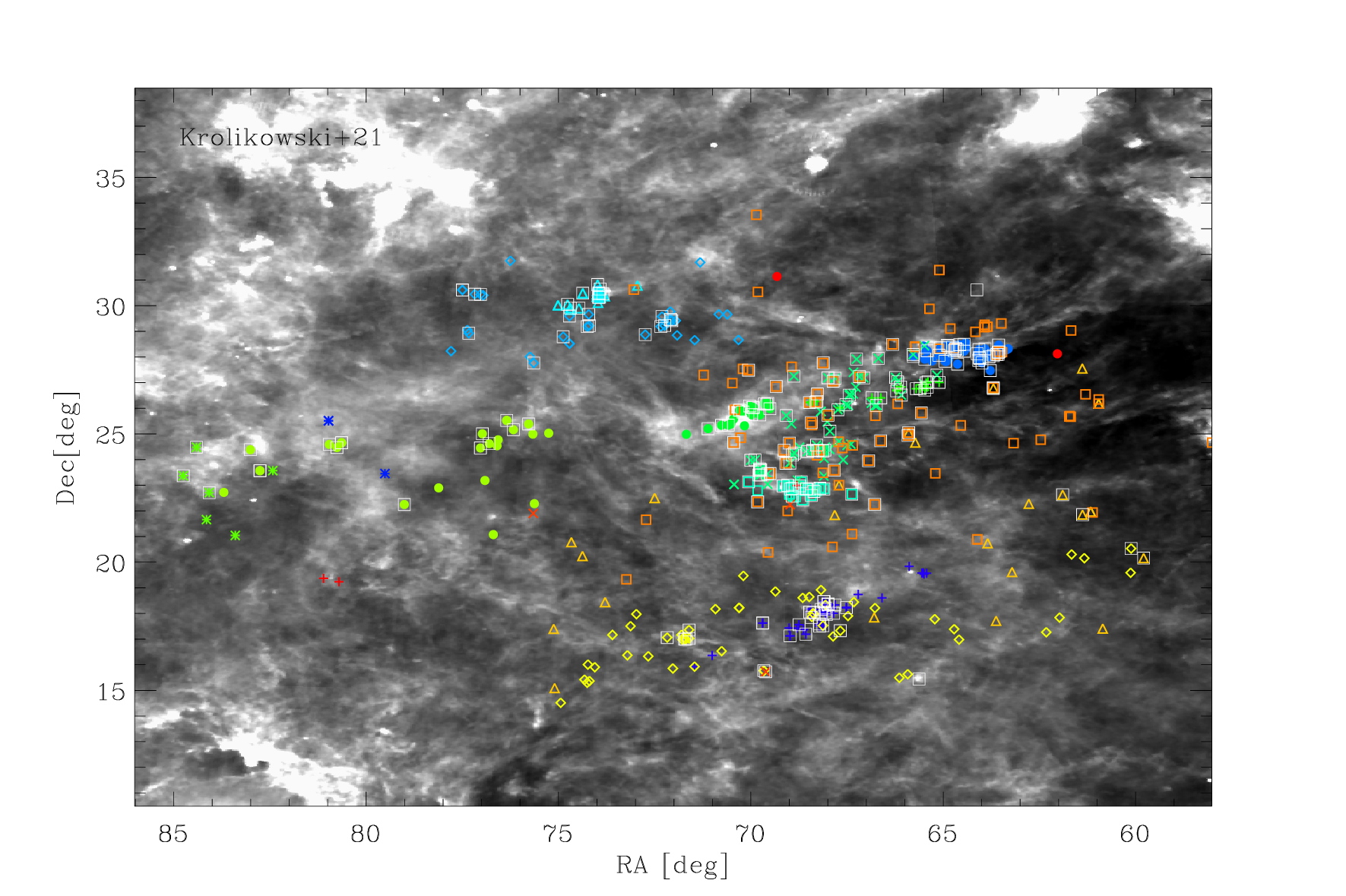}
\includegraphics[width=9cm]{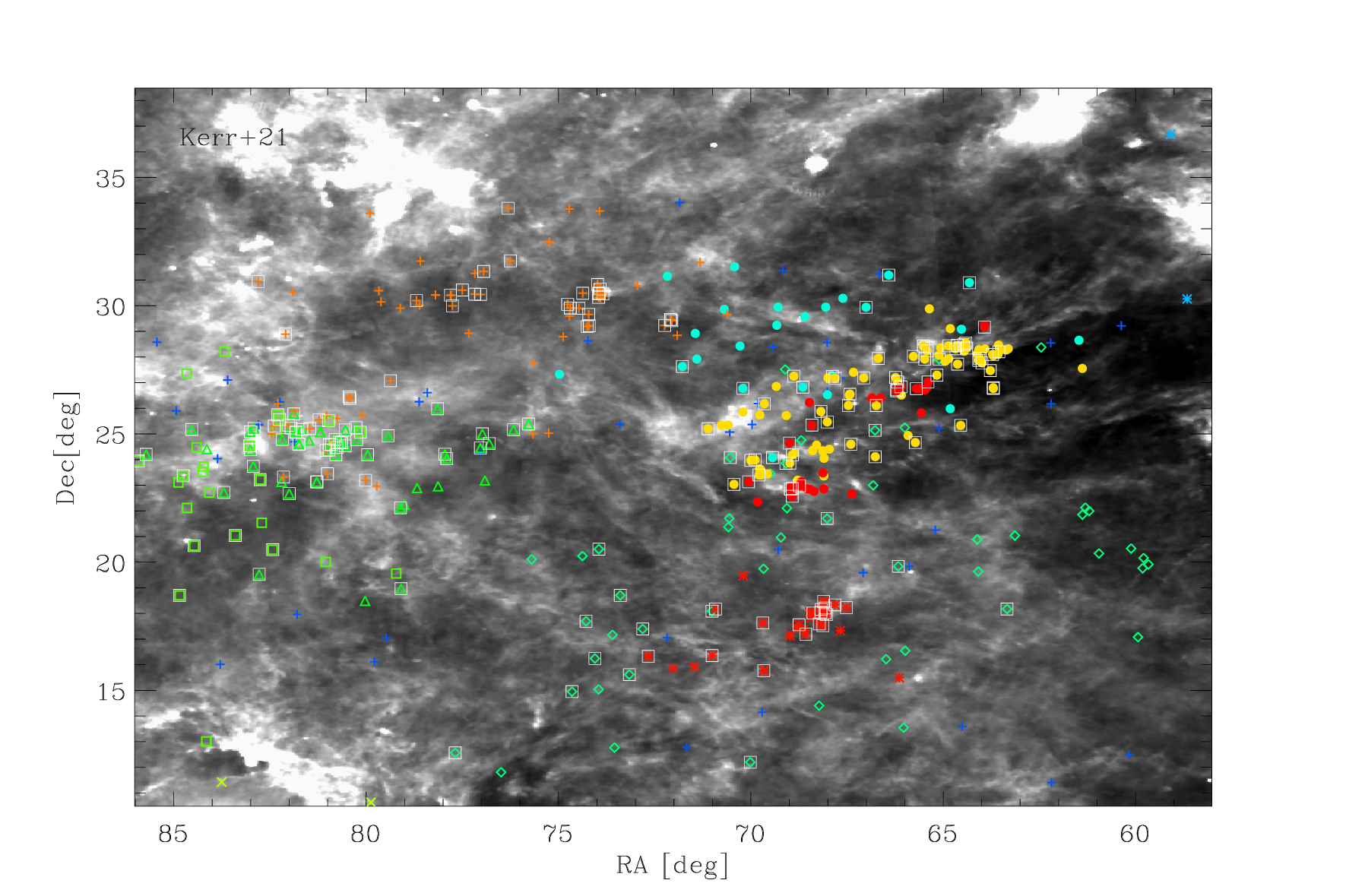}
      \caption{YSOs associated to the Taurus-Auriga Association, selected in this work (upper panel),
in  \citet{krol21} (middle panel) and \citet{kerr21} (lower panel). 
Colors and symbols indicate the substructures found by us with DBSCAN, those derived by the Gaussian
mixture model (GMM) in \citet{krol21} and those derived as EOM by \citet{kerr21}. 
White boxes in the middle and lower panels indicate the YSOs in common with our catalogue. YSOs are 
overplotted on a IRIS 100\,$\mu$m image \citep{mivi05}.}
         \label{radectaurus}
   \end{figure}

\begin{figure*}[!h]
\centering
\begin{minipage}{0.32\linewidth}
\includegraphics[scale=0.3,angle=0]{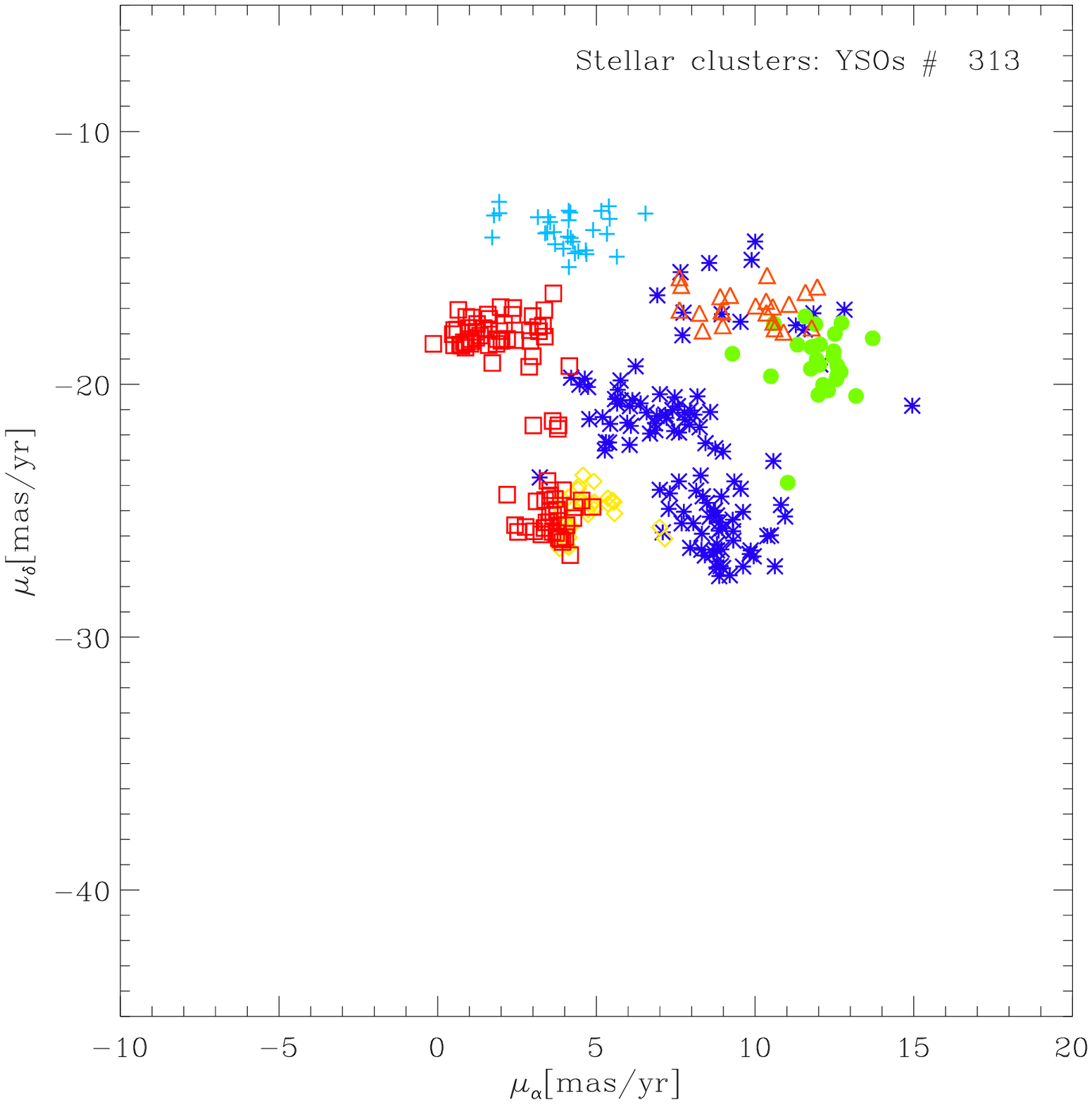} 
\end{minipage}
\begin{minipage}{0.32\linewidth}
\includegraphics[scale=0.3,angle=0]{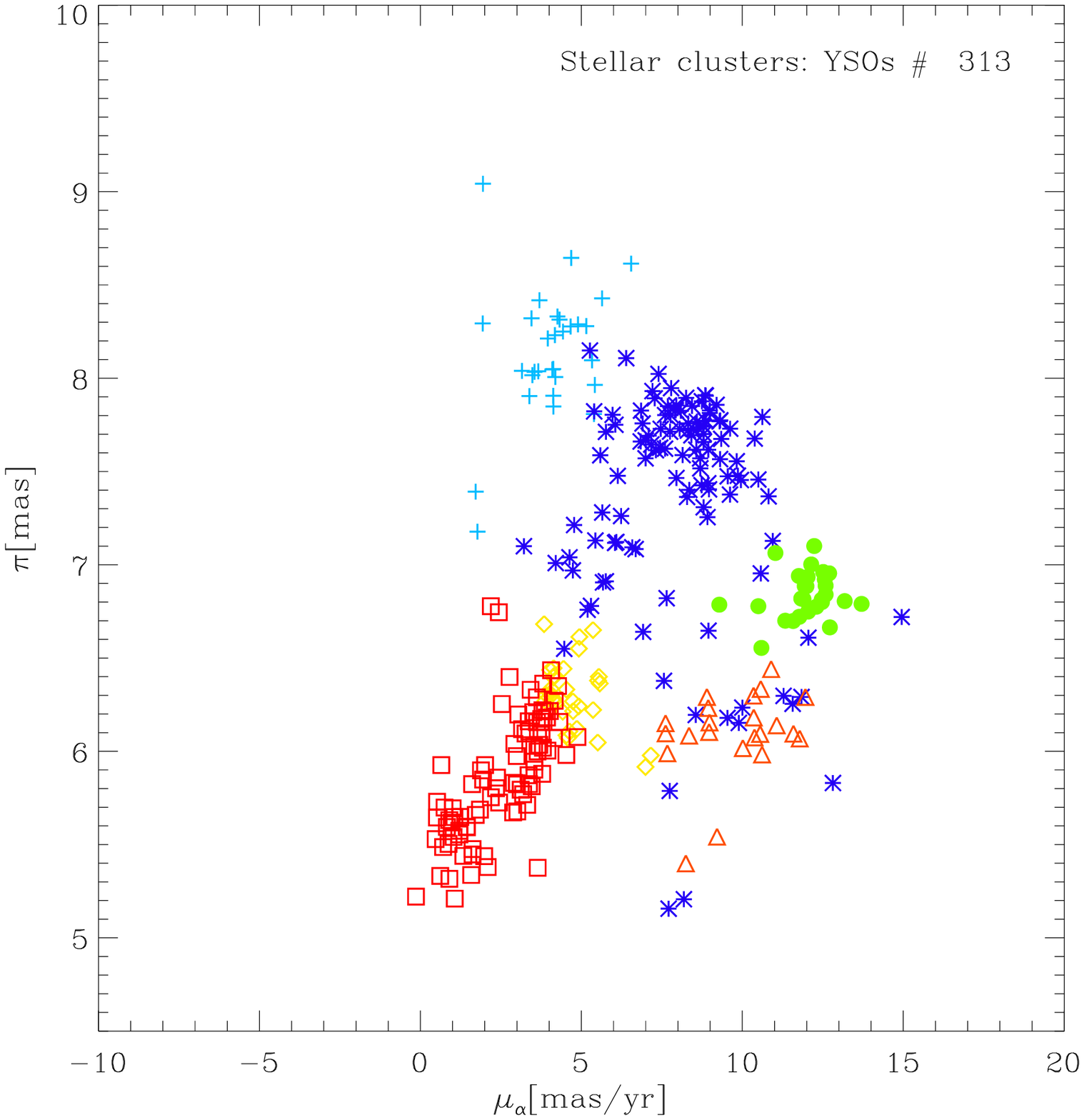} 
\end{minipage}
\begin{minipage}{0.32\linewidth}
\includegraphics[scale=0.3,angle=0]{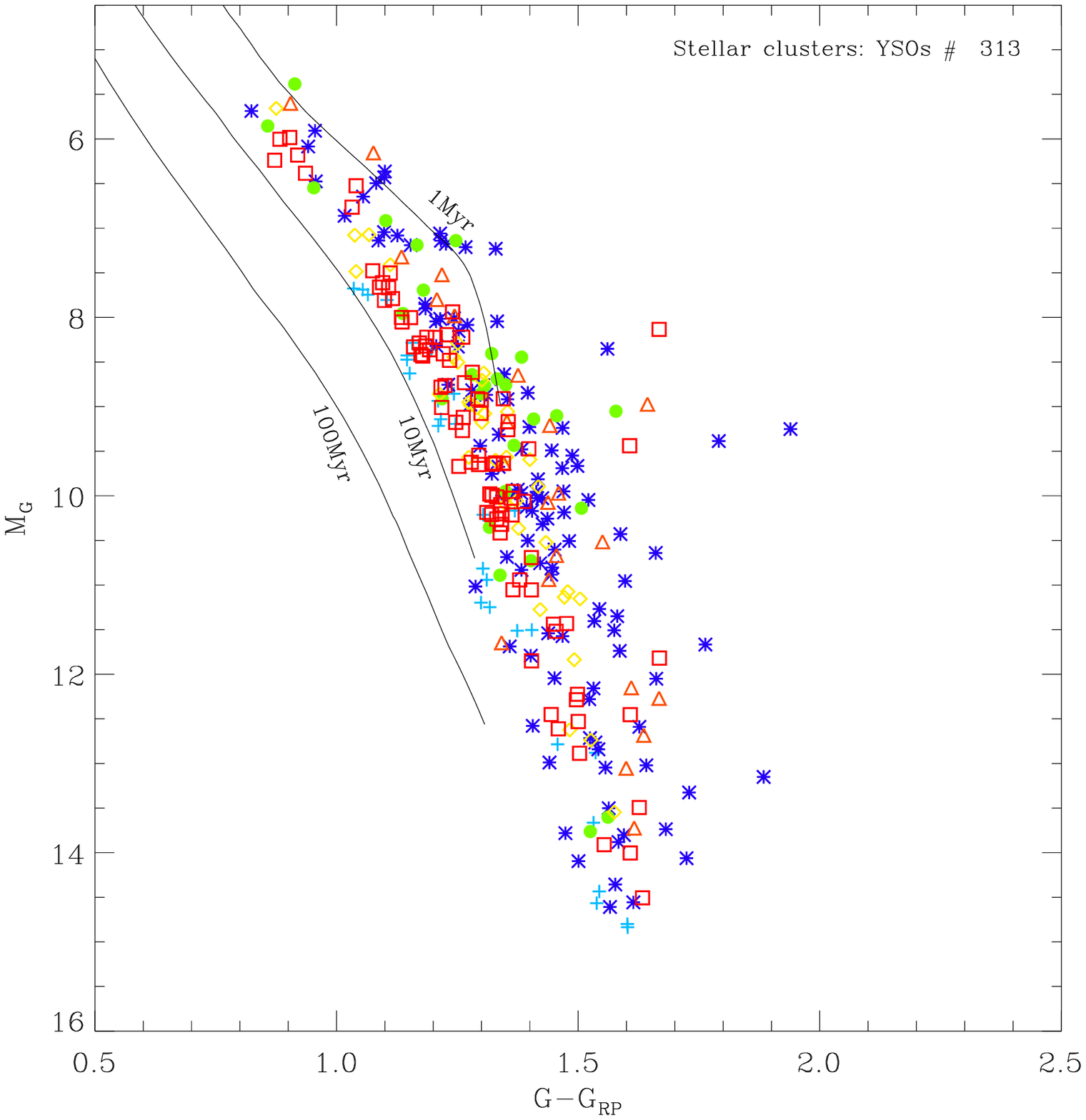} 
\end{minipage}
\caption{Proper motions in RA and Dec, parallaxes and CAMDs of the  clusters associated to the Taurus-Auriga
complex. Three representative  solar metallicity isochrones 
from the Pisa models are
also shown. Symbols and colors are as in Fig.\,\ref{radectaurus}.}
\label{taurusall}
\end{figure*}

The Taurus-Auriga complex is one of the nearest active SFRs of low mass stars to which
many works have been dedicated. 
In this region, we identified several substructures, as can be seen from Fig.\,\ref{radectaurus}.
In order to identify the YSOs associated to the Taurus-Auriga association,
we imposed the upper distance limit equal to 225\,pc, as done in \citet{krol21},
and restricted  the spatial region in the range 58.0\deg$<RA<$86.0\deg\, and 10.5\deg$<Dec<$38.5\deg.
We  considered the  sub-structures whose members are all within these
 limits.
With these conditions we identified a total of 313 YSOs associated to 6 substructures.
% see radec_taurus.pro output ind_taurus, cl_dist_mch__
The spatial distributions are  shown in Fig.\,\ref{radectaurus}, while PMs, parallaxes and the CAMD
are shown in Fig.\,\ref{taurusall}.  

The members in the southwest subregion (light blue plus symbols in the Figures) are distributed quite close to the
10\,Myr isochrone and then they could be part of an older population unselected by us for the photometric cut
we used. But, with the exception of this,
the members of the other substructures show the typical distribution of stars in PMS. 
The PMs of the substructures are quite well distint, as well as parallaxes, suggesting a complex 
3D structure with the known core including members in the region 63.0\deg$\lesssim RA\lesssim$70.0\deg\, and 
23\deg$\lesssim Dec\lesssim$28\deg (blue star symbols in the Figures), 
being also on the close side (median distance equal to 132\,pc ). The easternmost and most populated
substructure  (red square symbols in the Figures) is, instead, the most distant (median distance equal to 171\,pc).
A marginal evidence of age spread, as found in \citet{krol21}, is also found with our analisys but 
our results cannot be considered conclusive being based on reddening uncorrected photometry. 

\citet{krol21} compiled, very recently, the most complete
and inclusive census of members of this region found in the literature. Among these, 587 
objects have {\it Gaia} EDR3 counterparts, with 528 having a full astrometric solution.
% see mch_edr3_Krolikowski21.pro ; lore_sel 437
Using the {\it Gaia} EDR3 identification
number given in the  \citet{krol21} table, 
we matched the list of the 437 Taurus members in the  \citet{krol21} table
that are included in the photometric limits imposed in our work,
with the YSOs  with $t<10$\,Myr (i.e. classified  with
flag from 1 to 28),
% see mch_edr3_Krolikowski21.pro 
and found 202  objects in common,
that amount to about 46\% (202/437) of the \citet{krol21} list and 65\% (202/313)
% see radec_taurus.pro output ind_taurus,
 of our list of YSOs in this region. 
We note that the \citet{krol21} list has been obtained from the compilation of previous works, 
including local spectroscopic and IR data surveys that
 do not homogeneously cover the entire region as we have done  with {\it Gaia} data. 
For example, many  % stelle non matchate dei cluster_id_merge 774 e 2484 (579 e 572)  (265 e 260 in old paper)
% help,help,ind_sfr =313 ; matched are 202 quindi 313-102=111
  of the 111 YSOs not included in the  \citet{krol21} table
belong to the clusters 579 and 572 in Table\,\ref{printtexclusters} that includes 88 and 30 YSOs
(red squares and light blue symbols in Fig.\,\ref{radectaurus}, top panel), 
in two  sub-regions poorly covered by 
\citet{krol21}. 

We compared the list of YSOs in Taurus also with the list of members identified as
 excess of mass (EOM) by \citet{kerr21} using {\it Gaia} DR2 data. Details on the match
 with our catalogue are given in Sect.\,\ref{compareallsky}. As in our case,
 \citet{kerr21}  found substructures beyond the distance of known members.
 To perform a consistent comparison, we  restricted the \citet{kerr21} catalogue 
 in RA, Dec and distance, as done for our catalogue. Those identified as EOM in this region
  are 429.
  Among these, we considered the ones with $G>7.5$ 
 and $M_G>5$, to match the same photometric region adopted for our catalogue. 
Of the 409 \citet{kerr21} YSOs that % MCH_COMP in radec_taurus_kerr21
 meets these conditions, we found that 197  % aa in radec_taurus_kerr21
 (about 48\%) are in common 
 with our list of YSOs. We note that 
 a rigorous consistent comparison is very hard to perform,
since it strongly depends not only on the adopted clustering  techniques
but also on the subsample of {\it Gaia} data that is selected as  starting point of the subsequent analysis. 
 %see radec_taurus_kerr21.pro

The Taurus region is a well known complex structure in which the membership has been very hard to
achieve due to its  large spatial extent and  strong obscuration  by the  
nebula.  The comparison we have done is sufficient to assert 
 that about 50\% of the selected YSOs in this region are already found in other surveys
 and that they are distributed in sub-structures that are consistent with those 
 found in other works, in particular with the results presented by \citet{kerr21},
 that homogeneously cover the entire region. 
 
%%%%%%%%%%%%%%%%%%%%%%%%%%%%%%%%%%%
  
%%###########################################%%%%%%%%%%%%%%%%%%%%%%%%%%%%%%%%%%%%%%%%%%%%%
\subsection{The Orion Complex\label{orionsect}}
     \begin{figure}
   \centering
\includegraphics[width=9cm]{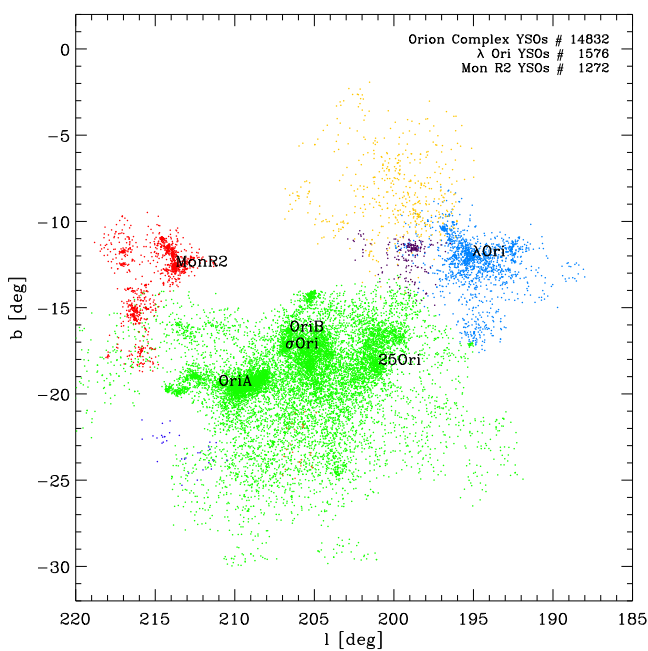}
      \caption{Spatial distribution in  Galactic coordinates of the 
       YSOs associated to the Orion Complex. YSOs identified in the 7 substructures
        are drawn with different symbols and colors.} 
        %Symbol sizes in the different 
        %substructures are weighted to the number of
       %cluster members.}
         \label{orionlb}
   \end{figure}
\begin{figure*}[!h]
\centering
\begin{minipage}{0.32\linewidth}
\includegraphics[scale=0.3,angle=0]{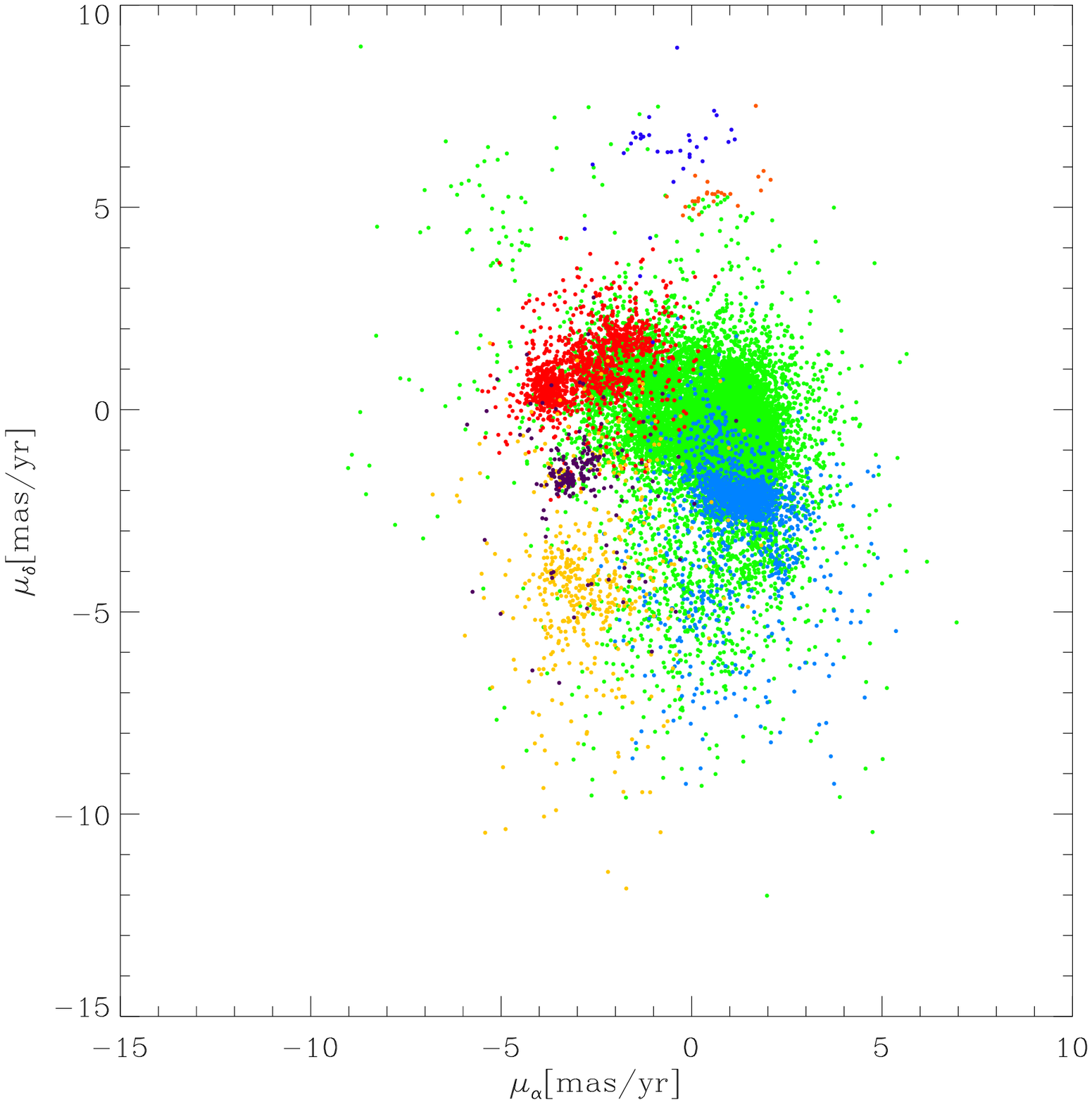} 
\end{minipage}
\begin{minipage}{0.32\linewidth}
\includegraphics[scale=0.3,angle=0]{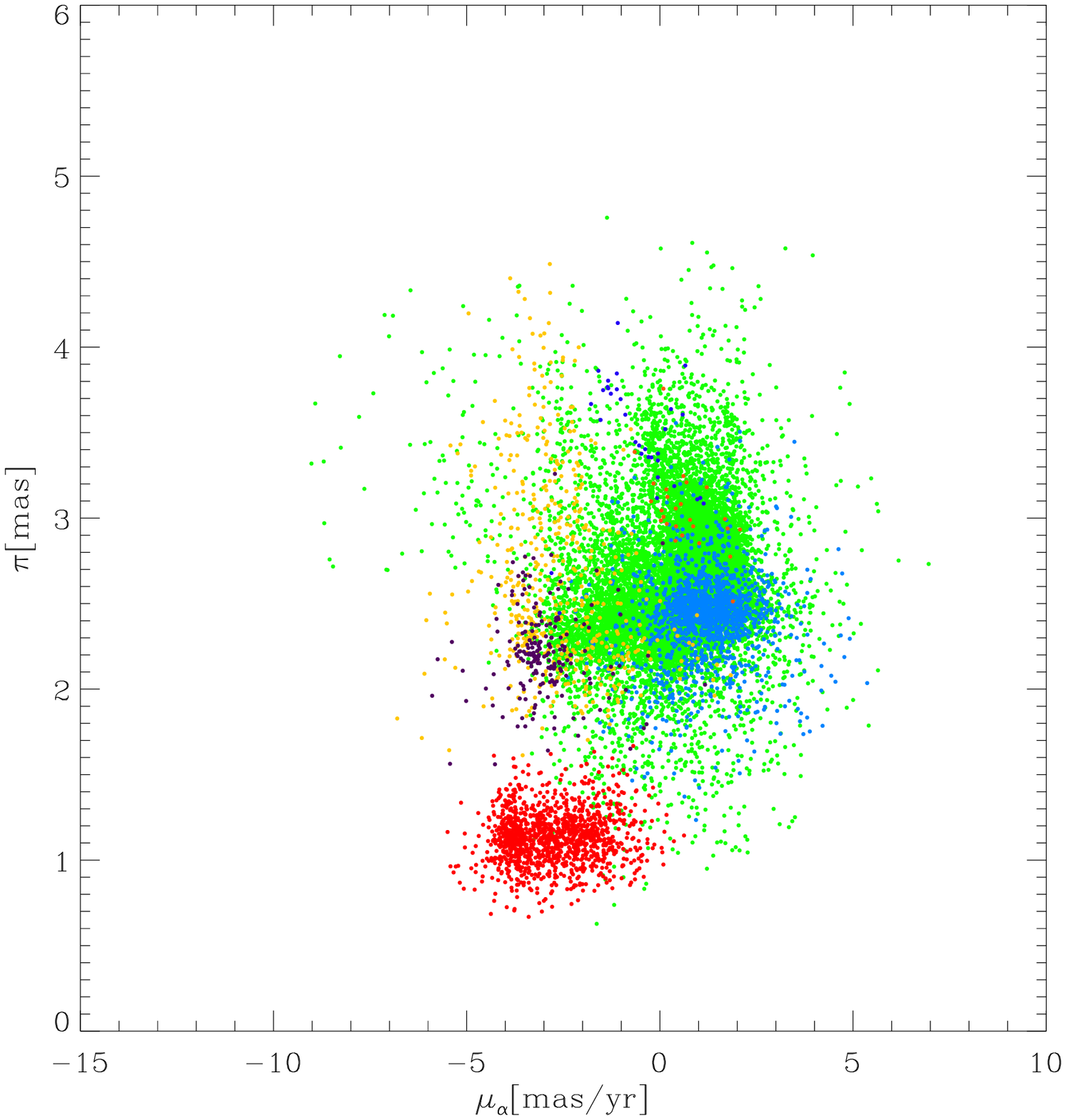} 
\end{minipage}
\begin{minipage}{0.32\linewidth}
\includegraphics[scale=0.3,angle=0]{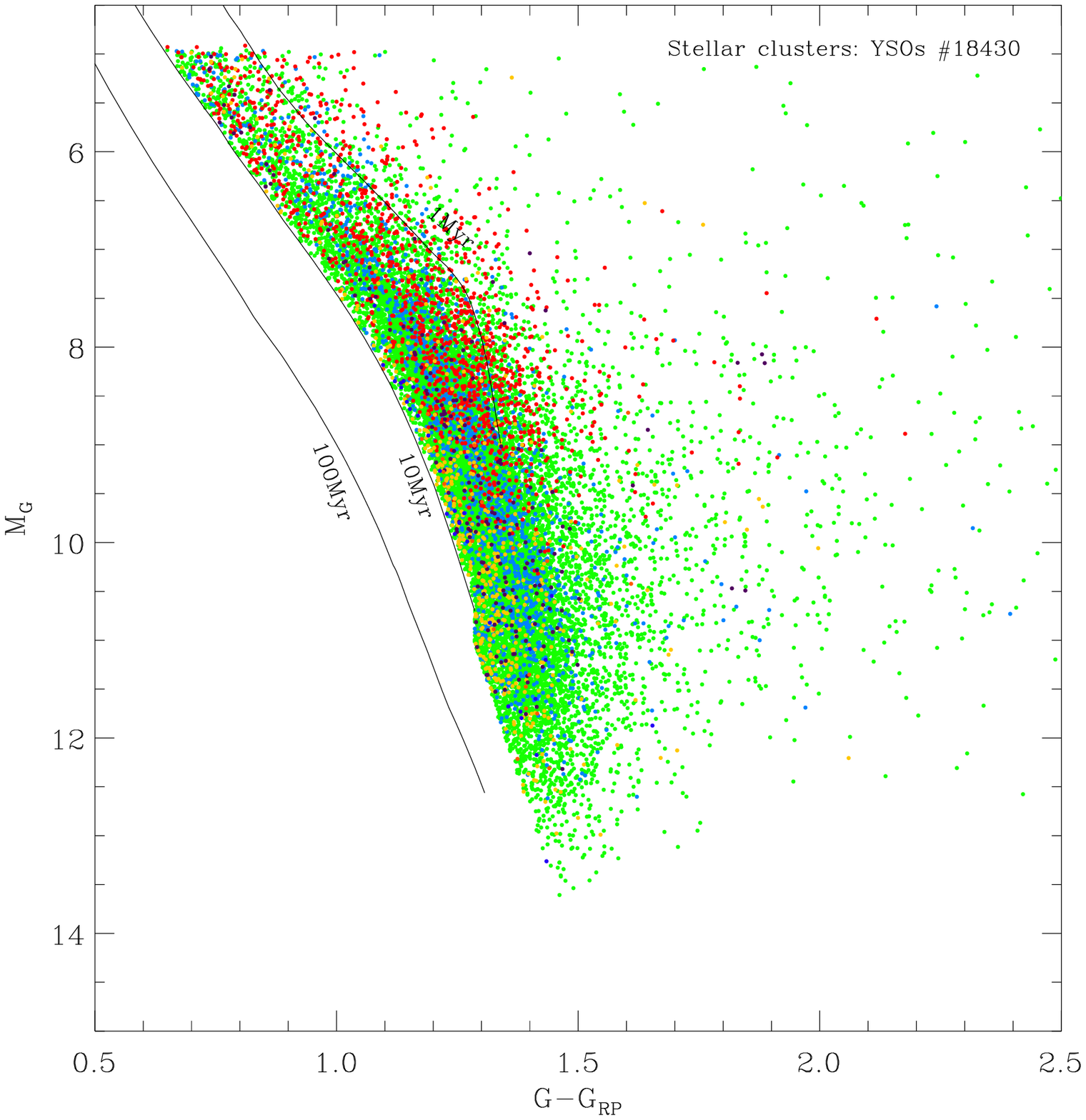} 
\end{minipage}
\caption{Proper motions in RA and Dec, parallaxes and CAMDs of the YSOs associated to the Orion
complex. The symbol colors of the sub-clusters are as  
in Fig.\,\ref{orionlb}. Three representative  solar metallicity isochrones 
from the Pisa models are also shown.}
\label{orionall}
\end{figure*}  

\begin{figure}
  \centering
\includegraphics[width=9cm]{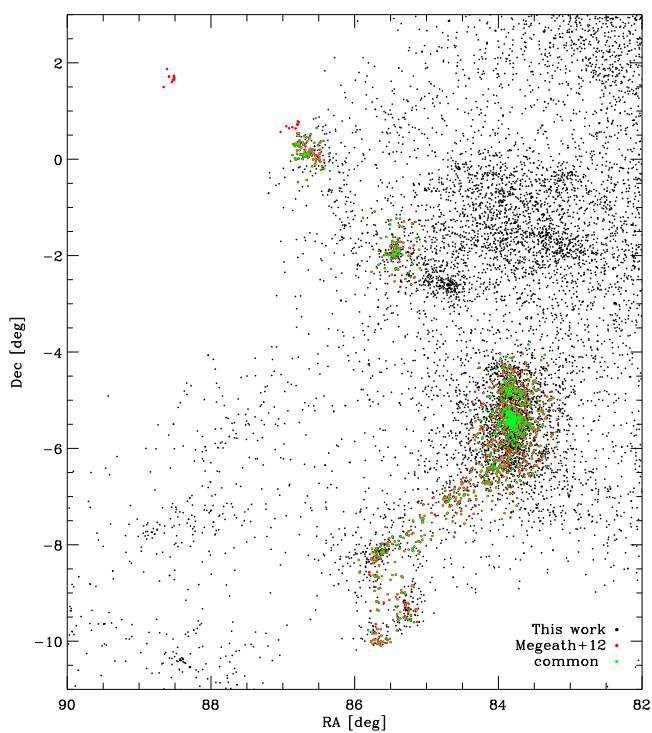}
      \caption{Spatial distribution of  the Orion YSOs
         compared to YSOs found in \citet{mege12}, indicated as black and red symbols, respectively.
          YSOs in common to the two catalogues
        are drawn as green symbols.}
         \label{radecorionmegeath}
   \end{figure}
YSOs associated to the  Orion complex have been identified  by selecting  objects with
  $75$\deg$<RA<90$\deg\, and $-11$\deg$<Dec<10$\deg.
 % and, as in \citet{koun19},
% distance $d<500$\,pc.
%radec_orion.pro
%xmin=75
%xmax=90
%ymin=-11
%ymax=10
%Then we extracted the list of clusters associated to these YSOs and  retained only those
%for which more than 90\% of the members. %  have distance $d<500$\,pc. 
In this way, 
%we found 17\,680 YSOs % see radec_orion.pro ind_sfr
we found 18\,840 YSOs % see radec_orion.pro ind_sfr
associated to 7  substructures with $t\lesssim10$\,Myr. % help,cl_id_mch
%, the main Orion Complex and $\lambda$\,Ori.
These are shown in Fig.\,\ref{orionlb}
 \footnote{ For a direct visual comparison,
spatial limits of the figure are the same used in Fig.\,1 of \citet{koun19}.},
 where we note
the presence of already known substructures  such as $\lambda$ and $\sigma$\,Ori, ONC, 25\,Ori.
All the main sub-structures covering the Orion A and B Nebulae have been merged by our procedure
in a single complex including \orionab\, YSOs and further  \orionlambda\, YSOs 
in the $\lambda$\,Ori cluster.
%Additional more extended substructures in the east part of the region are also identified.
The most distant cluster associated to Monoceros R2 (Mon-R2) is not part of the close Orion Complex
 and includes 1\,272 % see radec_orion.pro 
 %usando dist_lim=2000 ; come in kounkel 2019
%dist_inf=500
YSOs with a mean distance of 897\,pc ($\sigma=$123\,pc).

Figure\,\ref{orionall}  shows proper motions and parallaxes of the
substructures found in the Orion area. In particular, the proper motions  show
a very complex kinematic pattern of 
the subclusters in this region. %, with a distint velocity pattern of the more extended eastern
% populations. 
%  In addition we note how several small subclusters are actually well separated from the
% Orion complex kinematic pattern and some of them are significantly closer to the Sun ($d\sim 200-250$\,pc).
However, a detailed analysis of the Orion kinematics is beyond the aims of this work.

Figure\,\ref{orionall}  also shows  the CAMD of the populations associated to Orion.
Even though we can not rigorously interpret it, being 
 our data  not corrected for reddening,
we note  an apparent large age spread for all the populations.
%However, members of $\lambda$\,Ori and Mon R2 appear to be, on average, younger than YSOs in the main Orion Complex.

%%%%%%%%%%%%%%%%% Kounkel
We compare our findings in the Orion Complex region with the \citet{koun19} catalogue. 
Details on the match between the two catalogues are given in Sect.\,\ref{compareallsky}.
To retrieve the YSOs identified by \citet{koun19} in the Orion Complex,
we considered from their Table\,2, the 16 structures (Theia groups) falling in the Orion 
region as defined above. The YSOs  in the \citet{koun19}   % see orion_kounkel19.pro
and \citet{koun20} catalogues   % see mch_edr3_kounkel20.pro
associated to the Theia groups of the Orion Complex are 11\,882 and 10\,373, respectively.
Those in common with the list of Orion members found in this work  are
% 8\,256  (69\%)  % see orion_kounkel19.pro
 7\,983  (67\%)  % see orion_kounkel19.pro
 and 7\,822 (75\%). % see mch_edr3_kounkel20.pro
 
%%%%%%%%%%%%%%%%% MEGEATH
The Orion Complex has been extensively investigated with Spitzer IR data.
For example, the \citet{mege12} catalogue includes 3\,479 YSOs
 stars\footnote{retrieved at
 http://astro1.physics.utoledo.edu/$\sim$megeath/Orion/The\\
 \_Spitzer\_Orion\_Survey.html} that cover a quite extended region of the Orion A and B nebulae.
 Using the   cross-match service provided by CDS, Strasbourg, 
 and a matching radius of 1\sec, we found that 2\,612 IR sources from the \citet{mege12} catalogue 
have a {\it Gaia} EDR3 counterpart. From this sample, we considered only those with photometric and astrometric
 restrictions  given in equation \ref{reqdata}, with $G$-$G_{\rm RP}>0.58$ and in the range 
 203\deg$<l<216$\deg\, and $-30$\deg$<b<30$\deg, that amount to 1\,667 YSOs. % see ext radec_orion.pro
Of these, 1\,561 ($\sim$94\%) are identified by us members of the Orion Complex.
 The spatial distributions of our members and
 those found in \citet{mege12}
are shown in Fig.\,\ref{radecorionmegeath}. 
This high percentage proves that {\it Gaia} data have an efficiency in accurately diagnosing membership of YSOs in SFRs
%in optical bands of the adopted DBSCAN method,
comparable to that of IR  data.
%for which accurate membership was determined only with IR or X-ray data or spectroscopic data. 
If we consider the subsample of 2\,612 \citet{mege12} objects 
 with Gaia counterparts, and assume that it includes only genuine YSOs (i.e. 0\% of contamination),
 we can conclude that  our completeness level is about 60\%. This value
 is the result of the restrictions we imposed to our initial data set to reduce the contamination level.
 We note that we can have a significant bias against (missed) binary stars. In fact, if we only 
 discard the condition $RUWE<1.4$, and retain the other conditions,
 the \citet{mege12} YSOs Gaia counterparts are 1\,953 and this implies that  
 286 YSOs (1953-1667),
 i.e. about 14\% of the total sample (very likely binary systems), are missed in our data set.
We do not attempt to estimate the fraction of false positives that could be included in our sample
by considering the \citet{mege12} catalogue since it includes mainly Class\,II stars, i.e. YSOs with 
IR excess emission from the circumstellar disk and it is therefore incomplete for the Class\,III stars,
that do not show excess emission in the IR.

\subsection{The interstellar dust free SFR NGC\,2362 }
\begin{figure}
   \centering
\includegraphics[width=9cm]{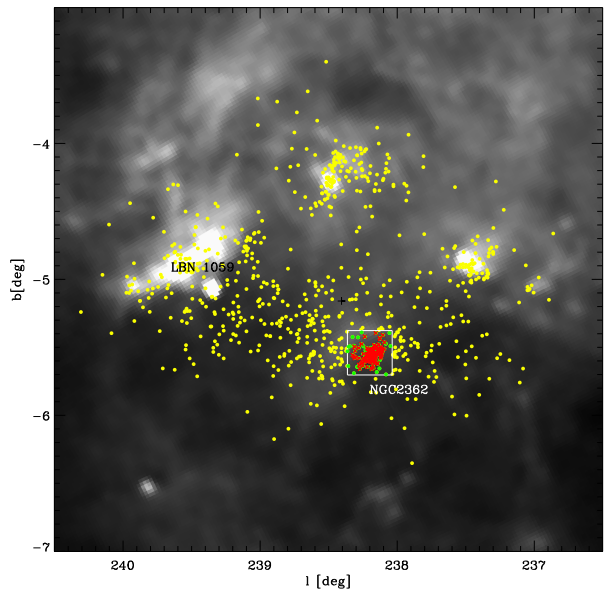}
      \caption{Spatial distribution in  Galactic coordinates of the 
       YSOs associated to NGC2362 (yellow symbols).  YSOs falling in the box of 
       16.9\sec$\times$16.9\sec\, equal to the Chandra-ACIS field (white box) 
       used in \citet{dami06a}
       are indicated as green symbols. YSOs in common with \citet{dami06a} X-ray detections 
       are indicated as red symbols. Objects are 
overplotted on a IRIS 100\,$\mu$m image. }
         \label{ngc2362lb}
   \end{figure}

At a distance of 1\,354$\pm$192\,pc, NGC\,2362 is a SFR characterised by a very low and
uniform reddening, estimated to be E(B-V)=0.1 \citep{moit01}.
For this reason, the cluster  shows a small
spread in the optical V vs. V-I diagram, as found by \citet{moit01} and confirmed by \citet{dami06a}
and this enables to constraint  the duration of
the star formation process that in this region has been about 1-2\,Myr  \citep{dami06a}.
This result has been derived on the basis of a  Chandra-ACIS X-ray observation,
pointed in the cluster, from which a list of very likely members has been obtained.
As for the case of NGC\,2264, this cluster has been found with our procedure
in a region  more extended than that investigated by \citet{dami06a}. 
The 879 YSOs compatible with being members of NGC\,2362 are  plotted in 
Fig.\,\ref{ngc2362lb}. Within the nominal cluster center  l=238.2\deg, -5.54\deg\,
\citep{dami06a},
we found 150 candidate members while the others are mostly concentrated 
around the three  bumps visible in the IR image. A further subgroup of cluster members
shows an aligned spatial distribution roughly going from NGC\,2362 to the
 \hii region LBN\,1059.
 
 To compare our data, with the list of 387 X-ray members by \citet{dami06a},
 we cross-matched this list with the {\it Gaia} EDR3 catalogue, by using the
 cross-match service provided by CDS, Strasbourg, adopting a matching radius of 0.5\sec.
 We find that 294 of them have a single {\it Gaia} EDR3 counterpart but
% dami=mrdfits('../TABLES/damiani06_gaiaedr3.fits',1)
%help,dami 
%see ngc2362_lb.pro
those that are compliant with our initial data set restrictions 
and falling in the PMS region of the CAMD compatible with ages $<10$\,Myr
are 129. Among these, %see ngc2362_lb.pro 
118, i.e. $\sim91$\%, 
are in common with our list of YSOs.  This fraction confirms that,
even though our list of YSOs is incomplete, due to the significant 
fraction of  members discarded a priori with the adopted data 
 restrictions, in the adopted photometric ranges, the efficiency 
of our method in detecting very likely members is very high,
if we consider
that  X-ray detections select YSOs without any bias based on the
stellar evolutionary status (Class\,II or III YSOs), with an high efficiency
in the spectral types (G to M) we are working on. 

Within the Chandra-ACIS field of view, we selected a total of 150 YSOs, 
% see ngc2362_lb.pro acis
and 32 of them (21\%) are not X-ray detected. % (acis=150, damiani_comm=118, 150-118)
X-ray  detections found in \citet{dami06a}  are complete 
%down to V=19 that, at the cluster distance, roughly corresponds to $M_V\sim8.34$
%and $M_G\sim7.7$. 
for masses larger than 0.4\,M$_\odot$, that, assuming the cluster age of 4-5\,Myr \citep{mayn08},
corresponds to $M_G\simeq 7.5$. % see glim in ngc2362_mg_grp.
%By considering that YSOs in NGC\,2362 have been detected
%by us down to $M_G\sim9$, i.e. more than 1 magnitude deeper, 
By considering that more than 50\% of these X-ray undetected YSOs are fainter than
this limit % see   ngc2362_mg_grp perc median at  7.4
and that most of them are located far from the cluster center, where the
Chandra-ACIS  spatial resolution is lower, 
we are confident that 
the 32 X-ray undetected YSOs classified by us are likely members. 
\begin{figure*}[!h]
\centering
\begin{minipage}{0.32\linewidth}
\includegraphics[scale=0.3,angle=0]{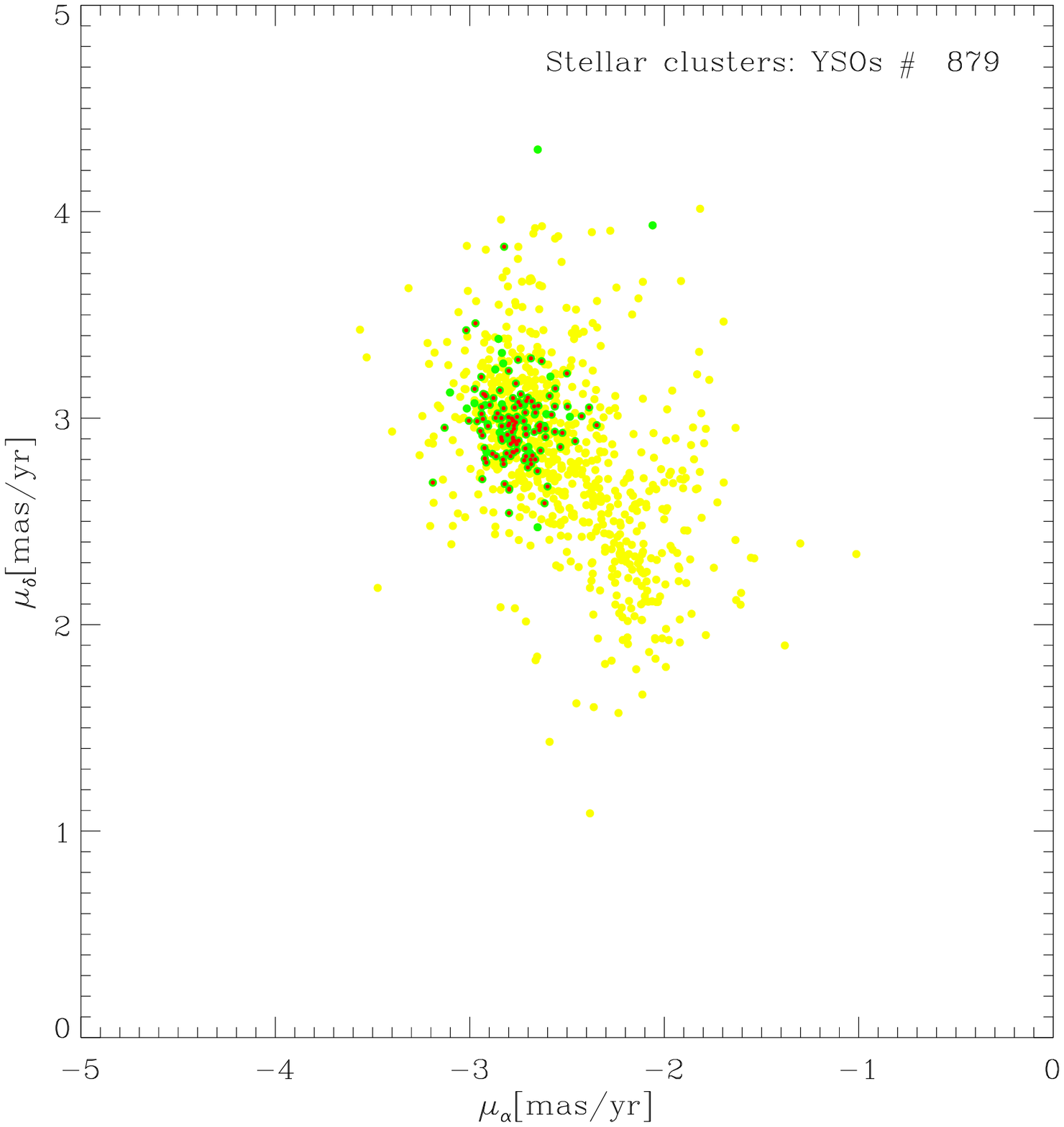} 
\end{minipage}
\begin{minipage}{0.32\linewidth}
\includegraphics[scale=0.3,angle=0]{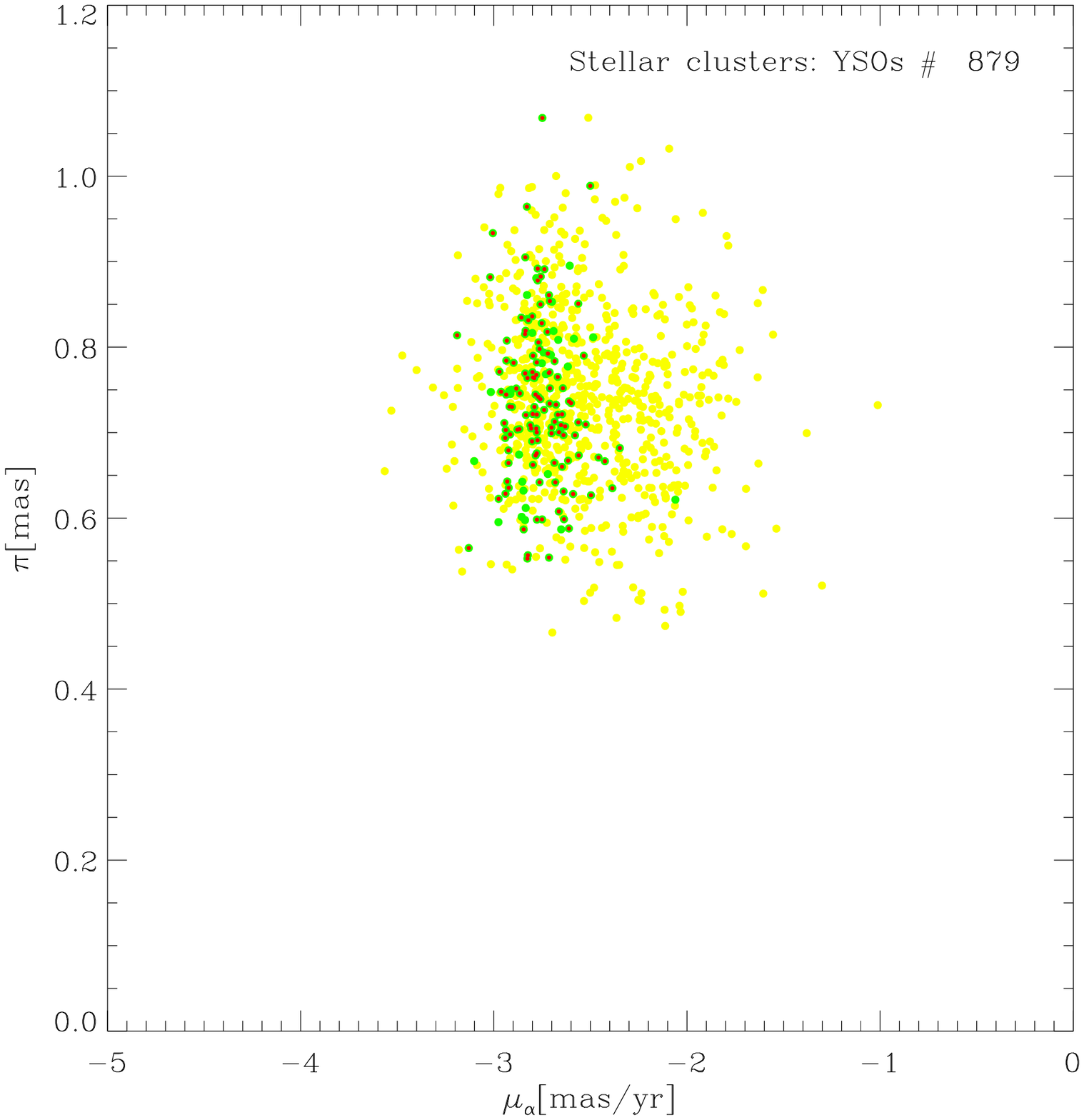} 
\end{minipage}
\begin{minipage}{0.32\linewidth}
\includegraphics[scale=0.3,angle=0]{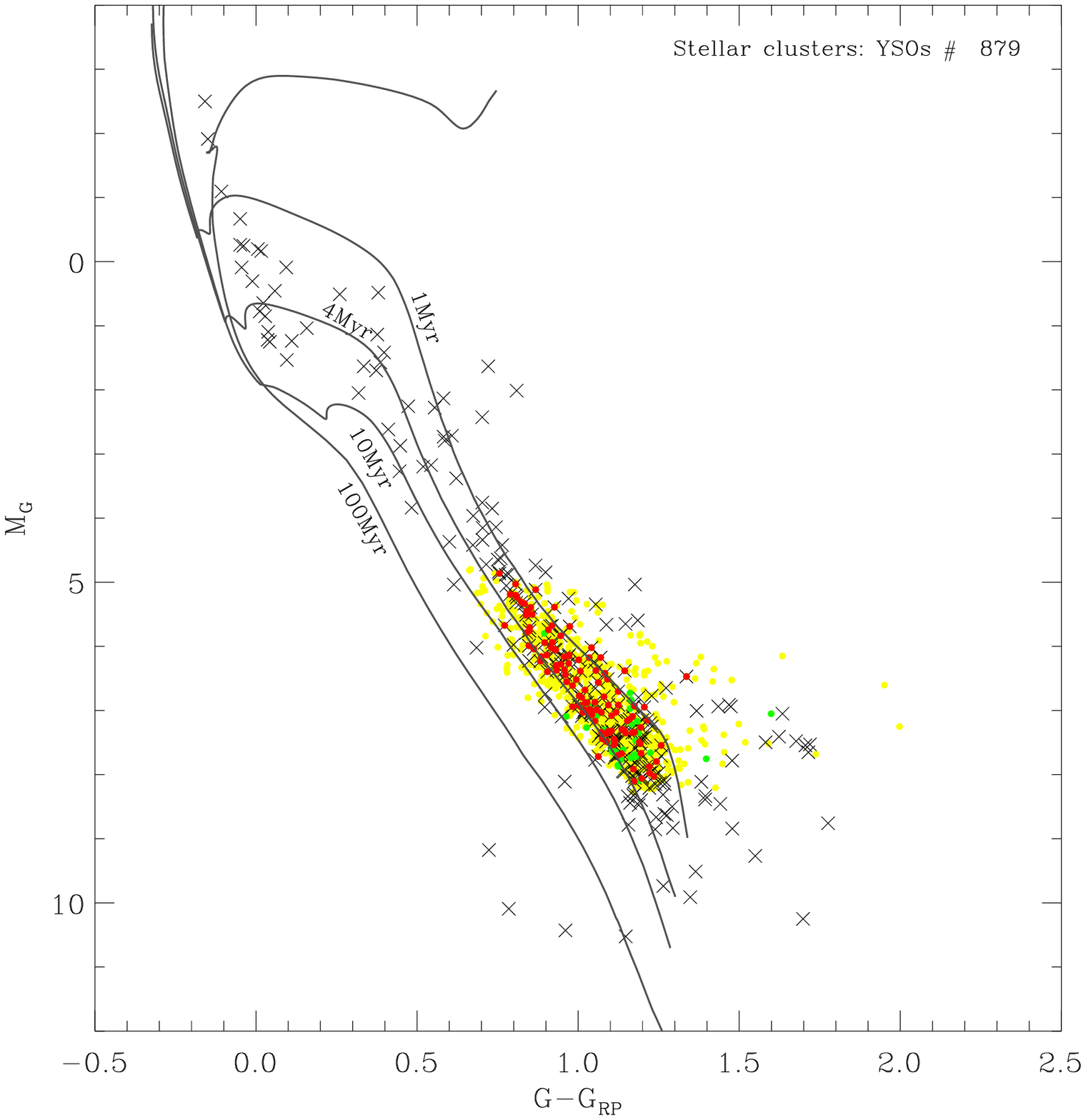} 
\end{minipage}
\caption{Proper motions in RA and Dec, parallaxes and CAMDs of the
YSOs associated to NGC\,2362. Symbol colors are as in Fig.\,\ref{ngc2362lb}.
Black x symbols are the X-ray detected YSOs by \citet{dami06a}.  
Four representative solar metallicity isochrones from the Pisa models 
are also shown. }
\label{ngc2362all}
\end{figure*}

As for the other clusters, we investigated proper motions, parallaxes and CAMD,
that are shown in Fig.\,\ref{ngc2362all}. The proper motion scatter plot
indicates that actually the distribution of YSOs falling in the Chandra-ACIS
is more concentrated than that of the overall cluster, which shows an inclined trend.
 This confirms  that the entire cluster is characterised by a kinematic structure
 slightly more complex than that of the subgroup of YSOs falling around the
  known cluster center. The parallax values indicate that all the detected
  YSOs are located at consistent distances. 

%The small luminosity spread observed in the V vs. V-I diagram of \citet{dami06a}
%is not, instead, observed  in the CAMD , where also
%the same YSOs detected by \citet{dami06a} with a {\it Gaia} EDR3 counterpart
%are not distributed along a tight sequence but show a quite large spread
%as the YSOs identified in this work. 
We note that to reduce the observed spread in the $M_G$ vs. $G-G_{\rm RP}$ diagram 
shown in Fig.\,\ref{ngc2362all}, in the computation
of $M_G$, we used the median cluster distance, rather than the individual member distances.
The residual observed luminosity spread 
 in the $M_G$ vs. $G-G_{\rm RP}$ diagram 
is likely due to reddening effects not corrected here and that, on the contrary, are
very small in the
 V vs. V-I diagram, where the reddening vector is almost parallel to the cluster sequence
 in the low mass range (see Fig.\,4 in \citet{dami06a}).

%suggests that other observational factors, such as large uncertainties 
%in the Gaia parallaxes\footnote{In the computation of
%$M_G$ for the \citet{dami06a} sample, we did not applied the  parallax zero point 
%correction recommended in \citet{lind21},
%as we did for our data, and this could contribute to enhance the observed luminosity
%spread for the \citet{dami06a} members.},
%or the low sensitivity of the $G-G_{\rm RP}$ to stellar ages, or the metallicity effects
% can be at the origin of the observed spread.

\subsection{Comparison with literature all-sky star cluster catalogues\label{compareallsky}}

Using the {\it gaiaedr3.dr2\_neighbourhood} table in the Gaia archive, 
we retrieved the {\it Gaia} DR2 identification
number of the candidate YSOs selected in our work and thus, using these IDs, we performed the
match with the \citet{kerr21} list, including 30\,518 YSOs within 333\,pc and 
selected with {\it Gaia} DR2. We found
a total of \kerrmch\,  objects in common.
Among these, \kerrmchsfr\, are associated to clusters with $t\lesssim 10$\,Myr 
and \kerrmchmidy\, are 
associated to clusters with 10\,Myr$\lesssim t\lesssim 100$\,Myr of our catalogue. 

Using the same procedure for the 
\citet{koun19} and \citet{koun20}  catalogues, that include 288\,370 entries
up to 1\,Kpc and 987\,376 entries, up to 3\,Kpc,
respectively,
 we find a total of  \kounkelamch\, and \kounkelbmch\,  YSOs in common.
  Those associated to SFRs with $t\lesssim10$\,Myr 
 (young  clusters with 10\,Myr$\lesssim t\lesssim100$\,Myr) are 
   \kounkelamchsfr\,  (\kounkelamchmid) for the \citet{koun19} list,
 and \kounkelbmchsfr\, (\kounkelbmchmid) for the \citet{koun20} list.
 The remaining common stars have been discarded by us since they do not belong to
 the young age range.
We note that, while in the contest of the entire all-sky 
catalogue  the fraction of   objects in common is very low 
 ($\sim$13\% % 38567/288370.=0.13374138
 and $\sim$4\%), %42350./987376= 0.042891461
in the region of the Orion Complex   it is 67\% and 75\% (see Sect.\,\ref{orionsect}). 
However, we note that our catalogue does not include the string-like massive clusters
at $\gtrsim1$\,kpc,  with spatial distribution aligned to the GP, 
that we discarded in the cluster validation phase 
(see Sect.\,\ref{validationsect}).
Instead, the \citet{koun19}  and \citet{koun20} lists include
 many of these objects
and this could explain the  low fraction of  objects in common with respect to the entire catalogue. 
In addition, the restrictions to the initial data set are very different. 
For example, we imposed a photometric selection in the extinction uncorrected
 $M_G$ vs. $G-G_{\rm RP}$ CAMD,
aimed to select mainly objects with age $<10$\,Myr. On the contrary, in the \citet{koun19}
and \citet{koun20}
catalogues, no photometric selection has been applied and in fact these catalogues
includes up to $\sim1$\,Gyr old clusters. 

We compared our results also with the list of 2\,017 
%can_cluster=mrdfits('../TABLES/cantat20.fits',1)  
%help,can_cluster 2017
% help,where(can_cluster.ageNN GT 0. )   ; 1867
clusters recently 
published by \citet{cant20} that includes 234\,128 cluster members.
They used the most complete list of clusters from the literature and assigned them cluster membership 
using the UPMASK procedure \citep{kron14}, that is based on the compactness of 
the groups in the positional space and it is constrained to a fixed field of view.
For 1\,867 of these clusters, reliable  parameters have been derived. 

We find that the  members presented  by \citet{cant20} in common with our
catalogue are 12\,438. Those associated to SFRs ($t\lesssim 10$\,Myr), 
young  (10\,Myr$\lesssim t\lesssim 100$\,Myr) and old ($t\gtrsim 100$\,Myr) clusters  are
 6\,788, 2\,519 and 2\,109, respectively, corresponding to 66, 38 and 76 clusters in our catalog,
 in the same age ranges.
%restore,'../SAVEFILES/cantat_mch.save',/v
%help,ind_mch_catalog,ind_mch_data,sub1_28,sub29_34,sub35_51
%IND_MCH_CATALOG LONG      = Array[12438]
%IND_MCH_DATA    LONG      = Array[12438]
%SUB1_28         LONG      = Array[6788]
%SUB29_36        LONG      = Array[2351]
%SUB37_52        LONG      = Array[2936]
 They belong to 311 clusters of the   \citet{cant20} list\footnote{This apparent discrepancy is due to
 the fact that our catalogue includes merged clusters that
 can include more than one cluster in the \citet{cant20} list.}
%IDL> restore,'../SAVEFILES/cantat_mch.save',/v
%IDL> help,IND_MCH_CATALOG
%IND_MCH_CATALOG LONG      = Array[12438]
% plot_cluster_cantat.pro
% help,help,un_can_mem
%UN_CAN_MEM      LONG      = Array[311]
%print,minmax(can_cluster_comm.plx)  
%     0.617000      7.41500
with parallaxes $>0.617$\,mas, that, approximatively, corresponds  to the maximum distance of YSOs
identified in our work. In the \citet{cant20} catalogue, the cluster members with $\pi>0.617$,
$G>7.5$ and $M_G>5.0$  
%can_cluster=mrdfits('../TABLES/cantat20.fits',1)    
%can_mem=mrdfits('../TABLES/cantat20_members.fits',1)
%can_mem.cluster=strtrim(can_mem.cluster,2)
%mg=can_mem.gmag-5*alog10(1000./can_mem.plx)+5.
%help,where(can_mem.plx GT 0.671 and can_mem.gmag GT 7.5 and can_mem.gmag LT 20.5 and mg GT 5)
%<Expression>    LONG      = Array[49074]
are in total 49\,074 and therefore we find that only $\sim25$\% % 0.25345397
of YSOs detected by us are in common with  \citet{cant20}.
Using the ages derived in \citet{cant20}, we find that 226 of the matched
% see plot_cluster_cantat with the condition agenn GT 7
 clusters are older than 10\,Myr. 

\begin{figure}
   \centering
\includegraphics[width=9cm]{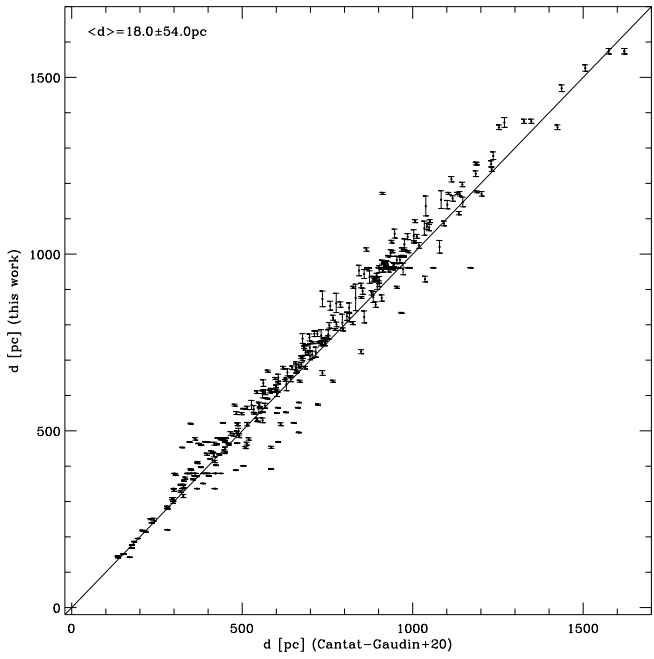}
      \caption{Comparison between the cluster distances derived by \citet{cant20}
      and those derived in this work. The line with slope one is shown for guidance.}
         \label{comparedistcantat20}
   \end{figure}

For the 331  clusters in common, we compared the distances assigned by \citet{cant20} computed
as the inverted parallaxes  of the value given for each cluster and the mean distance obtained
by us, computed from the  weighted mean parallaxes. Errors on the parallaxes were
computed as the error on the mean. The comparison is shown in Fig.\,\ref{comparedistcantat20},
where the mean and standard deviation of the  residuals between the two measurement sets 
are also given.  
The two determinations are consistent, even
 though there is a bias  due to
the different {\it Gaia} data realises adopted in our work (EDR3)  and in \citet{cant20} (DR2).
%\end{appendix}
%\begin{appendix}

%
\end{appendix}

\end{document}